\documentclass[11pt,aps,prd,nofootinbib,superscriptaddress,preprintnumbers]{revtex4}

\usepackage{graphicx,epsfig}
\usepackage{amsmath}
\usepackage{amssymb}
\usepackage{slashed}
\usepackage{color}
\usepackage{multirow}

\newcommand{\bea}{\begin{eqnarray}}
\newcommand{\beq}{\begin{equation}}
\newcommand{\eea}{\end{eqnarray}}
\newcommand{\eeq}{\end{equation}}

\newcommand{\lsim}{\raise0.3ex\hbox{$\;<$\kern-0.75em\raise-1.1ex\hbox{$\sim\;$}}}
\newcommand{\gsim}{\raise0.3ex\hbox{$\;>$\kern-0.75em\raise-1.1ex\hbox{$\sim\;$}}}

\newcommand{\unity}{{\hbox{1\kern-.8mm l}}}

\newcommand{\ord}[1]{\mathcal{O}({#1})}

\begin{document}

\title{Flavour and Collider Interplay for SUSY at LHC7}
\author{L.~Calibbi}
\email{calibbi@mppmu.mpg.de}
\affiliation{Max-Planck-Institut f\"ur Physik (Werner-Heisenberg-Institut),
F\"ohringer Ring 6, D-80805 M\"unchen, Germany}
\author{R.~N.~Hodgkinson}
\email{robert.hodgkinson@uv.es}
\affiliation{Departament de F\'{\i}sica Te\`orica and IFIC, Universitat de 
Val\`encia-CSIC, E-46100, Burjassot, Spain}
\author{J.~Jones~P\'erez}
\email{joel.jones@lnf.infn.it}
\affiliation{INFN, Laboratori Nazionali di Frascati, Via E.~Fermi 40, I-00044 Frascati, Italy}
\author{A.~Masiero}
\email{antonio.masiero@pd.infn.it}
\affiliation{Dipartimento di Fisica, Universit\`a di Padova, via F.~Marzolo 8, I--35131, Padova, Italy}
\affiliation{INFN, Sezione di Padova, via F.~Marzolo 8, I--35131, Padova, Italy}
\author{O. Vives}
\email{oscar.vives@uv.es}
\affiliation{Departament de F\'{\i}sica Te\`orica and IFIC, Universitat de 
Val\`encia-CSIC, E-46100, Burjassot, Spain}

\preprint{IFIC/11-61, FTUV-11/1028, MPP-2011-126}

\begin{abstract}
The current 7 TeV run of the LHC experiment shall be able to probe gluino and squark masses up to values larger than 1 TeV. Assuming that hints for SUSY are found in the jets plus missing energy channel by the end of a 5 fb$^{-1}$ run, we explore the flavour constraints on three models with a CMSSM-like spectrum: the CMSSM itself, a Seesaw extension of the CMSSM, and Flavoured CMSSM. In particular, we focus on decays that might have been measured by the time the run is concluded, such as $B_s\to\mu\mu$ and $\mu\to e\gamma$. We also analyse constraints imposed by neutral meson bounds and electric dipole moments. The interplay between collider and flavour experiments is explored through the use of three benchmark scenarios, finding the flavour feedback useful in order to determine the model parameters and to test the consistency of the different models.
\end{abstract}

\maketitle

%\tableofcontents

\section{Introduction}

The current run of the LHC, operating at $\sqrt{s}=7$ TeV (LHC7 from now on), has been very successful
and new results with significant improvements on our previous knowledge of the
electroweak symmetry breaking (EWSB) energy region have been presented in the
summer conferences and published afterwards. Although no Higgs or new physics (NP) signals
have been found so far, LHC experiments have started to provide stringent test
to the relevant extensions of the Standard Model (SM), in particular, to low-energy Supersymmetry (SUSY).
The most recent analyses based on $\sim 1$ fb$^{-1}$ of integrated luminosity already exclude squarks and gluinos, 
if close in mass, up to $\sim 1$ TeV \cite{cms-atlas}. Even though the relevant SUSY parameter space has not been fully explored yet, 
it is well known that naturalness arguments point towards a quite light spectrum of the SUSY particles.\footnote{After the first 
SUSY searches at the LHC, the fine-tuning price of the constrained MSSM is already at the percent level \cite{Strumia:2011dv}.} 
Therefore, if SUSY is indeed realised at low energies as a theory related to the EWSB, 
we can expect some SUSY partners to lie in the reach of LHC7. This assumption is the starting point of the present work.

The Minimal Supersymmetric Standard Model (MSSM), understood as the minimal 
supersymmetric version of the SM with respect to the number of fields, can be considered
 the main goal of the early SUSY searches at the LHC. 
However, to explore the parameter space in a
completely general MSSM is a formidable task given the huge number of unknown
parameters (the number of ``free'' parameters in the MSSM being bigger than one hundred). 
This is the main reason why most of the analyses of SUSY phenomenology at 
collider experiments are made in the framework of the
so-called Constrained Minimal Supersymmetric Standard Model
(CMSSM), where all the soft SUSY breaking terms are assumed to be degenerate
and only five parameters (beyond the already known SM parameters) are enough to
define completely the model. In fact, this simple structure of the soft breaking
terms can be found in some theoretical models, although complete flavour universality 
of the soft SUSY breaking is not generally true in most
supergravity and string-inspired models. 

On the other hand, the CMSSM can be hardly considered a satisfactory NP model, 
since it does not account neither for the peculiar hierarchy pattern we observe in
the fermion masses and mixing nor for the generation of light neutrino masses, which
call for extension of the CMSSM.  
The existence and smallness of neutrino masses make it natural to extend the CMSSM in the direction of a supersymmetric seesaw mechanism \cite{typeI}, while the flavour puzzle of the Standard Model suggests a flavour model based on horizontal flavour symmetries \cite{Froggatt:1978nt}.
It is however very unlikely that, even if SUSY is indeed
discovered, the LHC alone can shed light on the flavour structure of the SUSY partners sector. Such extensions of the CMSSM
(and to some extent the CMSSM itself) can be better probed and discriminated from other scenarios by means of the interplay between the LHC
and the experiments dedicated to measure rare flavour changing or CP violating processes. While we are witnessing a rich 
experimental activity with experiments already running like MEG and LHCb, with others under construction or development like SuperBelle, 
the Super Flavour Factory, $\mu\to e$ conversion experiments, on the theoretical side it is crucial to address the question 
of the complementarity of direct and indirect searches as a tool for discriminating among different SUSY models. For a detailed discussion
of this problem we refer to \cite{Raidal:2008jk}. Here we concentrate on the specific case of the present bounds and the expected near future
sensitivity of LHC7.

In the view of what discussed above, in this work we mainly aim at answering the following questions: 
(i) what is the impact of the present limits on the SUSY spectrum provided by the LHC experiments on the capability
of flavour experiments to observe deviations from the SM predictions?  
(ii) assuming that SUSY is indeed in the reach of the LHC7 searches, what are then the most promising channels probed at low-energy 
experiments where to look at in order to get more information about the fundamental SUSY parameters and discriminate among models?

In this paper, we try to address the above questions considering the interplay between direct SUSY searches at LHC7 and indirect searches in flavour and CP violation (CPV) experiments, within the CMSSM and some classes of phenomenologically motivated extensions of the CMSSM, which usually predict larger flavour and CPV effects than the CMSSM itself, namely SUSY seesaw and a Flavoured CMSSM, i.e.~an extension of the CMSSM with non-trivial flavour structures in the sfermion sector.
In particular, we are going to study the relevant flavour observables as predicted in the above mentioned models within the supersymmetric parameter space accessible at LHC7 up to 5 fb$^{-1}$ of integrated luminosity, that is the amount of data which should be collected and analysed between 2011 and 2012. Even though it is likely that by the end of 2012 both ATLAS and CMS might collect up to 10 fb$^{-1}$, we consider this conservative value since at the moment an increase of the centre of mass energy up 8 TeV during 2012 is still under consideration. Notice we do not intend to do a fit of LHC data within these models, but to point out that all of these give similar collider signatures, and show that indirect experiments can help us differentiate between MSSM versions.

The rest of the paper is organised as follows: in the next section the current status and prospects of SUSY searches at LHC7 are briefly reviewed;  
the models we are going to study are discussed in section \ref{models}; the numerical analysis is presented in section \ref{interplay}; conclusions are drawn in
section \ref{conclusion}.

\section{SUSY searches at LHC7: current bounds and prospects}
\label{sec:LHC}
As a proton-proton collider, the search for SUSY at the LHC hinges upon the production of coloured squarks or gluinos through strong interactions.  These heavy new particles are then expected to decay rapidly through a decay chain ending in one or more jets (by conservation of QCD colour), possibly leptons, and the lightest supersymmetric particle (LSP), which is stable under the assumption of R-parity conservation. 
To have escaped astrophysical and cosmological observations, the LSP must be a neutral particle and will hence escape the detector unseen.  This leads to an apparent imbalance in the measured part of the final state, so that the characteristic signature of a SUSY process at the LHC is the ``missing energy'' (more accurately, missing momentum), $\slashed E_T$, associated with the unobserved LSP leaving the detector.

We have only a limited knowledge of the partonic initial state in proton-proton collisions, in particular the boost of the partonic centre-of-mass frame relative to the lab-frame is not known and it is therefore impossible to reconstruct and measure the longitudinal momentum of any escaping neutral particle, such as the LSP.  For this reason, LHC searches for SUSY look for an excess of events with two or more high transverse energy jets along with significant missing transverse energy.

In typical SUSY scenarios, the lightest coloured SUSY particle is usually either a gluino or a stop.  The cross-section for gluino pair-production at the LHC depends at leading order on only a single unknown parameter, the mass of the gluino itself.  Indeed, the gluino production cross-section 
at LHC7 is known to be large, $\sigma(pp\to\tilde{g}\tilde{g})\gtrsim 100$~fb for gluinos of mass $m_{\tilde g}\lesssim 700$~GeV ~\cite{Baer:2010tk}.  
By contrast, the stop pair production cross-section is around two orders of magnitude smaller for stop masses $m_{\tilde t}\sim m_{\tilde g}$ and so plays essentially no role in early LHC SUSY searches.

Although typically significantly heavier than the top- and bottom-squarks, the squarks of the first generation \textbf{do} play a role in the search strategy.  The cross-section for production of the up- and down-squarks is boosted by the presence of a t-channel gluino exchange diagram, with the valence quarks of the incoming protons in the initial state.  The production cross-section for these first-generation squarks therefore depends on two parameters at leading order, the mass of both the exchanged gluino and that of the outgoing squarks themselves ($m_{\tilde q}$).
The most stringent bounds on these parameters
are already due to LHC results, which have overtaken the previous Tevatron and LEP bounds. 
Both ATLAS and CMS have published searches based on their initial $1$~fb$^{-1}$ data sets analysing the multi-jets plus missing energy and 0-lepton final state \cite{cms-atlas}. The exclusion limits of both experiments are currently comparable. They set a mass limits of about 
1 TeV for  $m_{\tilde g}\simeq m_{\tilde q}$.

In the event of a discovery, of course, we can also expect to learn something more about the scale of SUSY.  Inclusive searches in the $0$-lepton channel look for events with two or more high-energy jets plus missing transverse energy. 
The analysis of the excess in this channel can give us information on the SUSY 
spectrum responsible for the signal, or more exactly on the gluino and first generation squarks. The main observable for this at LHC7 is the 
``effective mass'' $M_{\rm eff}$, defined as:
\begin{equation}
M_{\rm eff} \equiv \sum_i p_T^i + \slashed E_T\ ,
\end{equation}
where the sum of transverse momenta runs on the four most energetic jets in the event. 
It has been shown that the peak of the $M_{\rm eff}$ distribution is correlated to the mass of the parent SUSY particles produced in the initial
hard scattering, in particular to the mass parameter $M_{\rm SUSY}$, defined as $M_{\rm SUSY} \equiv {\rm min}(m_{\tilde g}, m_{\tilde q})$. 
It results that typically $M^{\rm peak}_{\rm eff}~ \simeq~ 1.6 \times M_{\rm SUSY}$~\cite{Hinchliffe:1996iu,Tovey:2000wk}, so that the $M_{\rm eff}$ distribution can be used to extract information on the mass scale of the SUSY particles produced at the LHC. As we are going to discuss, this in combination 
with the information from indirect SUSY searches can be crucial to constrain the parameter space in case of a positive signal of SUSY is 
observed at LHC7.

\section{Models}
\label{models}
\subsection{CMSSM}

The first model we consider is the so-called
Constrained MSSM, which assumes perfect universality of the soft-breaking terms. 
The CMSSM is completely defined by four new parameters and one sign:
the universal scalar mass $m_0$, the common gaugino mass $M_{1/2}$, the universal trilinear coupling $A_0$, the sign of the Higgsino mass parameter $\mu$ and ratio of the two Higgs vacuum expectation values (vevs) $\tan\beta$.   
These parameters are specified at a large scale, that is usually taken to be the Grand
Unification scale, $M_{\rm GUT}$, and the low-energy SUSY spectrum is then obtained by solving the renormalisation group equations (RGEs).

As discussed in the previous section, the relevant quantities for the jets + missing energy searches at LHC
are the mass of first generation squarks $m_{\tilde q}$ and the gluino mass $m_{\tilde g}$.
Within the CMSSM, $m_{\tilde q}$ and $m_{\tilde g}$
are essentially governed by the high scale parameters $m_0$ and $M_{1/2}$ only, whilst remaining relatively insensitive to $A_0$, $\tan\beta$ and the sign of $\mu$. For this reason, the hadron collider bounds on the SUSY particle masses, as well as the LHC7 discovery prospects, can be conveniently interpreted as contours in the $m_0$--$M_{1/2}$ plane, independent of the values of the other CMSSM parameters. In our numerical study
presented in section \ref{interplay}, we are going therefore
to define a band in the $m_0$--$M_{1/2}$ plane to be explored at LHC7 and we study such region of the parameter space by 
randomly varying the remaining CMSSM parameters.

The renormalisation group (RG) running from $M_{\rm GUT}$ down to the electroweak scale generates a flavour structure for the sfermion masses, 
in the form of Minimal Flavour Violation (MFV)~\cite{D'Ambrosio:2002ex}. In this scenario, all flavour-violating terms are determined by the CKM elements and the third generation Yukawa couplings.
In principle, we should also add
two CP violating phases $\varphi_\mu$ and $\varphi_A$ if $\mu$ and
$A_0$ are complex parameters. These are usually constrained by the electric dipole moments (EDMs) of the electron and the neutron, and are studied separately in Section~\ref{CMSSM-Pheno}.

The CMSSM contributions to flavoured processes have two sources. First, as we have said, even though the CMSSM is completely universal at $M_{\rm GUT}$, small flavour off-diagonal entries in the soft mass matrices, proportional to the fermionic Yukawa couplings, are generated through the RG evolution from $M_{\rm GUT}$ to low energies. This shall generate small mixings for left-handed squarks. On the other hand, we always have chargino and charged Higgs contributions with flavour transitions controlled by the CKM matrix. This means that each point on the parameter space, apart from being subject to the LHC constraints, is also subject to flavour constraints. The most important processes that the CMSSM can contribute to are BR($b\to s\gamma$) and the muon anomalous magnetic moment, $(g-2)_\mu$. Another process that is becoming very important with the improvement of the experimental
constraints is $B_s \to \mu^+ \mu^-$. We will discuss all these processes and how they affect the CMSSM parameter space in the next section.\footnote{Another important constraint
comes from the comparison of the predicted dark matter abundance,
assuming is given by the neutralino, with the observed value by the
WMAP experiment, as well as from direct and indirect dark matter searches. For recent discussions of these constraints, 
see \cite{Profumo:2011zj,Buchmueller:2011ki}.}

\subsection{SUSY seesaw}
\label{sec:seesaw}
As widely discussed in the literature, the field content of the MSSM has to be extended, in order to account 
for neutrino masses and mixing. Here we are going to consider the simplest possibility, 
the so-called type-I seesaw mechanism \cite{typeI}, 
which requires the introduction of right-handed (RH) Majorana neutrinos ($N_i$) in the MSSM superpotential:
\begin{equation}
 W_{\mathrm{MSSM_{RN}}} = W_{\mathrm{MSSM}}
 + (Y_{\nu})_{ij}  N_i L_{j}  H_u  + (M_R)_{ij} N_i N_j 
\end{equation}
Since there is no gauge symmetry that
protects them, the RH neutrinos can get large Majorana masses
$(M_R)_{ij}$, breaking
the conservation of lepton number. When integrated out, they will give
rise to an effective light neutrino Majorana mass matrix:
\begin{equation}
\label{see-saw}
m_\nu = - Y_\nu^T M_R^{-1} Y_\nu \langle H_u \rangle^2\,,
\end{equation}
where $\langle H_u \rangle$ is
the vev of the up sector Higgs field.

We do not expect the SUSY searches at the LHC to be considerably modified with respect to the CMSSM by RG effects
induced by the presence of RH neutrinos. In fact, the gluino mass running is not modified (up to two loops) and 
squark masses can be affected only indirectly through 
modification of the RG running of $A_t$ and $m^2_{H_u}$. However, for particular regions of the parameter space,
these modifications can change the EWSB conditions.\footnote{Another consequence of introducing RH neutrinos is that this might destabilise the regions which provide a neutralino lightest supersymmetric particle consistent with the WMAP bounds on DM relic density, such as the focus point and A-funnel 
regions \cite{Calibbi:2007bk}.}

The main effect of the presence of RH neutrinos is given by the modification of the RG running of 
the left-handed slepton soft mass matrix $m^2_{\tilde L}$, such that, even starting with diagonal and universal 
scalar matrices at high energy, off-diagonal flavour mixing entries of $m^2_{\tilde L}$ are generated by the running above the 
RH neutrino mass scale \cite{Borzumati:1986qx}. 
In the basis where charged leptons are diagonal, these off-diagonal entries
are approximately given by the following expression:
\begin{equation}
\label{eq:m2l}
(m^2_{\tilde L})_{i\neq j} \simeq
- \frac{3m^2_0 + A^2_0}{8\pi^2} \sum_k (Y_{\nu})_{ik}^\dagger
(Y_{\nu})_{kj}
\ln \left(\frac{M_U}{M_{R_k}} \right)\,,
\end{equation}
where  $M_{R_k}$ is the mass of the $k$-th RH neutrino,
$M_U$ the energy scale at which universality conditions are imposed (in our case the GUT scale).
Terms in Eq.~(\ref{eq:m2l}) clearly determine a misalignment among lepton and slepton mass eigenstates
in the flavour space, inducing a contribution to lepton flavour violation (LFV) processes, such as $\ell_i\to\ell_j \gamma$, via
slepton-neutralino (or sneutrino-chargino) loops. 
Therefore, in the context of a SUSY seesaw model, we can consider in addition as key observables 
the rates of LFV processes, which are suppressed to vanishing values within the SM (and the CMSSM).
The rate for the processes $\ell_i\to \ell_j \gamma$ is given by \cite{hisano}
\begin{equation}
 \frac{{\rm BR}(\ell_i\to \ell_j \gamma)}{{\rm BR}(\ell_i\to \ell_j \nu \bar{\nu})} =
\frac{48 \pi^3 \alpha_{em}}{G_F^2} \left(|A_L^{ij}|^2 + |A_R^{ij}|^2\right)\,,
\label{eq:LFV-BR}
\end{equation}
where the amplitudes can be easily estimated 
in terms of the mass-insertion parameters (MIs), 
as usually defined as $\delta^f_{ij}= (m^2_{\tilde f})_{i\neq j}/ \bar{m}^2_{\tilde f}$ 
(with $\bar{m}^2_{\tilde f}$ being the average sfermion mass);
in SUSY seesaw models the main contributions are approximately given by the following expression~\cite{Hisano:2009ae}:
\begin{equation}
A_L^{ij} \simeq \frac{\alpha_2}{60 \pi} \frac{\tan\beta}{\tilde{m}^2} (\delta^e_{LL})_{ij}\,,
\label{eq:AL}
\end{equation}
where $\tilde{m}$ is the typical mass of the SUSY particles in the loop. On the other hand, the $A_R^{ij}$ amplitudes turn out to be negligible since $(\delta^e_{RR})_{ij}$ are vanishing in SUSY seesaw models.
From these equations, we can see that the ratio between the two of the most promising channels,
$\tau\to \mu\gamma$ and $\mu\to e\gamma$, is given by:
\begin{equation}
R_{\tau\mu} \equiv 
\frac{{\rm BR}(\tau \to \mu\gamma)}{{\rm BR}(\mu\to e\gamma)} \simeq
0.17\times \frac{|(\delta^e_{LL})_{23}|^2}{|(\delta^e_{LL})_{12}|^2}\,.
\label{eq:Rtm}
\end{equation}

It is clear that any estimate of $(m^2_{\tilde{L}})_{i\neq j}$ (and thus of $(\delta^e_{LL})_{ij}$)
would require a complete knowledge of the neutrino Yukawa matrix
$(Y_\nu)_{ij}$ which is not fixed by the seesaw equation, even with an
improved knowledge of the neutrino oscillation parameters,
as in Eq.~(\ref{see-saw}) there is a mismatch between the number
of unknowns and that of low energy observables.
For definitiveness, we are going to study two well-motivated scenarios.

{\bf Scenario (a).} In the first one we assume $Y_\nu \sim Y_u$, as expected 
in presence of an underlying Pati--Salam or SO(10) unification \cite{Masiero:2002jn}. 
A consequence of this assumption is that at least one entry in $Y_{\nu}$ results 
as large as the top Yukawas, so that sizeable
LFV entries might be generated from RG evolution as in Eq.~(\ref{eq:m2l}).
However, even if the eigenvalues of $Y_{\nu}$  are then related to the ones of $Y_u$,
the size of mixing angles in $Y_\nu$ are still uncertain. A way to bypass
the ignorance about the mixing is considering two extremal benchmark cases
\cite{Masiero:2002jn,Calibbi:2006nq}.
As a minimal mixing case we take the one in which the neutrino and the
up Yukawa unify at the high scale, so that the mixing is given by the
CKM matrix, in the basis where the lepton Yukawa $Y_e$ is diagonal; 
we refer to this case as `CKM-case'. As a maximal mixing scenario
we take the one in which the observed
neutrino mixing is coming entirely from the neutrino Yukawa matrix, so that
$Y_\nu = U_\mathrm{PMNS}^{\dagger} \cdot Y^{\mathrm{diag}}_u$, where $U_\mathrm{PMNS}$
is the neutrino mixing matrix. This is what we are going to call `PMNS-case'.

{\bf Scenario (b).}
The second case we are going to study can be better understood in terms of the Casas-Ibarra parametrisation \cite{Casas:2001sr}:
%
%\begin{equation}
% Y_\nu = \frac{1}{\langle H_u \rangle} U_{\rm PMNS} \mathcal{D}_{\sqrt{m_\nu}} R\, \mathcal{D}_{\sqrt{M_R}}\,,
%\end{equation}
\begin{equation}
 Y_\nu = \frac{1}{\langle H_u \rangle} \mathcal{D}_{\sqrt{M_R}}\,R\,\mathcal{D}_{\sqrt{m_\nu}} U_{\rm PMNS}^\dagger ,
\end{equation}
where $\mathcal{D}_{\sqrt{m_\nu}}$ and $\mathcal{D}_{\sqrt{M_R}}$ are diagonal matrices of the square roots of 
light and heavy neutrino masses respectively and the complex orthogonal matrix $R$ accounts for the mismatch
between seesaw and low-energy parameters.
For simplicity we are going to consider $R=\mathbf{1}$, which corresponds to a trivial flavour structure of $M_R$,
i.e. the leptonic mixing $U_{\rm PMNS}$ entirely provided by $Y_\nu$.
With this assumption, the mixing structure of 
$Y_\nu$ is fixed and the LFV effect of Eq.~(\ref{eq:m2l}) depends on the overall size of the Yukawas, 
namely on the $M_R$ scales. As a consequence, the experimental limits on LFV processes, such as $\mu\to e\gamma$ are
going to constrain the RH mass scales. For definitiveness, we are going to consider both cases of very hierarchical and almost 
degenerate RH neutrinos.  
\begin{table}[t]
\begin{center}
\begin{tabular}{|lc|cc|}
\hline
 \multicolumn{2}{|c|}{Scenario}  & $(\delta^e_{LL})_{ij}$ &  $R_{\tau\mu}$ \\  
\hline\hline
  \multirow{2}{*}{(a) $Y_\nu \sim Y_u$} & CKM case & 
$-{1  \over 8 \pi^2}~\frac{3 m_0^2 + A_0^2}{\bar{m}^2_{\tilde \ell}}~y_t^2 V_{ti} V_{tj} \ln{M_U \over M_{R_3}}$  & 
$0.17 \times \left| \frac{V_{tb}}{V_{td}}\right|^2 \simeq 2\times 10^3 $
\\
& PMNS case & $-{1  \over 8 \pi^2}~\frac{3 m_0^2 + A_0^2}{\bar{m}^2_{\tilde \ell}}~
y_t^2 U_{i 3} U_{j 3} \ln{M_U \over M_{R_3}}$  &  
$0.17 \times \left| \frac{U_{\tau 3}}{U_{e 3}}\right|^2 \simeq 1\div 15$
\\
\hline
\multirow{2}{*}{(b) $R=\mathbf{1}$} & Degenerate $M_R$ & $-{1  \over 8 \pi^2}~\frac{3 m_0^2 + A_0^2}{\bar{m}^2_{\tilde \ell}}~\left(\sum_k
y_{\nu_k}^2 U_{i k} U_{j k}\right) \ln{M_U \over M_{R}} $  &  $\simeq 0.5 \div 10$\\
 & Hierarchical  $M_R$ & $-{1  \over 8 \pi^2}~\frac{3 m_0^2 + A_0^2}{\bar{m}^2_{\tilde \ell}}~
y_{\nu_3}^2 U_{i 3} U_{j 3} \ln{M_U \over M_{R_3}} $  &  $0.17 \times \left| \frac{U_{\tau 3}}{U_{e 3}}\right|^2 \simeq 1\div 15$\\
\hline
\end{tabular}
\end{center}
\caption{\label{tab:seesaw-deltas} Estimated values of the LFV parameters $(\delta^e_{LL})_{ij}$ and 
the ratio $R_{\tau\mu}$ for different SUSY seesaw scenarios. The LH slepton masses is approximately 
$\bar{m}^2_{\tilde \ell}\simeq m_0^2 + 0.5 M_{1/2}^2$. 
In the $Y_\nu \sim Y_u$ scenario, the mass $M_{R_3}$ of $N_3$ is given by:
$M_{R_3} \approx {m_t^2(M_U)}/{4\, m_{\nu_1}}$ (CKM~case), 
$M_{R_3} = {m_t^2(M_U)}/{m_{\nu_3}}$ (PMNS~case),
where $m_t(M_U)\approx 0.5 \,m_t(m_t)$. In the $R=\mathbf{1}$ scenario $y_{\nu_3} = \sqrt{m_{\nu_3} M_{R_3}}/v_u$.
To estimate  $R_{\tau\mu}$, we take $U_{e 3}\simeq 0.08\div 0.28$ and $m_{\nu_1} \simeq 0\div 0.1$ eV. }
\end{table}

In Section~\ref{SUSYSeeSaw-Pheno}, we are going to study the interplay among SUSY searches at the LHC and
searches for LFV processes (in particular $\mu\to e\gamma$ at the MEG experiment) for
the two scenarios described above. 
In Table~\ref{tab:seesaw-deltas}, we summarise the estimates for the LFV parameters 
$\delta^e_{ij}$ we can obtain from the leading-log expression of Eq.~(\ref{eq:m2l}) in the different scenarios
we are going to consider in the full numerical analysis. 
We also display the resulting values for the ratio $R_{\tau\mu}$ defined in Eq.~(\ref{eq:Rtm}). 
To estimate  $R_{\tau\mu}$, we use for $U_{e3}$ the recently reported 95\% CL range of T2K $U_{e 3}\simeq 0.08\div 0.28$ \cite{Abe:2011sj},
and $m_{\nu_1} \simeq 0\div 0.1$ eV. We see that, for the cases where LFV is directly related to the $U_\mathrm{PMNS}$ entries,
the $U_{e3}$ values preferred by T2K make $R_{\tau\mu}$ at most of $\mathcal{O}(10)$, so that the present limit on
$\mu\to e\gamma$ prevents ${\rm BR}(\tau\to\mu\gamma)$ to be within the future experimental sensitivity, 
as we are going to discuss in the numerical analysis.

\subsection{A Flavoured CMSSM}
\label{FlavCMSSM}

Although the simple structure of the soft breaking terms of the CMSSM can be found in some theoretical models, complete flavour universality of the soft SUSY breaking is a strong assumption and it is not generally realised in most of supergravity and string-inspired models of SUSY breaking. 
This fact encourages the evaluation of models departing from this complete flavour universality.

A completely general MSSM includes two types of departures from the CMSSM structure: (1) the non-degeneracy of flavour-blind masses for different matter representations of the gauge group ($m_{\tilde Q} \neq m_{\tilde u^c} \neq m_{\tilde d^c} \neq m_{\tilde L} \neq m_{\tilde e^c} \neq m_{H_1} \neq m_{H_2}$) or gaugino masses ($M_1\neq M_2\neq M_3$), and (2) the inclusion of more general flavour structures, i.e.\ different flavour off-diagonal entries in the sfermions mass matrices. Both departures would be expected in a generic MSSM, but as a first step, we consider them separately. That is, in (1) we would consider different masses for different SM representations but identical masses for the three generations, while in (2) we keep gaugino universality and equal soft mass matrices for different SM representation at $M_{\rm GUT}$, as could be expected if we have a GUT symmetry at higher scales.

In the first kind of deviations it is natural to expect to have a spectrum that can be very different from the one of the CMSSM. Only if the various initial values of the scalar and gaugino masses are of similar magnitude, the CMSSM spectrum might still be a good approximation of the spectrum of this generic MSSM. However, for a more general case, one has to be careful when using the ATLAS and CMS data, and establish specific bounds for each particular model. Defining such bounds is outside the scope of this work, so in the following we shall not consider MSSM models with this type of deviations.

In contrast, from the second kind of deviations one can expect a spectrum very close to that of the CMSSM. This is due to the fact that the stringent flavour-changing neutral currents (FCNC) and CPV constraints force non-degeneracy and flavour-violating entries in the sfermion mass matrices to be very small, such that these additional parameters, though important in FCNC processes, are not very relevant for the sfermion masses themselves.\footnote{A similar approach can be found in~\cite{Nomura:2007ap}, where the deviations in the diagonal elements from the CMSSM expectations are described.} In this case, one can confidently apply the current ATLAS and CMS bounds on their parameter space, as they depend only on the gluino, lightest neutralino and first generation squark masses, but still get in the flavour sector a phenomenology that differs from the CMSSM expectations. This makes this sort of deviations particularly interesting for flavour physics, as one can concentrate on low-energy phenomenology without the necessity of performing a full collider simulation at the same time. We refer to any model exhibiting such kind of deviations as a ``Flavoured CMSSM,'' and shall study a particular example further ahead.

From a theoretical point of view we consider a Flavoured CMSSM as the first step towards a more ``realistic'' MSSM. Indeed, flavour
universality is not at all a feature of the SM Yukawa couplings and an analogous breaking of flavour universality is still allowed and should be expected in the SUSY soft breaking terms. Nevertheless, if we consider Grand Unification as a guiding principle of these models, different gaugino masses and the masses of the different SM representations are expected to unify.

Still, the main difficulty in defining a Flavoured CMSSM is to choose the flavour structure to assign to the soft-breaking terms. An attractive principle for this is to demand the mechanism that generates the structure in the observed Yukawa coupling to also generate the flavour structures in the soft-breaking terms. In this way, one can expect the structures in the latter to be related to the known Yukawa couplings. Notice that even when accepting this principle there is a host of different possibilities, simply because we do not know the full structure of the Yukawa matrices (only masses and left-handed mixing angles are observable). In the following, we shall follow this principle, and assume that the mechanism for generating flavour is based on a flavour symmetry.

The addition of a new flavour symmetry represents an interesting attempt to explain the mass hierarchies and mixings already found in nature. In the limit of the exact symmetry, under which the flavoured SM fields transform, the Yukawa couplings are usually forbidden, and need to be generated through the spontaneous breaking of the symmetry. This breaking is usually carried out through the introduction of new scalar particles, called flavons, which acquire a vev. The masses and mixings are then generated through effective couplings between the SM particles and the flavons. The hierarchy is interpreted as an effect of the ratio between the flavon vev and the scale of the operator, which acts as a suppression parameter (an incomplete list of examples can be found in~\cite{Froggatt:1978nt,Leurer:1992wg,Barbieri:1995uv,Altarelli:2005yp,King:2001uz}).

In a SUSY scenario, both scalar and fermion components of the flavoured superfields transform under the new symmetry. Thus, in the same way as for the Yukawa couplings, the flavon vevs generate a structure for the flavoured soft SUSY-breaking terms. In many cases, the generated structures are suppressed enough in order to satisfy the strict bounds coming from FCNC processes, but are larger than those predicted by a MFV framework
(see for instance \cite{Lalak:2010bk}). Thus, flavour symmetry models represent a testable solution to the so-called SUSY flavour problem.

The model proposed in~\cite{King:2001uz, King:2003rf} is based on an $SU(3)$ flavour symmetry, and reproduces successfully the quark and lepton masses and mixings, following the structure outlined in~\cite{Roberts:2001zy}:
\begin{align}
Y_u=\left(\begin{array}{ccc}
 0 & \varepsilon^3 & \varepsilon^3 \\
\varepsilon^3 & \varepsilon^2 & \varepsilon^2 \\
\varepsilon^3 & \varepsilon^2 & 1
\end{array}\right)y_t & &
Y_d=\left(\begin{array}{ccc}
 0 & \bar\varepsilon^3 & \bar\varepsilon^3 \\
\bar\varepsilon^3 & \bar\varepsilon^2 & \bar\varepsilon^2 \\
\bar\varepsilon^3 & \bar\varepsilon^2 & 1
\end{array}\right)y_b,
\end{align}
where $\bar\varepsilon=0.15$ and $\varepsilon=0.05$ parametrise the ratio between the flavon vevs and the scale of the effective operators. In addition, the model also solved the SUSY flavour problem, as it implied almost degenerate sfermions and small flavour-violation terms. The structures, in the basis where $Y_d$ is diagonal, roughly followed:
\begin{align}
 m^2_{\tilde Q}=\left(\begin{array}{ccc}
 1+\varepsilon^2 & \varepsilon^2\bar\varepsilon & \bar\varepsilon^3 \\
\varepsilon^2\bar\varepsilon & 1+\varepsilon^2 & \bar\varepsilon^2 \\
\bar\varepsilon^3 & \bar\varepsilon^2 & 1
\end{array}\right)m_0^2 & & 
m^2_{\tilde d_R^c}=\left(\begin{array}{ccc}
 1+\bar\varepsilon^2 & \bar\varepsilon^3 & \bar\varepsilon^3 \\
\bar\varepsilon^3 & 1+\bar\varepsilon^2 & \bar\varepsilon^2 \\
\bar\varepsilon^3 & \bar\varepsilon^2 & 1
\end{array}\right)m_0^2
\end{align}
A further expansion in~\cite{Ross:2004qn} contemplated the spontaneous breaking of a CP symmetry, through which all phases become constrained within the flavour sector. This was shown to solve the SUSY CP problem in~\cite{Calibbi:2008qt}, and to help reduce the CPV tensions in the quark sector in~\cite{Calibbi:2009ja,Altmannshofer:2009ne}.

As an example of how the collider and flavour interplay works for Flavoured CMSSM models, we shall take two definite examples, taken from~\cite{Calibbi:2009ja}. In Table~\ref{Table:RVV-MI} we show the order of magnitude of their mass-insertions, at the GUT scale.\footnote{The renormalisation group running down to low energy modifies the mass-insertion parameters in the following way:
$\delta^{f}_{XY} (M_{\rm SUSY}) \approx \mathcal{R}\times \delta^{f}_{XY} (M_{\rm GUT})$,
with $\mathcal{R}\approx m_0^2 /(m_0^2 + 0.5 M_{1/2}^2)$ ($ \approx m_0^2 /(m_0^2 + 0.15 M_{1/2}^2)$) 
for $\delta^{e}_{LL}$ ($\delta^{e}_{RR}$) and $\mathcal{R}\approx m_0^2 /(m_0^2 + 6 M_{1/2}^2)$ for the hadronic mass-insertions.}
Here, $\Sigma_f$ is the ratio between the vev of a Georgi-Jarlskog field and the scale, oriented in the $(B-L+2T^R_3)$ direction, used to differentiate between the charged lepton and down quark masses. Notice that our intention is to show the collider and flavour interplay for a Flavoured CMSSM and not to study the exact details of these models. Further details of the models can be found in~\cite{Jones:2010}, as well as the original references.

\begin{table}[t]
\begin{center}
\begin{tabular}{|c|ccc|ccc|ccc|}
\hline
& $|(\delta^{d,e}_{LL})_{12}|$  & $|(\delta^{d,e}_{LL})_{13}|$& $|(\delta^{d,e}_{LL})_{23}|$
& $|(\delta^{d,e}_{RR})_{12}|$ & $|(\delta^{d,e}_{RR})_{13}|$ & $|(\delta^{d,e}_{RR})_{23}|$
& $|(\delta^{u}_{RR})_{12}|$ & $|(\delta^{u}_{RR})_{13}|$ & $|(\delta^{u}_{RR})_{23}|$\\
\hline \hline
Model 1 & $ \frac{1}{\Sigma_f}\varepsilon^2\bar\varepsilon $  &
$ y_t \bar\varepsilon^3 $ &  $ \Sigma_f y_t \bar\varepsilon^2 $
& $ \frac{1}{\Sigma_f}\bar\varepsilon^3 $  & $\frac{1}{\Sigma_f}\bar\varepsilon^3$ & $\bar\varepsilon^2$ 
& $\frac{1}{\Sigma_u}\varepsilon^3$ & $y_t\varepsilon^3$ & $\Sigma_u y_t\varepsilon^2$ \\
Model 2 & $ \frac{1}{\Sigma_f}\varepsilon^2\bar\varepsilon $  &
$  \frac{1}{\Sigma_f} \sqrt{y_t} \varepsilon \bar\varepsilon $ & $ \sqrt{y_t}\varepsilon $
& $ \frac{1}{\Sigma_f}\bar\varepsilon^3 $  & $\frac{1}{\Sigma_f} \sqrt{y_b}\bar\varepsilon^2$ & $\sqrt{y_b} \bar\varepsilon$ 
& $\frac{1}{\Sigma_u}\varepsilon^3$ & $\frac{1}{\Sigma_u}\sqrt{y_t}\varepsilon^2$ & $\sqrt{y_t}\varepsilon$ \\
\hline
\end{tabular}
\end{center}
\caption{\label{Table:RVV-MI} Magnitude of mass-insertions in the considered models, at the GUT scale. Here $\varepsilon=0.05$, $\bar\varepsilon=0.15$ and $\Sigma_e=3\Sigma_d$.}
\end{table}

In terms of the size of the mass-insertions, the main difference between the MFV framework and these models is the existence of a sizeable $\delta_{RR}$. This fact, as well as the large leading-order phases, motivate us to include these models within our analysis. Moreover, as they are meant to be embedded within a GUT framework, the flavour structures of the squark and slepton sectors are related. This means that interesting correlations, unavailable in the MFV framework, might arise.

In the following analysis, we shall include both example models. Our strategy will similar to that for the CMSSM analysis, with the following exceptions: for each point in the parameter space, we vary the $\ord{1}$ terms randomly between $0.5$ and $2$ (with arbitrary sign). In addition, as the flavon phases in the soft terms are related to the phases in the Yukawas, we perform a fit of all parameters entering the Yukawas at the GUT scale, using the masses and CKM parameters at that scale as constraints. The electroweak scale CKM parameters are taken from the Tree Level fit of~\cite{Ciuchini:2000de}, and are shown in Table~\ref{tab:ckmpar}. After the fit, we run each point down to the electroweak scale with two-loop Renormalisation Group Equations (RGEs), using a modified version of SPheno3.1.4~\cite{Porod:2003um,Porod:2011nf}. SPheno then calculates the threshold corrections to the Yukawas, obtaining tree-level Yukawas, which are later used to re-fit the parameters. As the tree-level Yukawas are dependent on the point of the parameter space, we need to perform an independent GUT-scale fit for every point in our scan. We do this such that the most general sample of phases compatible with the CKM structure is obtained.

\begin{table}[tbp]
\renewcommand{\arraystretch}{1.3}
 \begin{center}
\begin{tabular}{|c|c|c|c|}
\hline
$\lambda$ & $A$ & $\bar\rho$ & $\bar\eta$ \\
\hline
$0.22535\pm.00065$ & $0.804\pm0.010$ & $0.111\pm0.07$ & $0.381\pm0.03$ \\
\hline
 \end{tabular}
 \end{center}
\caption{EW scale CKM parameters used in the fit.}
\label{tab:ckmpar}
\end{table}

As these models present a richer phase structure than the CMSSM, in addition to the observables mentioned in the previous sections, we shall also check whether the CPV tension in the $\epsilon_K - S_{\psi K_s}$ sectors is ameliorated. Furthermore, we shall show the size of the $S_{\psi\phi}$, as well as the EDMs of both the electron and the neutron. All of these observables shall be briefly explained in further sections.

\section{Interplay of LHC with flavour experiments}
\label{interplay}
\subsection{CMSSM}
\label{CMSSM-Pheno}
Our starting point in the analysis is assuming that an excess will be found at LHC7 
in the jets plus missing energy channel, which should correspond to the production 
of new coloured particles. The information we will have from LHC measurements will be 
the number of non-SM events and some information on $M_{\rm eff}$ and $\slashed{E}_T$. 

In an MSSM context, the produced SUSY particles can be mostly a pair of first-generation squarks, a gluino pair or a squark-gluino pair. 
As discussed in section \ref{sec:LHC}, the cross section is basically fixed by the superpartner masses. In fact, if we are able to observe an excess in the jets plus missing energy channel with a certain amount of integrated luminosity, one can find an upper limit on the produced SUSY particle mass, and this translates in a well-defined region in the $m_0$--$M_{1/2}$ plane. The lower edge of this region is set by the present bounds from non-observation of SUSY particle at LEP and Tevatron experiments together with the recent results from
ATLAS and CMS with 1 fb$^{-1}$. The upper edge can be obtained by simulations of the number of signal and background events expected at LHC7 with the assumed integrated luminosity. Throughout this analysis, we make use of the results of~\cite{Baer:2010tk} to estimate the LHC reach with a luminosity of 5~fb$^{-1}$.\footnote{In practice, we made an extrapolation of the estimated 2~fb$^{-1}$ reach of~\cite{Baer:2010tk}, based on an analysis of the total production cross-section of coloured SUSY particles at LHC7 by means of the routine PROSPINO 2.0~\cite{prospino}. Moreover, we have checked that points in this region are indeed observable with 5 fb$^{-1}$.} We have chosen this integrated luminosity as a conservative measure for the 7 TeV run, considering that at the moment there is a possibility of increasing the collider energy by 2012.

Thus, we assume that the band in the $m_0$--$M_{1/2}$ plane, as defined above, represents the region of the CMSSM parameter space that will be explored by the LHC experiments during this $\sqrt{s}= 7$ TeV run and we study it by means of the code SPheno~\cite{Porod:2003um,Porod:2011nf}. 
Within this band, gluino masses span values between 560 and 1350 GeV, while the masses of the heaviest squarks (which generally correspond to the first generation) are in the 1 -- 3.5 TeV range.

\begin{figure}
\includegraphics[scale=.6]{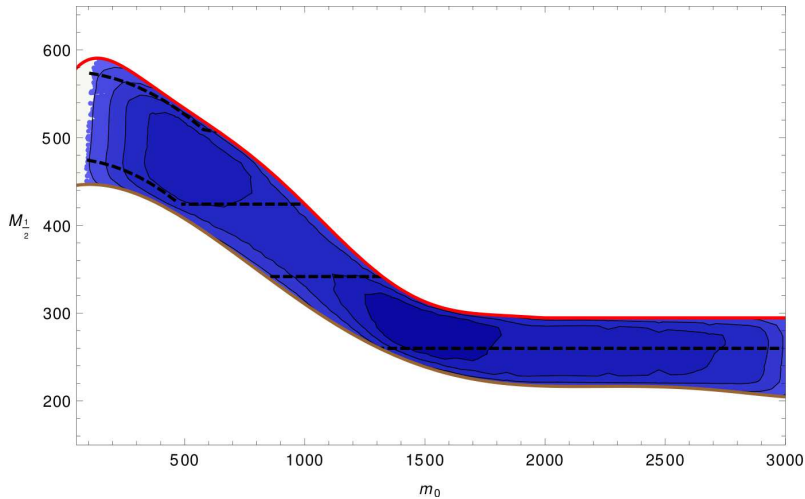}       
\caption{Area in the $m_0$--$M_{1/2}$ plane to be explored at LHC with 5 fb$^{-1}$ integrated luminosity at 7 TeV. The brown line indicates the current constraints from direct searches, while the red line shows our estimate for the reach. The black, dashed lines are contours of $M_{\rm SUSY}$, 
as defined in the text, with values $M_{\rm SUSY}=650, 850, 1050, 1250$~GeV from bottom to top. Coloured contours indicate the density of points.}
\label{fig:scenario}
\end{figure}

In Figure~\ref{fig:scenario}, we show the explored region in the $m_0$--$M_{1/2}$ plane, with dashed contours corresponding to different values of  $M_{\rm SUSY} \equiv {\rm min}(m_{\tilde g}, m_{\tilde q})$. The solid contours represent the density of allowed points, with the darkest colours representing a larger number of points. We show these contours to give the reader an idea of how much populated a particular region is, and do not intend to give any statistical significance to each contour. In this way, we differentiate our philosophy from that of fitting collaborations (for example~\cite{arXiv:1103.0969,arXiv:1103.1197,arXiv:1110.3568,arXiv:1111.6098}): in our point of view, every single point surviving the imposed bounds is equally valid.

The only constraints imposed in Figure~\ref{fig:scenario} are the SUSY and Higgs direct search constraints, together with the requirements of correct EWSB, absence of tachyons and neutral LSP. From the coloured contours, we see a region of intermediate $m_0$--$M_{1/2}$ slightly disfavoured by light Higgs searches. Two other regions, one with very large and one with very low $m_0$, are also slightly disfavoured, this time respectively by EWSB and neutral LSP requirements. 
Notice that we are only plotting the possible values of $m_0$ and $M_{1/2}$. The remaining CMSSM parameters, $\tan\beta$, $A_0$ and sgn($\mu$) are unbounded by the considered LHC searches, although they are affected by the Higgs and EWSB constraints. For them, we take the usual ranges, $5 \leq \tan \beta \leq 55$, $-3 m_0 < A_0< 3 m_0$ and sgn($\mu$) positive.  

\subsubsection{Flavour constraints: $b\to s\gamma$ and $(g-2)_\mu$}

\begin{table}[tbp]
 \begin{center}
\begin{tabular}{|c|rl||c|rl|}
\hline
${\rm BR}(b\to s\gamma)$ & $(3.55\pm0.25)\times10^{-4}$ & \cite{Asner:2010qj} & ${\rm BR}(\mu\to e\gamma)$ & $<2.4\times10^{-12}$ & 
\cite{Adam:2011ch} \\
${\rm BR}(B_s\to\mu^+\mu^-)$ & $<1.08\times10^{-8}$ & \cite{BsmumuComb} & ${\rm BR}(\tau\to e\gamma)$ & $<3.3\times10^{-8}$ & \cite{:2009tk} \\
$R(B^+\to\tau^+\nu)$ & $1.57\pm0.53$ & \cite{Tisserand:2009ja, Altmannshofer:2009ne}& ${\rm BR}(\tau\to \mu\gamma)$ & $<4.4\times10^{-8}$ & \cite{babar} \\
$\delta a_\mu\equiv (a_\mu- a_\mu^{\rm SM})$ & $(2.61\pm0.8)\times10^{-9}$ & \cite{Hagiwara:2011af} & $\epsilon_K$ & $(2.228\pm0.011)\times10^{-3}$ & \cite{Nakamura:2010zzi} \\
%$S_{\psi K_s}$ & $0.673\pm0.023$ & \cite{Nakamura:2010zzi} & $S_{\psi\phi}$ & 
%$\left\{\begin{array}{c} 0.13\pm0.19 \\ -0.55\pm0.38\end{array}\right.$ &
%$\begin{array}{l} {\rm \cite{Spsiphi}} \\ {\rm \cite{Abazov:2011ry}}\end{array}$ \\
\multirow{2}{*}{$S_{\psi K_s}$} & \multirow{2}{*}{$0.673\pm0.023$} & \multirow{2}{*}{\cite{Nakamura:2010zzi}} &
\multirow{2}{*}{$S_{\psi\phi}$} & \multirow{2}{*}{$\left\{\begin{array}{c} 0.13\pm0.19 \\ -0.55\pm0.38\end{array}\right.$} &
\cite{Spsiphi} \\ 
& & & & &\cite{Abazov:2011ry} \\
$\Delta m_{B}$ & $(3.337\pm0.033)\times10^{-13}\,{\rm GeV}$ & \cite{Nakamura:2010zzi} &
 $\Delta m_{B_s}$ & $(117.0\pm0.8)\times10^{-13}\,{\rm GeV}$ & \cite{Abulencia:2006ze} \\
$d_n$ & $<2.9\times10^{-26}\,e\,{\rm cm}^{-1}$ &\cite{Baker:2006ts} & $d_e$ & $<2\times10^{-27}\,e\,{\rm cm}^{-1}$ & \cite{Regan:2002ta} \\
\hline
 \end{tabular}
 \end{center}
\caption{Flavour constraints imposed throughout our analysis. We show only the experimental errors.}
\label{tab:constraints}
\end{table}

Indirect searches of SUSY in low energy FCNC experiments provide very stringent constraints in the MSSM parameter space. Even in the CMSSM where the only non-trivial flavour structures are the usual Yukawa couplings, the experimental results are sensitive to supersymmetric contributions with CKM mixings. The
main observables in this context are BR($b\to s \gamma$) and the muon anomalous magnetic moment $a_\mu \equiv (g-2)_\mu/2$. Another interesting process
is $B_s\to \mu^+ \mu^-$ which is becoming more and more constraining because of the dedicated searches at LHCb and CMS \cite{BsmumuComb}. 
Another important observable, although less restrictive at present, is ${R}(B^+ \to \tau^+ \nu)\equiv{\rm BR}(B^+ \to \tau^+ \nu)/{\rm BR}(B^+ \to \tau^+ \nu)^{\rm SM}$.
The main constraints we impose (including those that are relevant in the seesaw and Flavoured CMSSM models) are shown in Table~\ref{tab:constraints}.

The  $b \to s \gamma$ process in the CMSSM receives contributions from chargino-stop loops and from charged Higgs-top loops.\footnote{We 
remind that in presence of large flavour mixing in the squark sector, squark-gluino loops can also be important. This is the case of the
Flavoured CMSSM models we are going to study below.} 
The expressions for these supersymmetric contributions are well-known in the literature (e.g.~in \cite{Lunghi:2006hc}). 
In the numerical analysis we use the SPheno prediction for the decay, which contains NLO expressions for MSSM contributions and NNLO contribution for the Standard Model.

The charged Higgs and the SM W-boson contributions have the same 
sign and interfere always constructively. In contrast, chargino
contribution can have either sign depending on the sign of the Higgsino mass parameter $\mu$.
The relative sign of the
chargino mediated diagram is given by $\hbox{sgn}(A_t \mu)$.
Since, in the CMSSM, $A_t$  
tends to be driven to negative values at low-energies by the gluino contribution in the RGEs, 
$\mu>0$ implies destructive interference, unless the initial condition $A_0$ is large and positive. 
Conversely, $\mu<0$ usually implies constructive interference of the chargino contribution. 
It is also important to remember that the chargino contribution increases with the value
of $\tan \beta$.
Therefore, for $\mu > 0$ and large $\tan \beta$, the chargino contribution, which has opposite sign to the SM contribution, can be sizeable.  
Taking into account that the SM prediction is in the lower part of the experimentally allowed range, this constraint requires relatively heavy supersymmetric masses unless this contribution is compensated by a sizeable charged Higgs contribution. Therefore, we find that a supersymmetric spectrum accessible 
at LHC with 5 fb$^{-1}$ tends to prefer also a not too heavy charged Higgs, $m_{H^\pm}\lesssim 1$ TeV, when
$A_t$ is large and negative ($A_t \lesssim - 500$ GeV).

Similarly, in SUSY theories, $a_{\mu}$ receives contributions via vertex diagrams with
$\tilde{\chi}^0$--$\tilde \mu$ and $\tilde{\chi}^\pm$--$\tilde \nu$
loops~\cite{moroi}.
The chargino diagram usually dominates in most of the parameter
space. The dominant $\tan\beta$ enhanced contribution approximately reads:
\bea
\delta a_{\mu}\approx {g_2^2\over 32 \pi^2}
{m_\mu^2 \over m_{\tilde\nu}^2} {\hbox{Re} (\mu M_2) \tan \beta \over m_{\tilde\nu}^2}\,.
\label{gm2app}
\eea
The most relevant feature of Eq.~(\ref{gm2app}) is
that the sign of $\delta a_{\mu}$ is fixed by
$\hbox{sgn}[\hbox{Re}(\mu M_2)]$.

The latest
experimental measurement of $a_\mu$ presents 
a 3$\sigma$ discrepancy with the theoretical expectations in the SM~\cite{Passera:2008jk}, as can be seen in Table~\ref{tab:constraints}. Therefore, at present, a positive contribution $\delta a_\mu$ from slepton loops at
the level of $\sim 10^{-9}$ is required and so, this result
strongly favours the $\mu>0$ region in an MSSM scenario. 

\begin{figure}
\includegraphics[scale=.5]{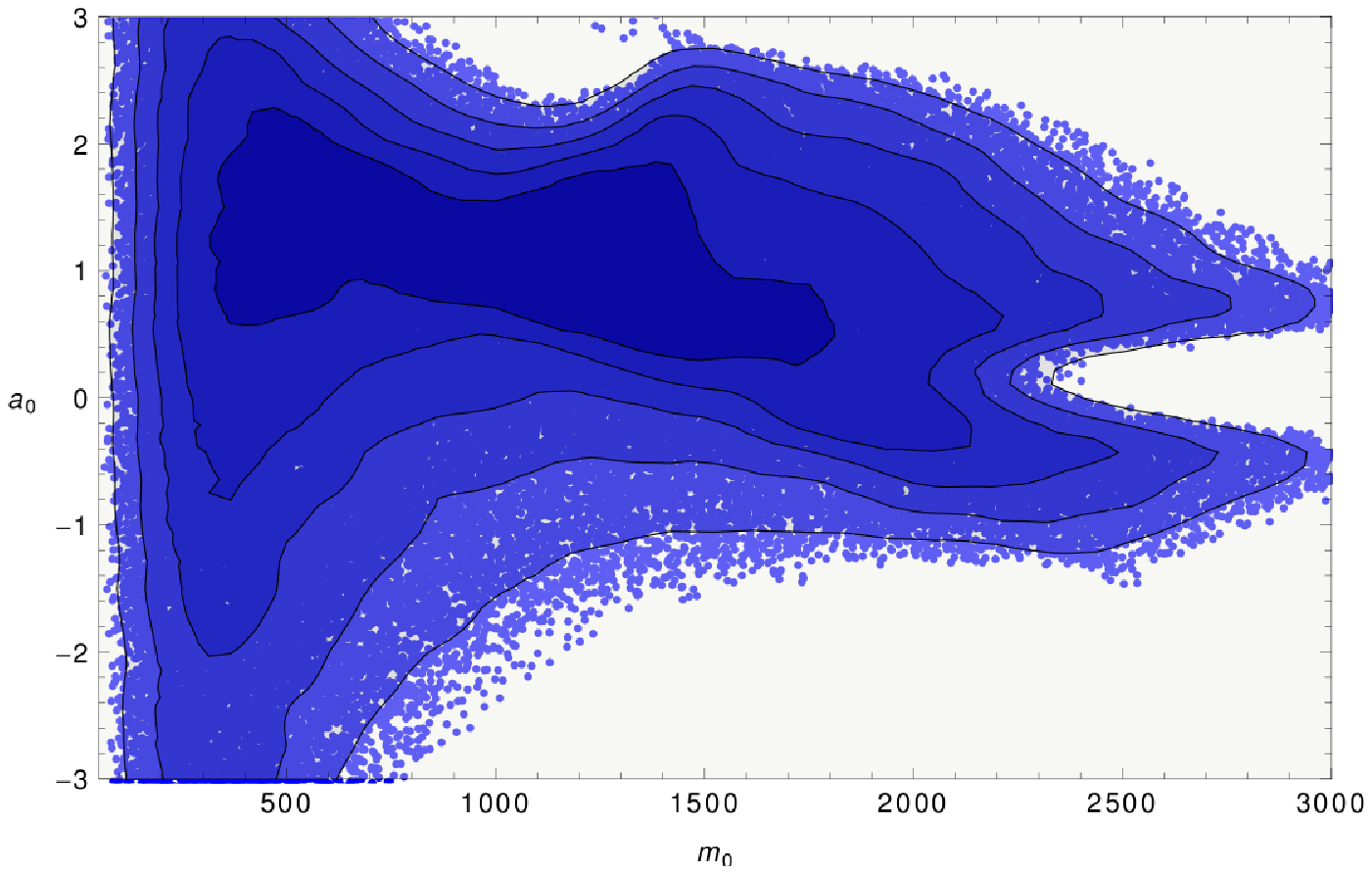} \qquad  \includegraphics[scale=.5]{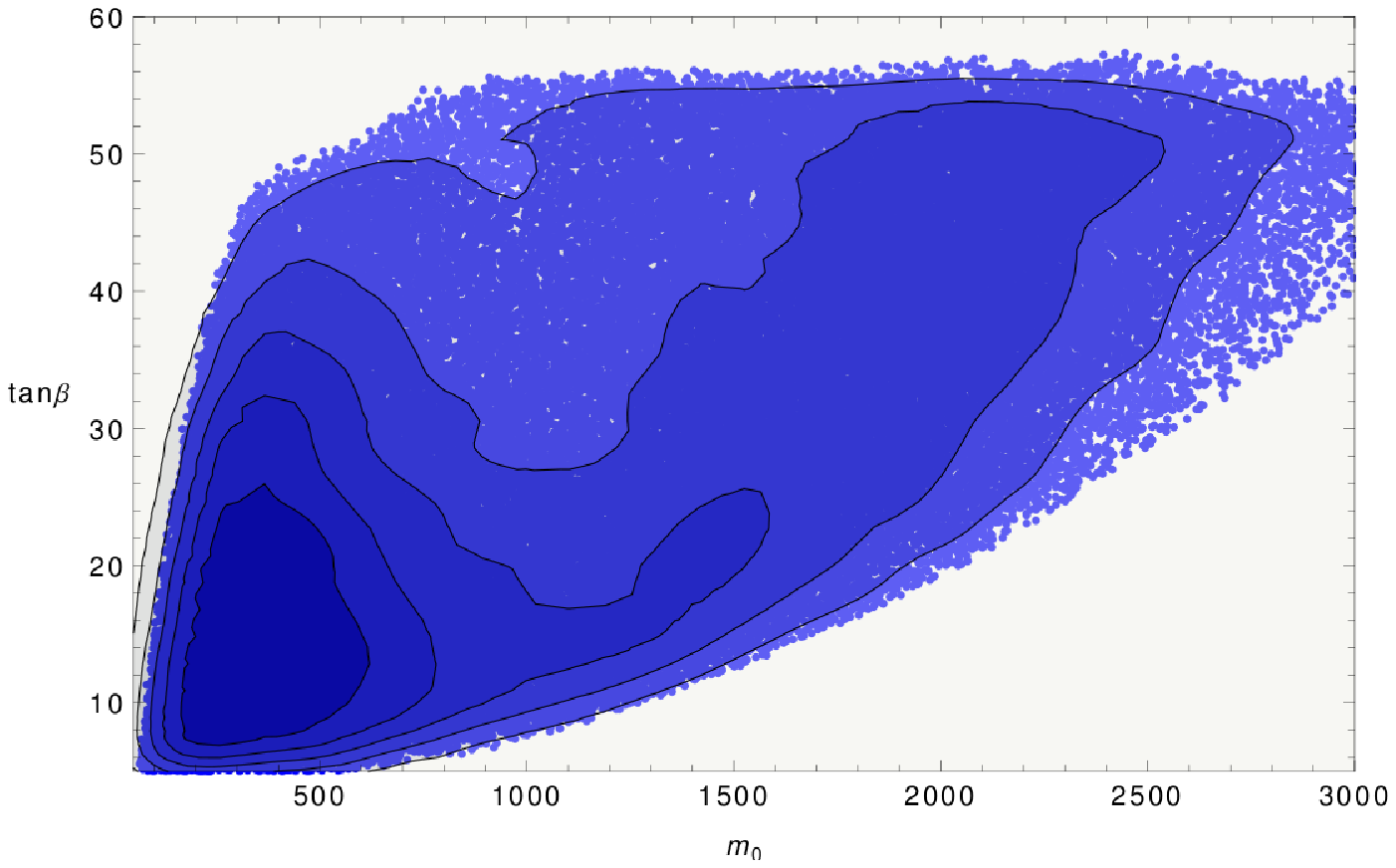}       
\caption{Area in the $m_0$--$a_0$ (left) and $m_0$--$\tan\beta$ planes (right) to be explored at LHC with 5 fb$^{-1}$ integrated luminosity at 7 TeV.}
\label{fig:indirect2}
\end{figure}

If we impose the $b\to s\gamma$ and $\delta a_\mu$ constraints at the 3$\sigma$ level the allowed region is modified, with the appearance of correlations with specific values of $\tan\beta$ and $A_0$. If we restrict ourselves to the $m_0$--$M_{1/2}$
plane, we shall not find large differences with respect to Figure~\ref{fig:scenario}, 
apart from a decrease in the density of points, particularly for large $m_0$. 
However, when examining the other planes, we find that the parameter space, although still big, is not so arbitrary anymore.
We can see this directly in Figure~\ref{fig:indirect2}.

On the left panel of Figure~\ref{fig:indirect2}, we show the $m_0$--$a_0$ plane, where $a_0\equiv A_0/m_0$. In this plane, we first see strong constraints due to EWSB and tachyons at large values of $m_0$ and $|a_0|$. These strong constraints are also applied for large values of $m_0$, when $a_0\sim0$. For large, positive values of $a_0$, we see a small, semi-circular excluded region, where the Higgs mass becomes too light. We also see a preference in the density of points disfavouring negative $a_0$. In these cases, the chargino contribution to $b\to s\gamma$ becomes too large, while for positive $a_0$ it has just the right size to be compensated by the charged Higgs contribution.

On the right panel of Figure~\ref{fig:indirect2} we show the $m_0$--$\tan\beta$ plane. The two most important constraints can be seen for large $m_0$, low $\tan\beta$, and for low $m_0$, large $\tan\beta$. The first one is due to $a_\mu$, where the SUSY contribution is not large enough to account for the anomaly. 
The second one corresponds to a neutral LSP. In addition, we get a significant reduction in the density of points for intermediate $m_0$ and large values of $\tan\beta$. Here, the $b\to s\gamma$ constraint rules out many points, as the chargino contribution again becomes too large.

\subsubsection{$B_s\to\mu^+\mu^-$ bounds and prospects}

In parallel with the direct searches ongoing at the ATLAS and CMS experiments, we can also expect improvements in measurements of flavour-violating observables during the $7$~TeV LHC run.  In particular, the LHCb experiment is designed for precision B-physics studies.  Of particular interest is the decay $B_s \to \mu^+\mu^-$.

In the CMSSM, the main NP contributions to the $B_s\to\mu^+\mu^-$ branching ratio are due to neutral heavy-Higgs mediated diagrams. The amplitude then acquires a $\tan^3\beta$ dependent contribution which, as for $b\to s\gamma$, is also proportional to $A_t\,\mu$.

This particular decay is of great interest at the times of the LHC. Although currently the LHC has only placed upper bounds on the branching ratio of this decay channel, as shown in Table~\ref{tab:constraints}, both the CMS and LHCb experiments shall be able to probe its value down to that of the SM $\sim3\times10^{-9}$. It is then particularly interesting to ask what would be the consequences if this particular decay is or is not observed. In particular, from a projection using $37$ pb$^{-1}$ of data~\cite{LambertMoriond}, LHCb claims that, after collecting an integrated luminosity of 2~fb$^{-1}$, it would be able to achieve a $5\sigma$ discovery for a branching ratio larger than $9\times10^{-9}$, or find $3\sigma$ evidence for a branching ratio larger than $5\times10^{-9}$. In the case of not seeing any signal, the same experiment claims to be able to rule out any branching ratio larger than $4\times10^{-9}$ with 95\% confidence.\footnote{For a study of the impact of the recent $B_s\to\mu^+\mu^-$ searches on various SUSY scenario and the discovery prospects, see also \cite{Akeroyd:2011kd}.}

The possibility of observing a large branching ratio for this decay should be taken seriously, as last summer CDF published a new analysis with a positive signal ${\rm BR}(B_s \to \mu^+\mu^-)=(1.8^{+1.1}_{-0.9}) \times 10^{-8}$~\cite{Aaltonen:2011fi}. Notice that this signal is still compatible at 1$\sigma$ with the upper bound of CMS and LHCb. We shall see that taking the central value of this positive signal at 1$\sigma$ would strongly restrict our parameter space, but at the 2$\sigma$ level it has practically no effect at all in our results. We hope this very intriguing situation is clarified in the near future with further analysis at CDF/D0 and CMS/LHCb.

\begin{figure}
\includegraphics[scale=.4]{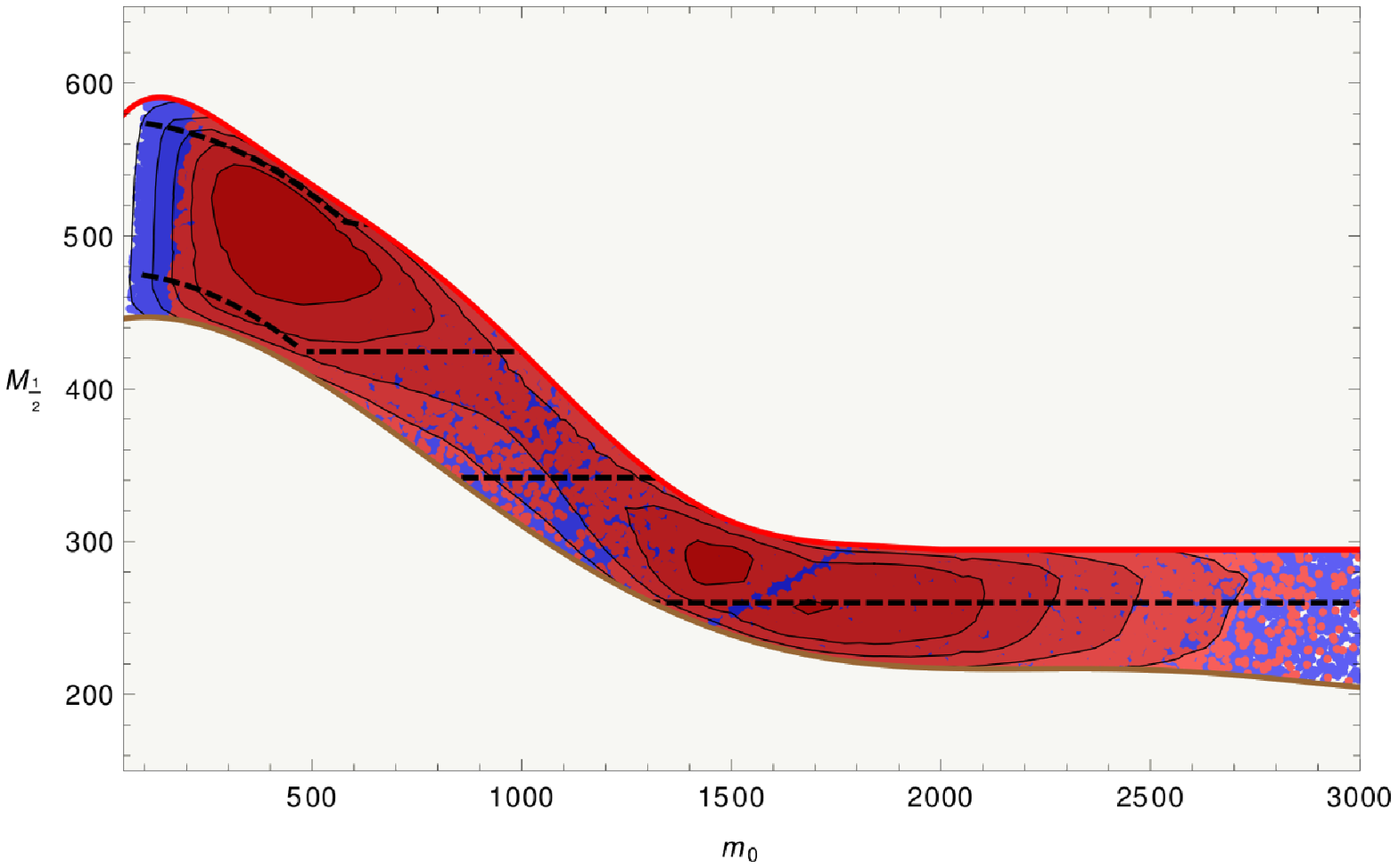} \qquad \includegraphics[scale=.4]{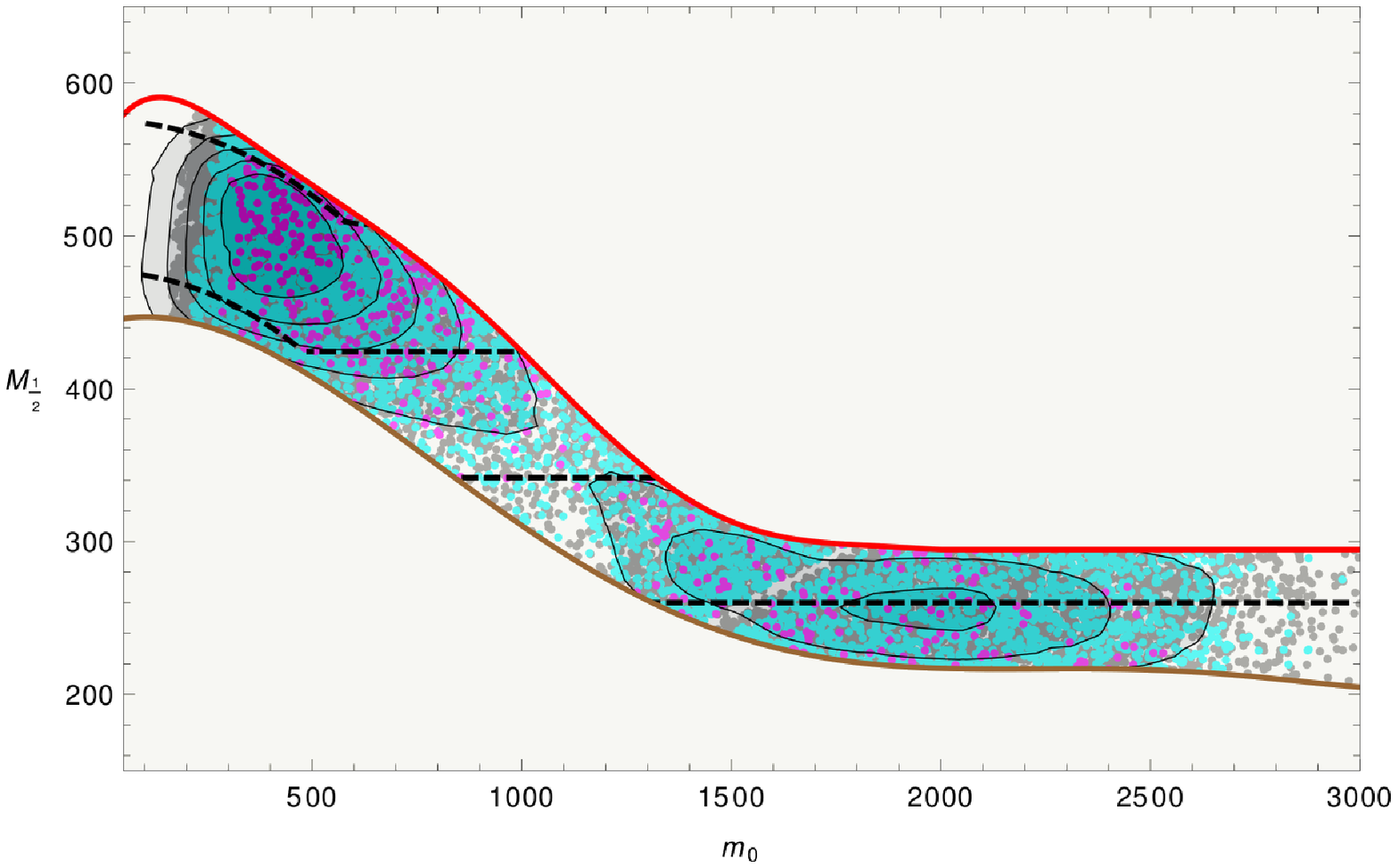}
\caption{$B_s\to\mu\mu$ signal in the $m_0$--$M_{1/2}$ plane. On the left, we show in red (blue) the points with a branching ratio larger (lower) than $4\times10^{-9}$. On the right, we show in magenta, cyan and grey the points with a branching ratio larger than $9\times10^{-9}$, $5\times10^{-9}$ and $4\times10^{-9}$, respectively.}
\label{fig:bsmm.m0m12}
\end{figure}

\begin{figure}
\includegraphics[scale=.4]{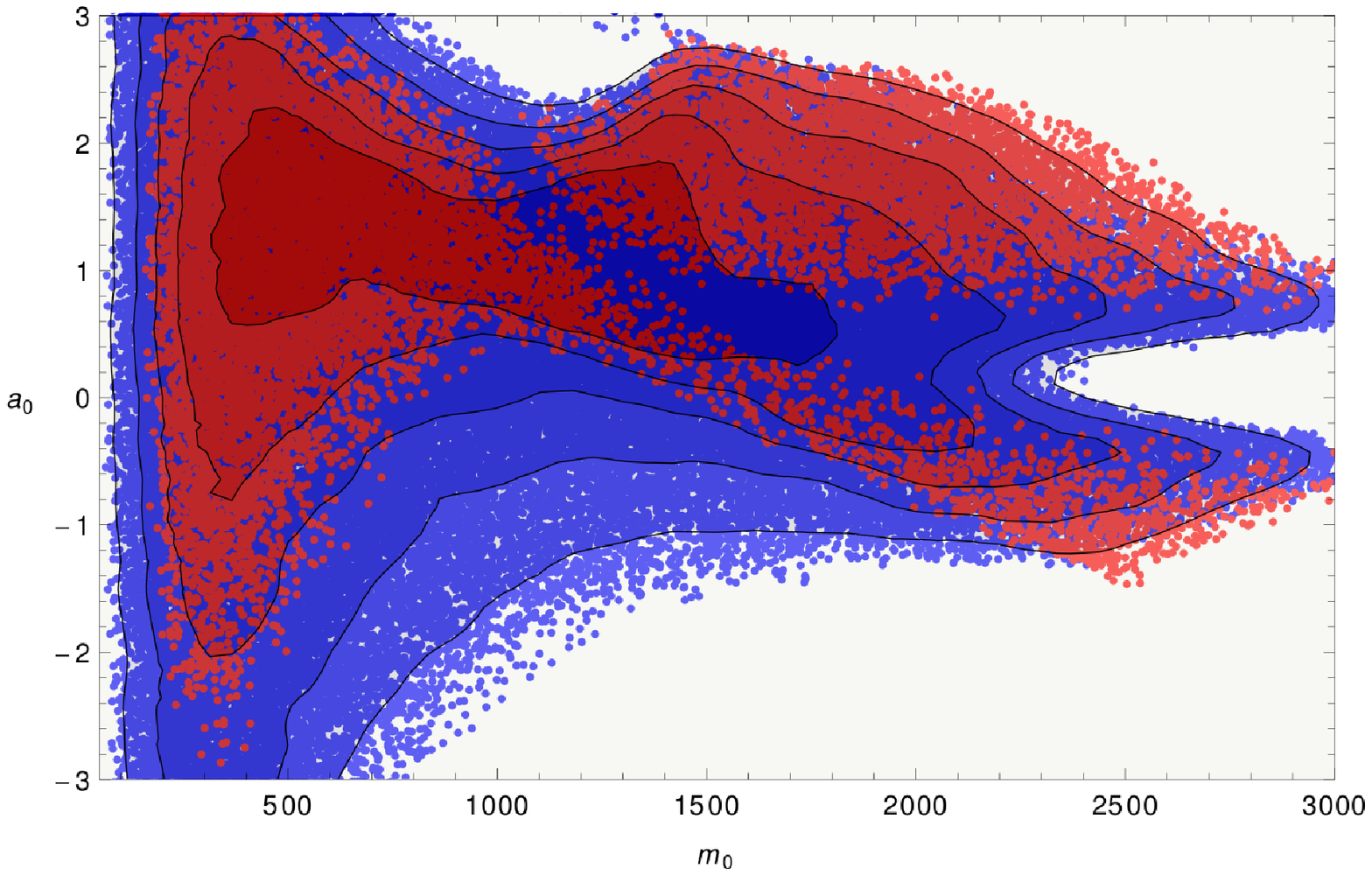} \qquad \includegraphics[scale=.4]{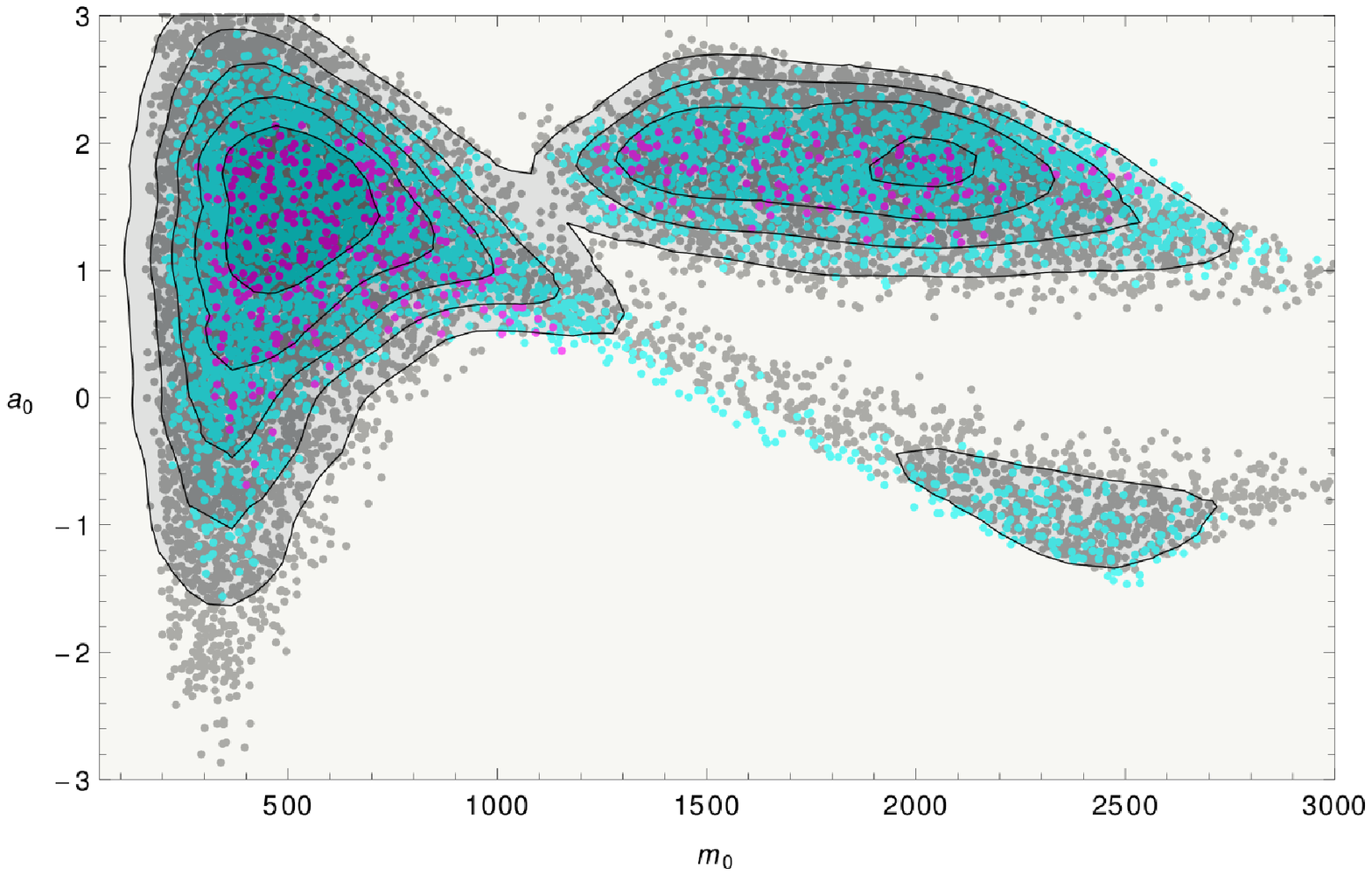} \\
\includegraphics[scale=.4]{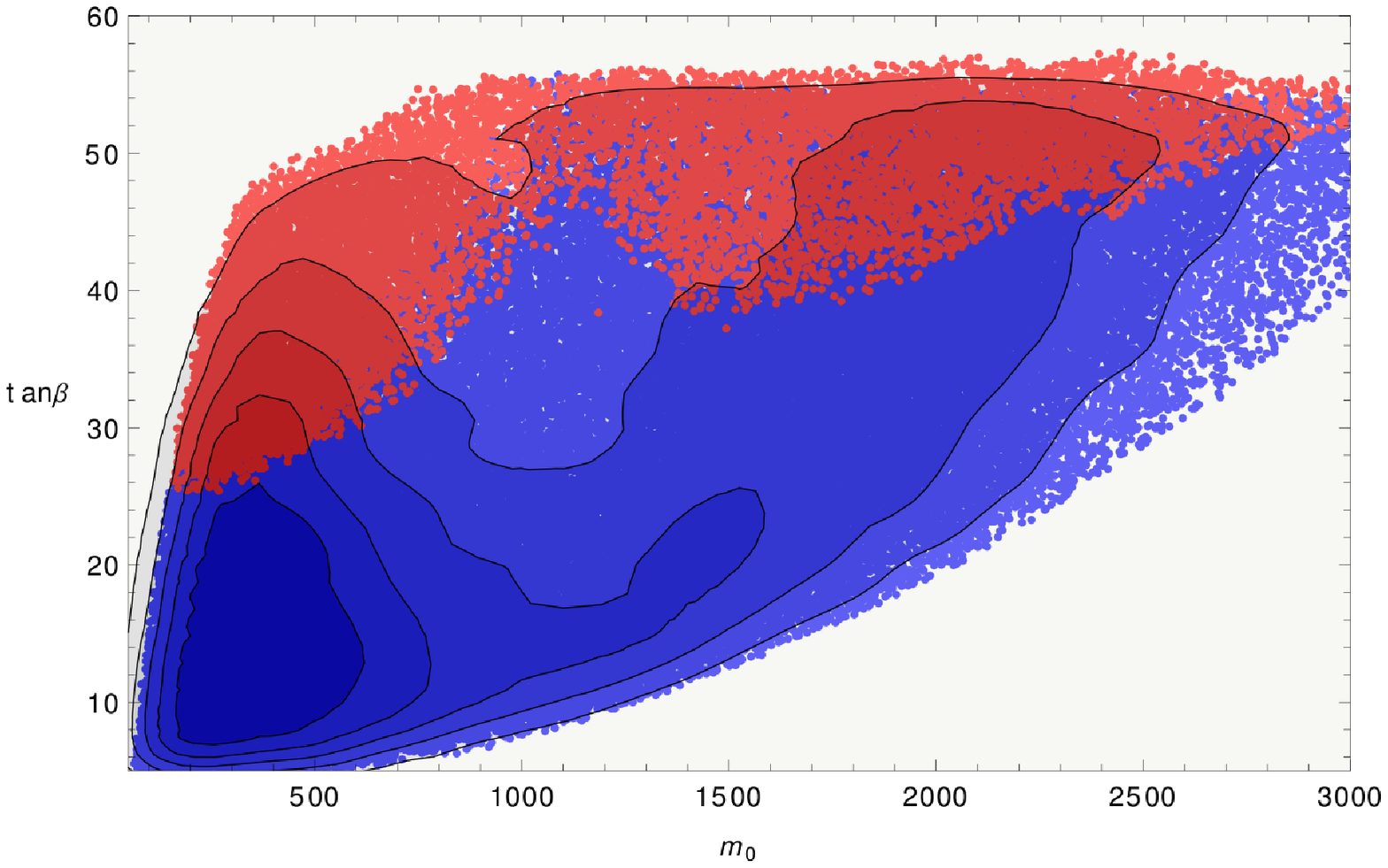} \qquad \includegraphics[scale=.4]{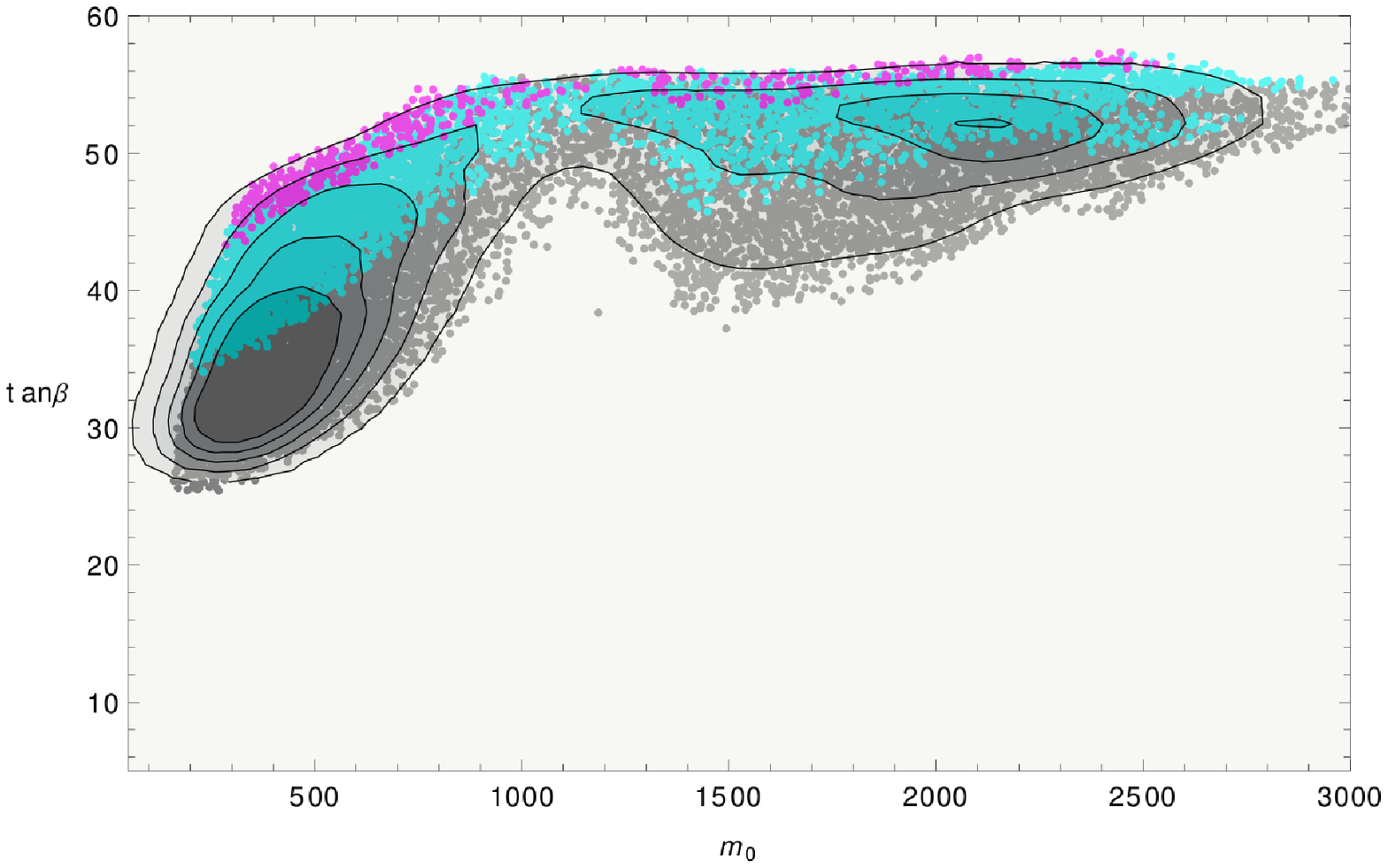} 
\caption{As in Figure~\ref{fig:bsmm.m0m12}, but for the $m_0$--$a_0$ (upper) and $m_0$--$\tan\beta$ (lower) planes.}
\label{fig:bsmm.m0other}
\end{figure}

In Figures~\ref{fig:bsmm.m0m12} and~\ref{fig:bsmm.m0other} we plot how the parameter space would be affected by an observation of $B_s\to\mu^+\mu^-$. In both Figures, we plot on the left in red those points that predict a branching ratio larger than $4\times10^{-9}$, and would thus be ruled out if LHCb does not see any signal for this decay with 2~fb$^{-1}$. We leave all points with a smaller branching ratio in blue (notice that in Figure~\ref{fig:bsmm.m0m12} and~\ref{fig:bsmm.m0other} there is an overlap between blue and red points). On the right part of each Figure, we plot in magenta (cyan) the points that would give a $5\sigma$ discovery ($3\sigma$ evidence), and leave in grey those that would give a weaker signal.

From the Figures, we see that, at this level, the observation of $B_s\to\mu^+\mu^-$ would not be crucial in the determination of $m_0$, $M_{1/2}$ or $a_0$, but could give an indication about the preferred regions of the parameter space. 
Larger values for the branching ratio tend to occur for positive values of $a_0$, since it turns out that equivalent points with negative values of $a_0$ are in conflict with $b\to s\gamma$. The reason is traced back to the correlation
between the dominant Higgs-mediated contribution to $B_s\to\mu^+\mu^-$ and the stop-chargino contribution to $b\to s\gamma$.
Points with a negative $a_0$ tend to give a large negative 
$A_t$, for which a sizeable BR($B_s\to\mu^+\mu^-$) is possible, but we get in turn a too large negative 
stop-chargino contribution to $b\to s\gamma$, that provides a value for the branching ratio below the experimental 
lower bound.
$B_s\to\mu^+\mu^-$ gives also very important information about the value of $\tan\beta$. Depending on the significance of the observation, along with the value of $m_0$, one can set a lower bound on the value of $\tan\beta$, favouring always values above 25.

%In addition, last summer CDF published a new analysis with a positive signal $BR(B_s \to \mu^+\mu^-)=(1.8^{+1.1}_{-0.9}) \times 10^{-8}$~\cite{Aaltonen:2011fi}. Notice that this signal is still compatible at 1$\sigma$ with the upper bound of CMS and LHCb. Clearly, taking the central value of this positive signal at 1$\sigma$ would strongly restrict our parameter space, but at the 2$\sigma$ level it has practically no effect at all in our results. We hope this very intriguing situation is clarified in the near future with further analysis at CDF/D0 and CMS/LHCb.

\begin{figure}
\includegraphics[scale=.45]{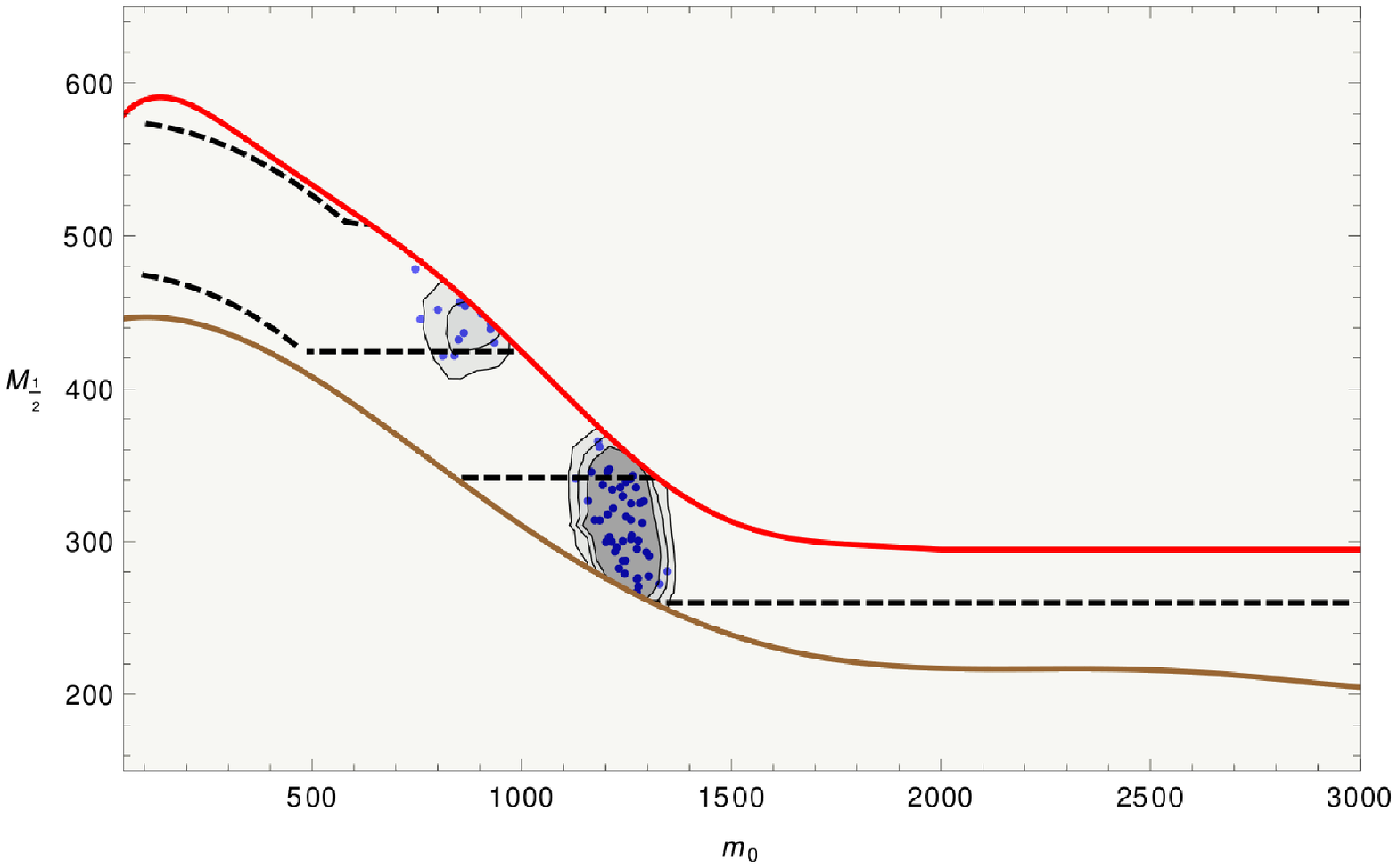}\quad\includegraphics[scale=.45]{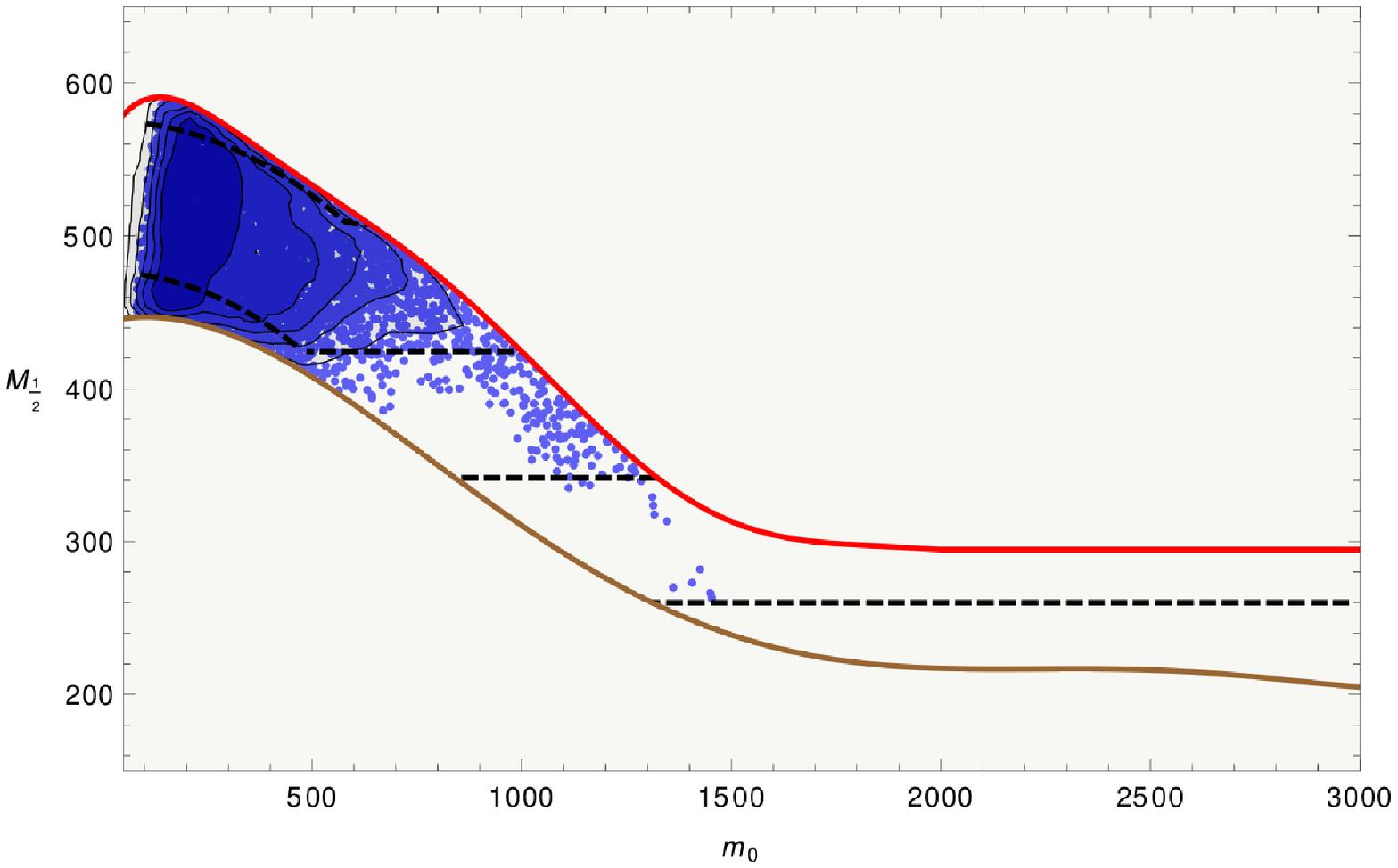} \\
$\,$ \\
\includegraphics[scale=.45]{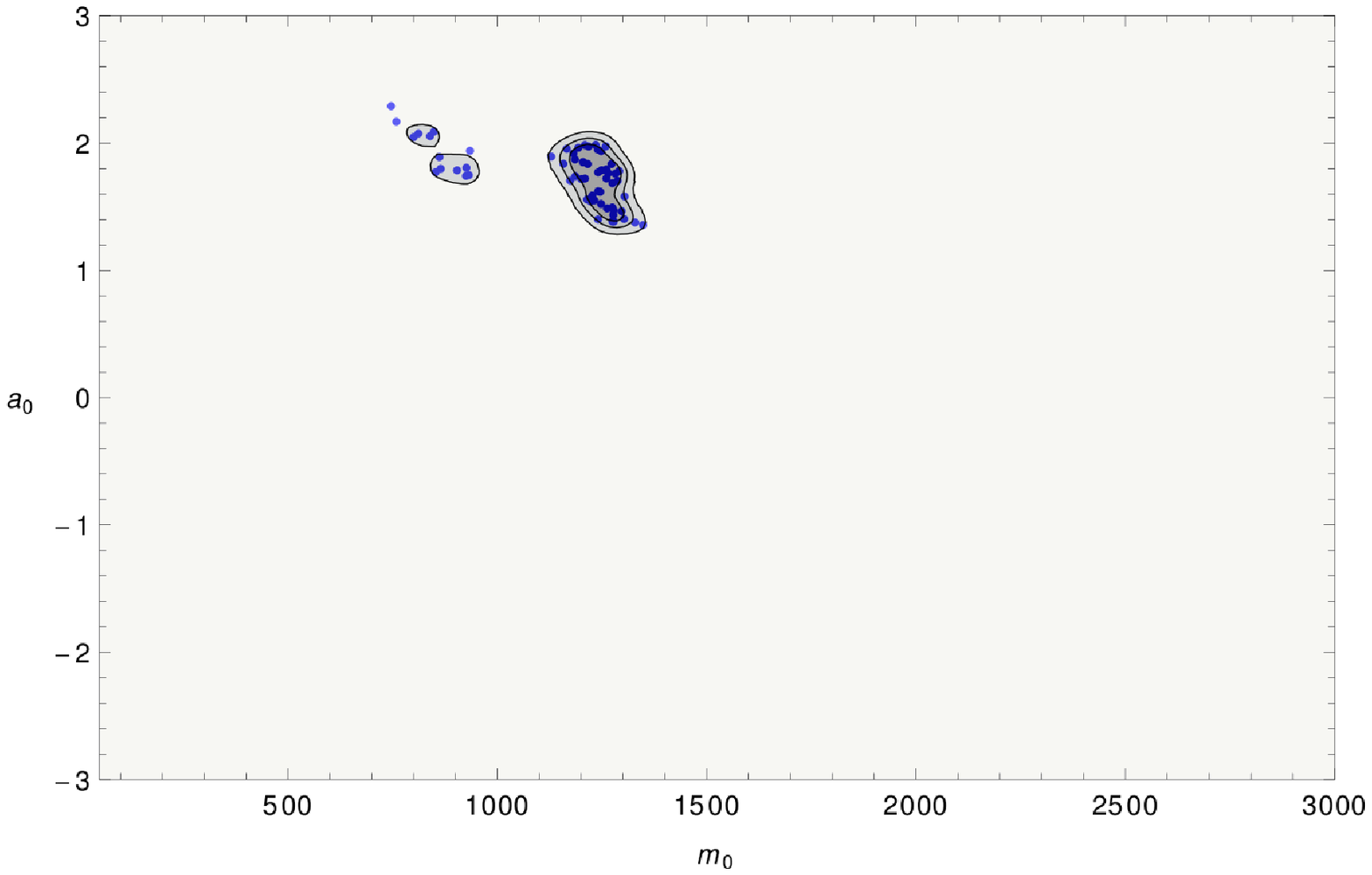}\quad\includegraphics[scale=.45]{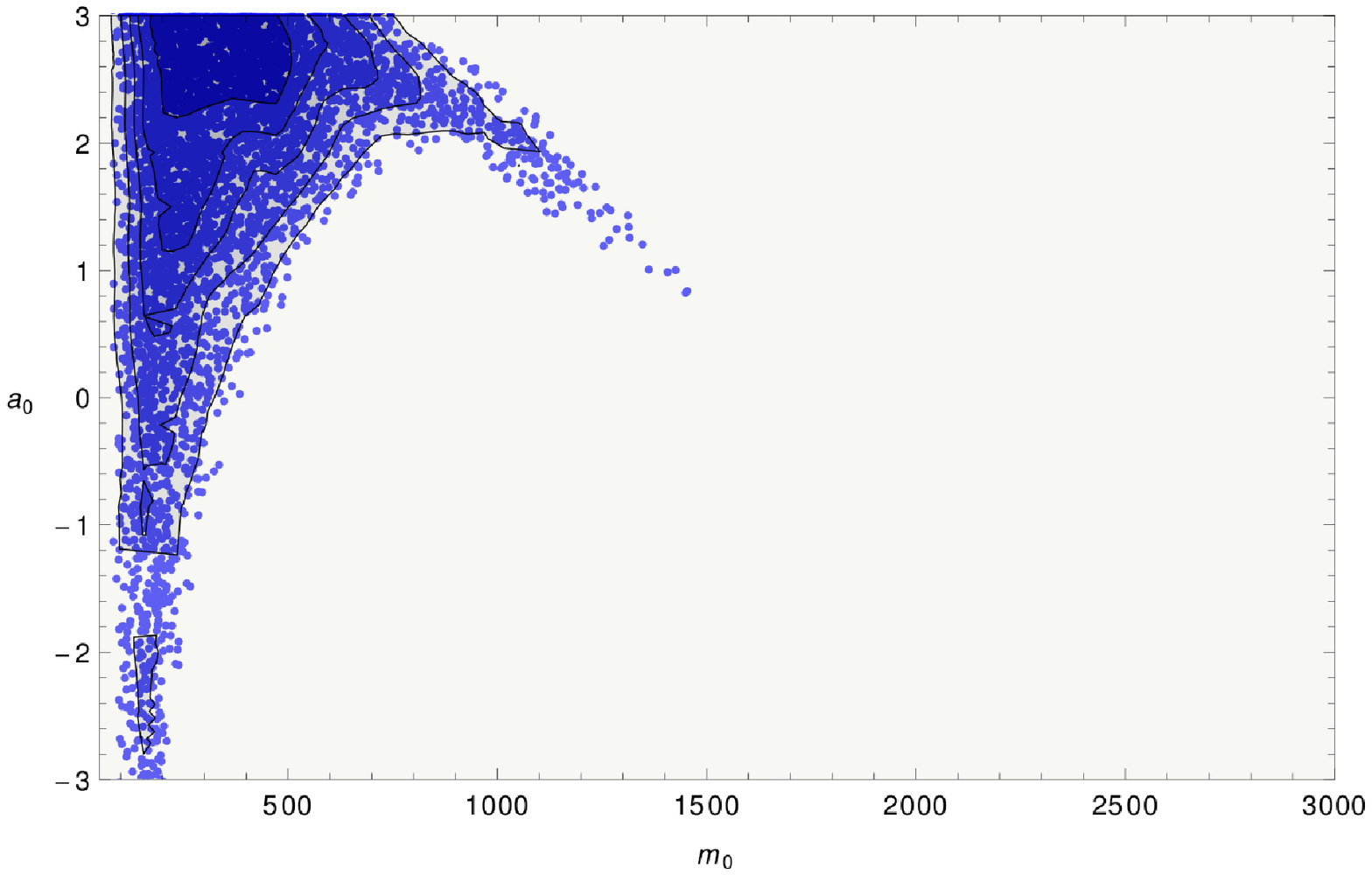} \\
$\,$ \\
\includegraphics[scale=.45]{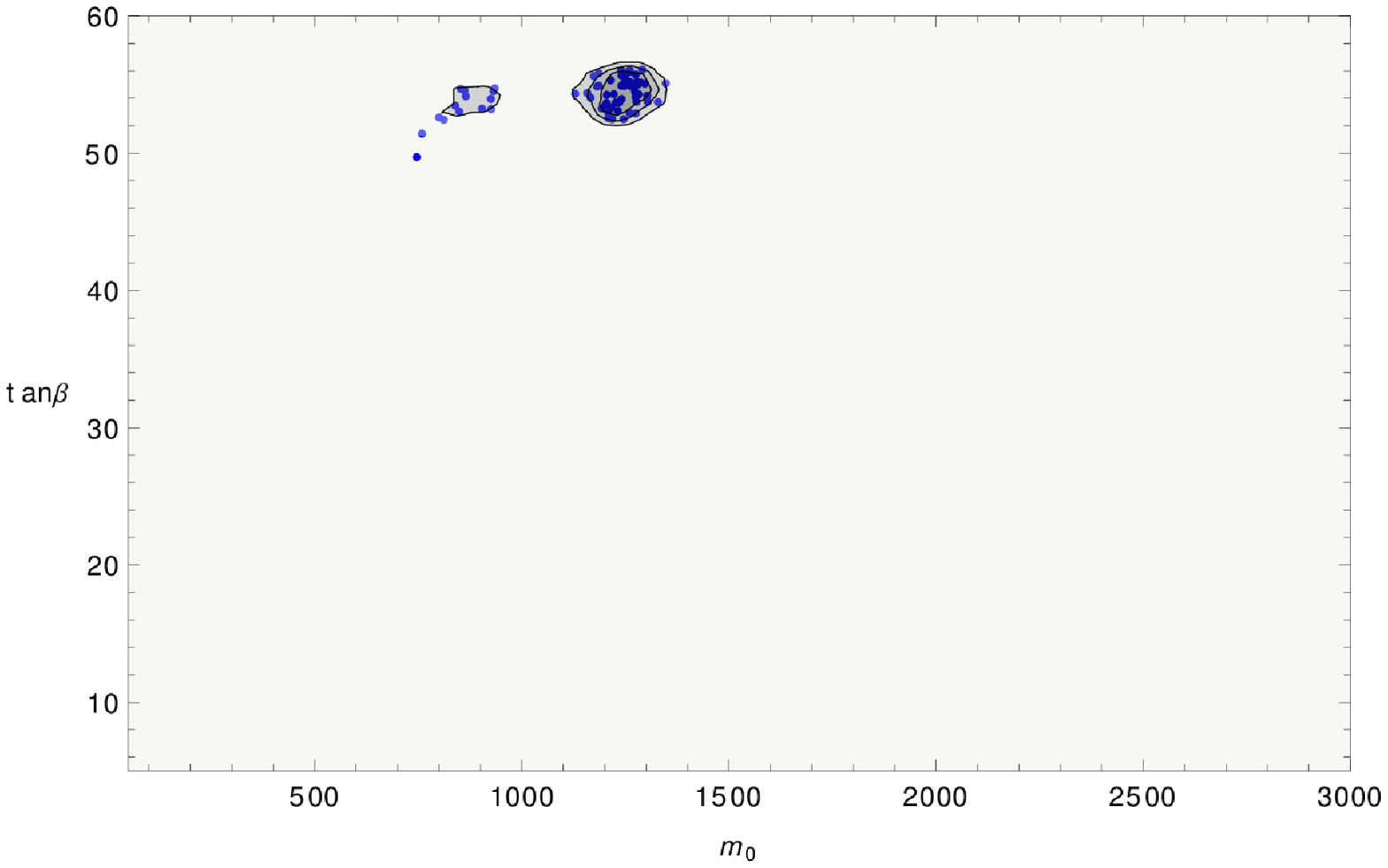}\quad\includegraphics[scale=.45]{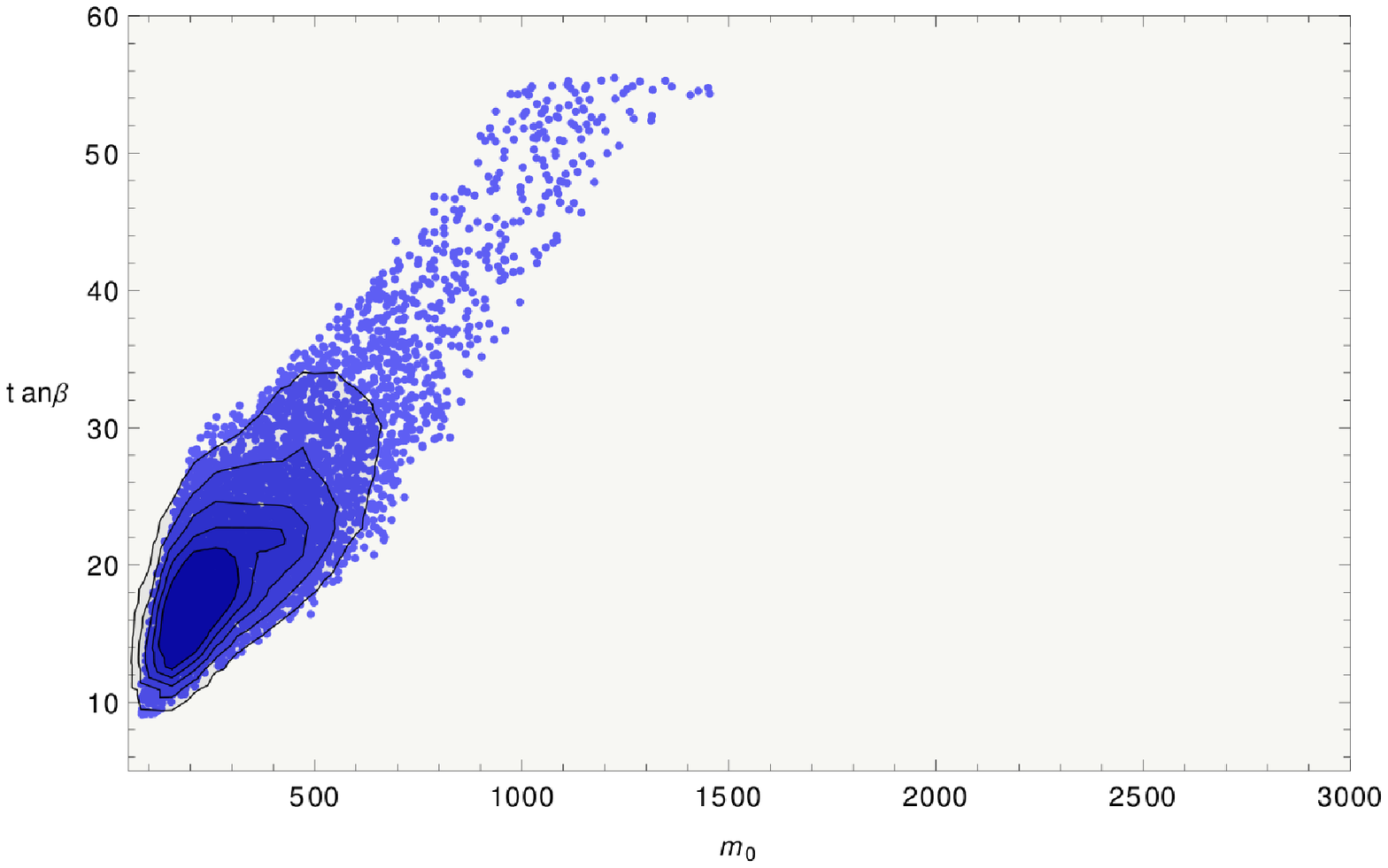}
\caption{Points in the parameter space satisfying experimental $2\sigma$ constraints. The plots on the left column are associated with a $3\sigma$ evidence signal of $B_s\to\mu^+\mu^-$ decay at LHCb with 2 fb$^{-1}$ of data, while the right column predict no such signal. We show the $m_0$--$M_{1/2}$, $m_0$--$a_0$ and $m_0$--$\tan\beta$ planes on the top, middle and bottom rows.}
\label{fig:sigma2}
\end{figure}

One can see more clearly the interplay between the collider and flavour observables by requiring stronger constraints on the latter. For instance, we can demand each point to satisfy flavour constraints at $2\sigma$, and, depending on what LHCb sees, ask for a $B_s\to\mu^+\mu^-$ signal of significance larger than $3\sigma$, or no signal at all. In Figure~\ref{fig:sigma2} we see the consequences of these requirements, with the upper plots corresponding to having at least a $3\sigma$ $B_s\to\mu^+\mu^-$ observation, while the lower plots correspond to not observing a signal at all. These Figures
show how the LHCb study of $B_s\to\mu^+\mu^-$, combined with the ATLAS/CMS results, can help to identify preferred 
regions in the SUSY parameter space.

The main conclusion from Figure~\ref{fig:sigma2} is that having a $3\sigma$ evidence for $B_s\to\mu^+\mu^-$ decay will most likely rule out a significant number of points with low values of $m_0$. In addition, $\tan\beta$ would be forced to remain large, and $a_0$ would only be positive. On the other hand, a lack of observation of $B_s\to\mu^+\mu^-$ would heavily disfavour points with large $m_0$, although it would not discard them completely. Also, $a_0$ would remain unfortunately unbounded (with a slight preference for large, positive values), while $\tan\beta$ would be preferred small, unless $m_0$ becomes large.

In the case where $B_s\to\mu^+\mu^-$ provides a significant signal, the correlation with $M_{\rm SUSY}$ (and thus $M_{\rm eff}$) would be a crucial check to determine if our signal is really due to SUSY. A $M_{\rm SUSY}$ lower than 850 GeV or larger than 1250 GeV would be incompatible with the observation of this process. In the case of no signal we can only make such a statement for low $M_{\rm SUSY}$, but, in addition, the correlation with this observable would help disentangle the more populated, low $m_0$ region from the less dense, medium $m_0$ region.

\subsubsection{Bounds from Electric Dipole Moments on the CPV phases}
\label{EDMinCMSSM}

The EDM of a fermion is the coupling constant related to a P- and T-violating interaction with a photon. As for LFV decays, the EDM of a particle like the electron ($d_e$) or the neutron ($d_n$) is vanishingly small in the SM, becoming thus an excellent probe of NP. Due to its CP violating nature, an EDM is connected to phases, and the non-observation of the former becomes an important constraint on the size of the latter. Currently, the strongest constraints come from Thallium~\cite{Regan:2002ta} (related to that of the electron), Mercury~\cite{Griffith:2009zz} and the neutron~\cite{Baker:2006ts} (both related to quarks).

In the MSSM, we find that the six relevant flavour-independent phases are related to the following invariants~\cite{Dugan:1984qf,Dimopoulos:1995kn}:
\begin{align}
 \arg\left(A_f^* M_i\right), & & \arg\left((B\mu)^*\mu A_f \right),
\end{align}
where $i=1,3$ and $f=u,d,e$. It is common to take the convention where $B\mu$ and the gluino mass $M_3$ are real parameters. Moreover, if gaugino universality is invoked, we can eliminate two further phases, and have only a phase for $\mu$ and a global phase for each $A_f$.

Further complex invariants can appear if one allows flavour-violating terms. For example, if one takes into account $\delta^f_{LL}$ and $\delta^f_ {RR}$ insertions, it is possible to construct the following invariants~\cite{Hisano:2008hn,Botella:2004ks}:
\begin{align}
\arg\left(Y_f\left[Y_f^\dagger\,Y_f\,,\,\delta^f_{LL}\right]\right), & &
\arg\left(\delta^f_{RR} Y_f\, Y_f^\dagger\, Y_f\right), & &
\arg\left(\delta^f_{LL}\,Y_f\,\delta^f_{RR}\right).
\end{align}
The dominance of one invariant over the other, may them be flavour-dependent or independent, is determined by the underlying flavour structure of the model. For instance, in the CMSSM, the $\delta^f_{RR}$ insertions vanish and the $\delta^f_{LL}$ phases are small, such that the main contribution to EDMs shall come from the flavour-independent terms. In contrast, in the flavour models we are taking as examples of a Flavoured CMSSM, flavour-independent phases are forbidden, while sizeable $\delta^f_{RR}$ are available, meaning that the only contributions to EDMs shall be those of the second type. For more details on the SUSY contributions to the EDM of the electron and neutron, for these models, we refer the reader to~\cite{Calibbi:2008qt,Calibbi:2009ja}.

For the CMSSM with and without right-handed neutrinos, it is possible to use $d_e$ to directly place upper bounds on the imaginary part of $\mu$ and $A_e$ at the EW scale. For $d_n$, as the prediction of the EDM depends on the neutron model one uses, the establishment of bounds is not so straightforward. In the following, we shall use the quark-parton (QP)~\cite{Ellis:1996dg} and chiral quark (CQ)~\cite{Manohar:1983md} models for $d_n$. By considering these models, it is possible to establish further bounds on the imaginary part of $\mu$, and new bounds on those of $A_u$ and $A_d$.

\begin{figure}
\begin{center}
\includegraphics[scale=.6]{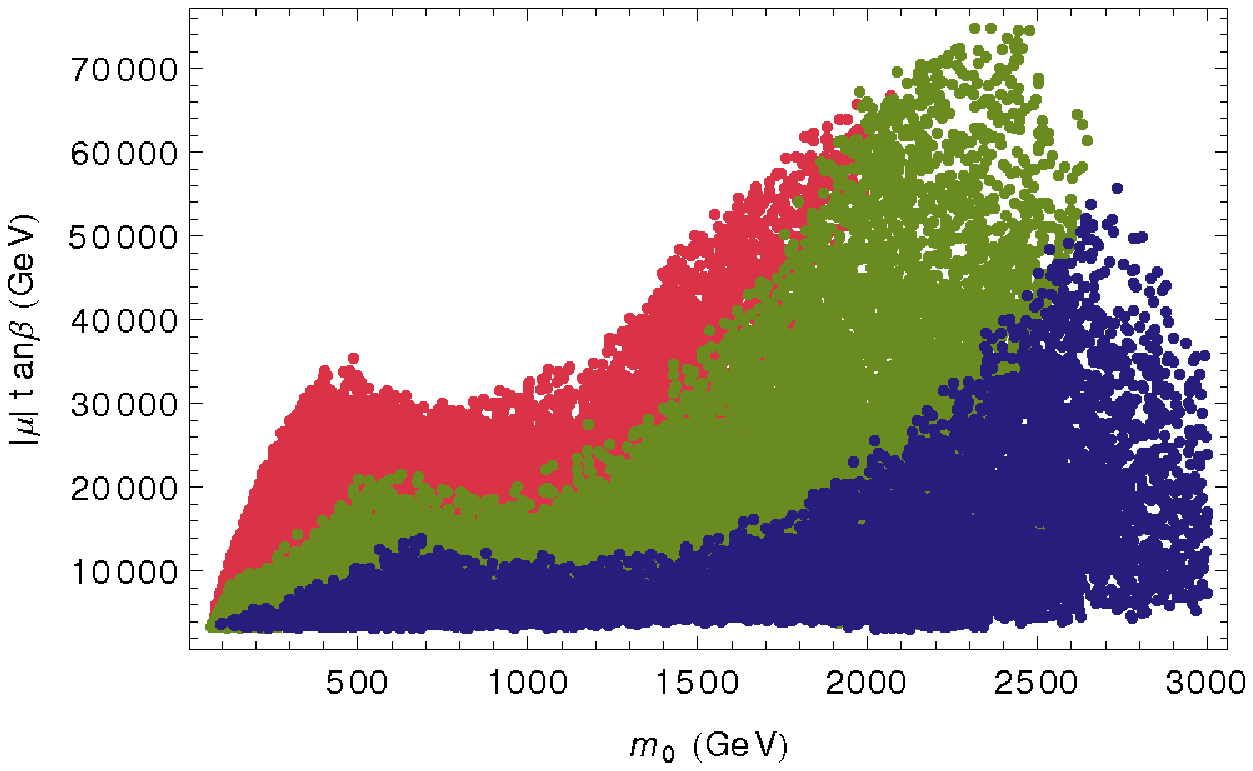} \qquad
\includegraphics[scale=.6]{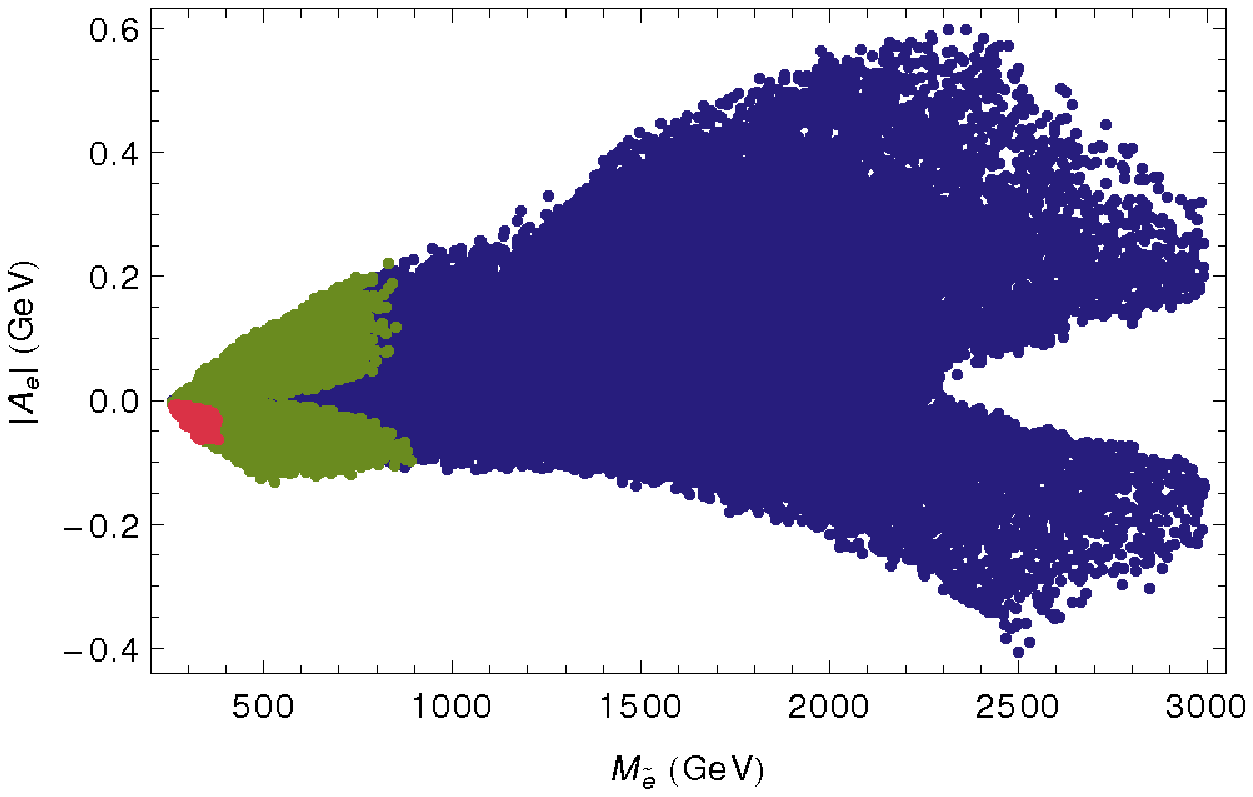} \\
\includegraphics[scale=.6]{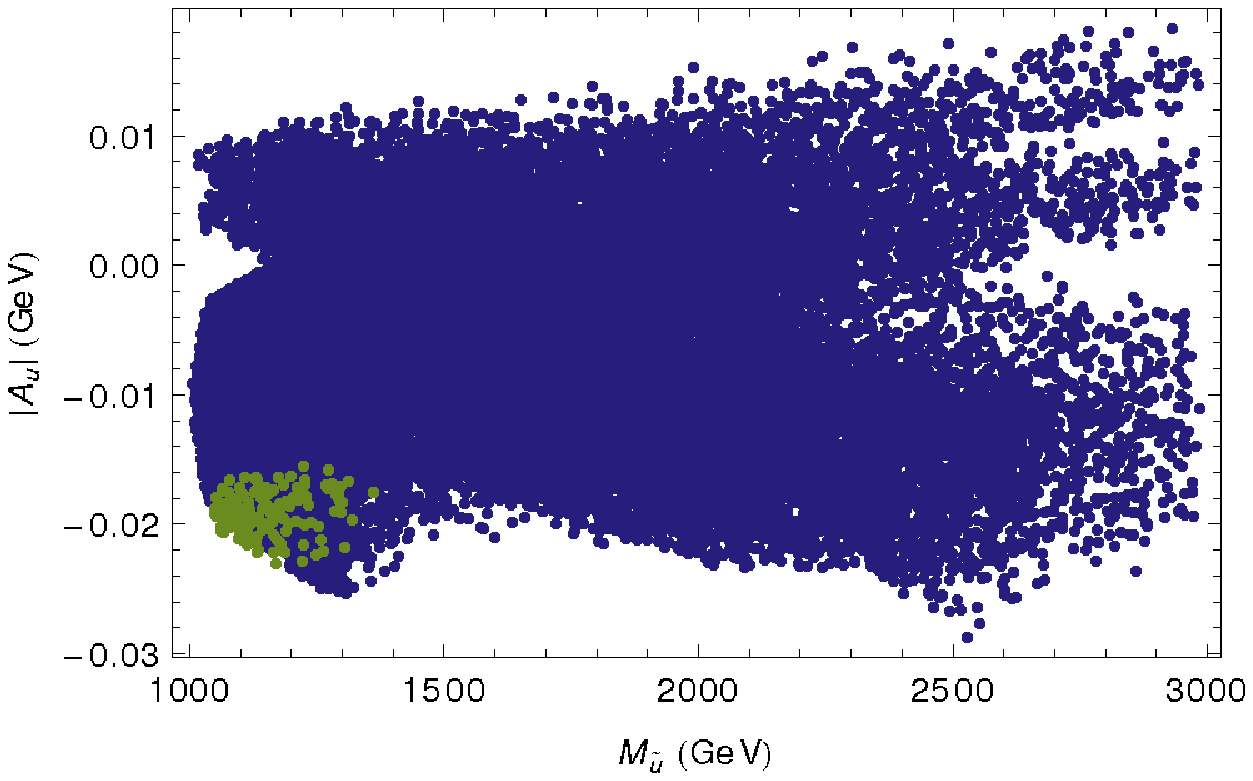} \qquad
\includegraphics[scale=.6]{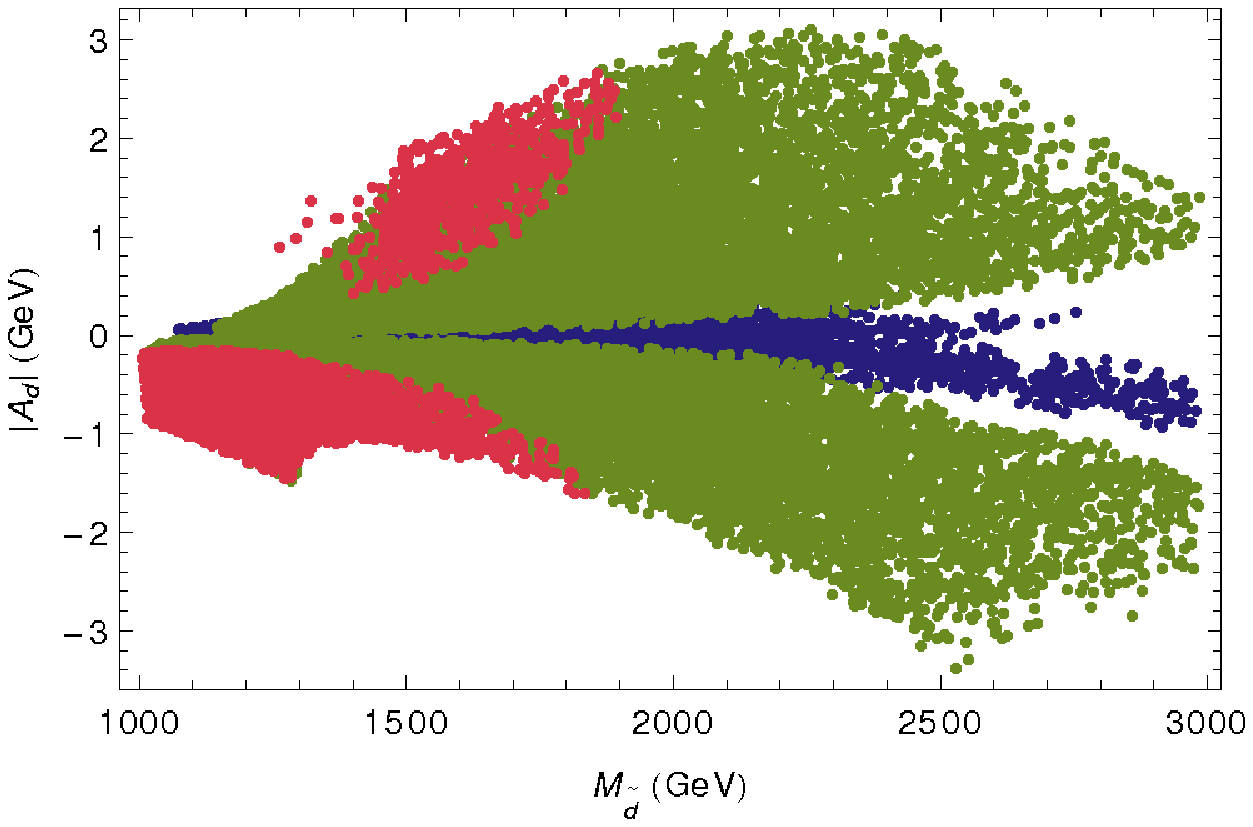}
\end{center}
\caption{From left to right, top to bottom, upper bounds on the global phase of $\mu$, $A_e$, $A_u$ and $A_d$ at the electroweak scale. For the trilinears, we specify the sign of their real part. The text explains the meaning of the colours.}
\label{fig:edmsCMSSM}
\end{figure}

Using only the points that survive direct search, Higgs search and flavour $3\sigma$ constraints, we have calculated the upper bound on the phases of $\mu$ ($\delta_\mu$) and the global phases of $A_f$ terms ($\delta_{A_f}$). Our results are shown in Figure~\ref{fig:edmsCMSSM}. The upper left panel shows the values of $\delta_\mu$ in the $|\mu|\tan\beta-m_0$ plane. We see in red the points where $\sin\delta_\mu<5\times10^{-3}$, in green those where $\sin\delta_\mu<10^{-2}$ and in blue those points where the phase can be larger. Within the data, we find a maximum possible phase $\sin\delta_\mu^{\rm max}=0.02$ rad.

The upper right panel shows the values of $\delta_{A_e}$, in the $|A_e|-M_{\tilde e}$ plane. Here, all blue points are unconstrained by the EDM data. The green points have $0.1<\sin\delta_{A_e}<1$, while the red points must have smaller values. We do not find any phase constraint smaller than $5\times10^{-2}$. We find that the constraints are weaker than those on the phase of $\mu$. This is due to the $\tan\beta$ enhancement that the $\mu$-insertion receives, as well as to an additional chargino contribution dependent on the imaginary part of $\mu$. Of course, as the EDM is proportional to the imaginary part of $A_e$, the size of its absolute value after the RG running also plays a role on how much the phase is constrained.

The lower left panel shows the constraints on $\delta_{A_u}$, in the $|A_u|-M_{\tilde u}$ plane. Again, the blue points are unconstrained, while the green points have some weak upper bound for $\sin\delta_{A_u}$, all of them larger than 0.5. Here we find that the absolute value of $A_u$ after the running is smaller than that for $A_e$, so it is reasonable to find even weaker constraints than in the former case.

Finally, the lower right panel shows the constraints on $\delta_{A_d}$, in the $|A_d|-M_{\tilde d}$ plane. This time, as the average size of $|A_d|$ is larger, we find somewhat stronger constraints. The red points have $\sin\delta_{A_d}<0.1$, the green points have $\sin\delta_{A_d}<1$, and the blue points are unconstrained. The maximum bound is never lower than $10^{-2}$.

To summarise, within the region accessible to LHC7 with 5 fb$^{-1}$, we find the requirement of a very strong suppression on the phase of $\mu$, and a somewhat milder one for those of $A_e$ and $A_d$. The phase of $A_u$ is much less constrained. The fact that the imaginary part of most flavour independent complex parameters need to be suppressed is commonly known as the SUSY CP problem.

\subsection{SUSY seesaw}
\label{SUSYSeeSaw-Pheno}

\begin{figure}[t]
\includegraphics[width=.55\textwidth]{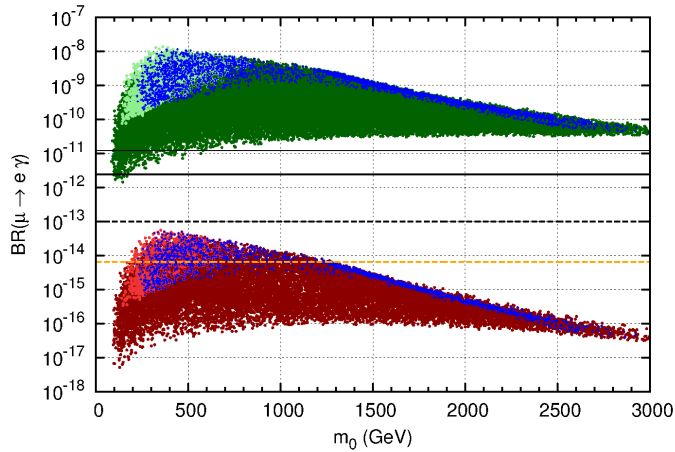}   
\caption{$Y_\nu \sim Y_u$ scenario: ${\rm BR}(\mu\to e\gamma)$ as a function of $m_0$ 
for the CKM (red) and the PMNS (green) case.}
\label{fig:msusy-meg}
\end{figure}

We present here the results for the SUSY seesaw scenarios introduced in section \ref{sec:seesaw},
starting with the $Y_\nu \sim Y_u$ case. 

In Figure~\ref{fig:msusy-meg}, we plot ${\rm BR}(\mu\to e\gamma)$ 
as a function of $m_0$ for the CKM-like (dark red) and the PMNS-like (dark green) mixing cases. All points satisfy the flavour constraints at $3\sigma$, as in the previous section.
The lighter points correspond to a contribution to $a_\mu$, which lowers the tension with the experiments
below the 1$\sigma$ level.
The blue points provide a sizeable BR($B_s\to\mu^+\mu^-$), such that a 3$\sigma$ evidence for such a decay is expected 
to be found at LHCb in the upcoming months (BR $\gtrsim 5\times 10^{-9}$). 
The thick horizontal line represents the current best limit recently published by the MEG experiment,
${\rm BR}(\mu\to e\gamma) < 2.4\times 10^{-12}$ at 90\% CL~\cite{Adam:2011ch} (for comparison we display the previous limit 
$1.2\times 10^{-11}$ as well~\cite{Ahmed:2001eh}). The black dashed line represents the expected final sensitivity of MEG, $\sim 10^{-13}$.
The meaning of the orange dashed line will be explained below.

From the Figure, we see that, within the parameter space region accessible to LHC7, 
the PMNS case seems to be completely ruled out by the recent MEG analysis.
This is a consequence of the large eigenvalue (we remind that in both cases $y_{\nu_3} \simeq y_t$ at the GUT scale)
and the large mixing angles in the neutrino Yukawa matrix. From Tab.~\ref{tab:seesaw-deltas}, we see that in the PMNS
case we have a dependence of BR($B_s\to\mu^+\mu^-$) on $U_{e3}$. For the plot in Figure~\ref{fig:msusy-meg}, we used 
a value of $U_{e3}$ in the lower side of the range preferred by recent T2K results, $U_{e3}=0.08$. 
If $U_{e3}$ will turn out to be smaller (but non-vanishing), we can still conclude that most of the parameter space in the PMNS case either is excluded or can be tested soon at MEG 
(we remind that in this case ${\rm BR}(\mu\to e\gamma)\propto |U_{e3}|^2$).\footnote{For vanishing small values 
of $U_{e3}$, on the other hand, running effects of $U_{e3}$ itself and double mass-insertion contributions 
become important and still guarantee a sizeable ${\rm BR}(\mu\to e\gamma)$ \cite{Calibbi:2006nq,Calibbi:2006ne}.}

On the contrary, the CKM case seems to escape the future MEG sensitivity. 
This is a consequence of the small mixing angles in the $Y_\nu$ combined with the present LHC bound on SUSY particle
masses, which leads to already quite heavy sleptons and gauginos within models with CMSSM-like boundary conditions.
There is however a way to test LFV in the $\mu$-$e$ sector beyond the sensitivity of MEG in the next years.
This is represented by experiments searching for $\mu\to e$ conversion in nuclei.
We remind that, within SUSY models, there is typically a striking correlation between $\mu\to e\gamma$ and the 
$\mu\to e$ conversion
in nuclei, namely the $\mu\to e$ conversion rate is well approximated by
\begin{equation}
 {\rm CR}(\mu\to e~{\rm in~N})\simeq \alpha_{\rm em}\times{\rm BR}(\mu\to e\gamma). 
\label{eq:cr}
\end{equation}
This means that our prediction for BR$(\mu\to e\gamma)$ can be easily translated in an estimate for 
${\rm CR}(\mu\to e\,{\rm in\, N})$. The proposed $\mu-e$ conversion experiments at Fermilab~\cite{Carey:2008zz}
and at J-PARC~\cite{comet} aim at sensitivities below $10^{-16}$ on ${\rm CR}(\mu\to e~{\rm in~ Ti})$. 
In Figure~\ref{fig:msusy-meg}, we show with an orange dashed-line how a sensitivity reach 
on ${\rm CR}(\mu\to e~{\rm in~ Ti})$ at the level of $5\times 10^{-17}$ would translate in terms of 
${\rm BR}(\mu\to e\gamma)$, using the trivial approximation in Eq.~(\ref{eq:cr}).
This shows that these experiments can actually access the CKM case parameter space 
and test at least the region favoured by $a_\mu$ (which, interestingly, 
might also provide $B_s\to\mu^+\mu^-$ above the SM prediction).
\begin{figure}[t]
\includegraphics[width=.45\textwidth]{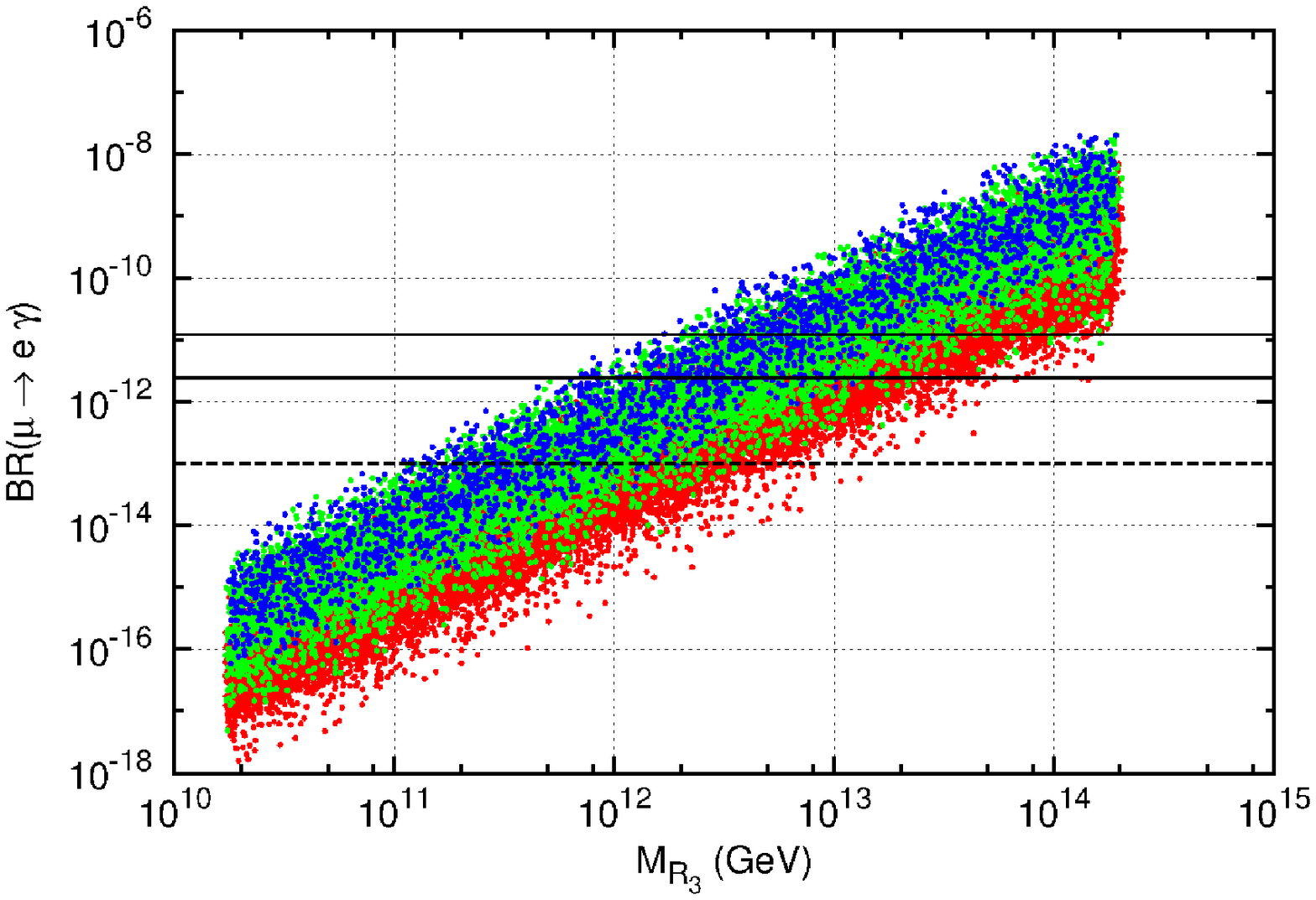} \qquad
\includegraphics[width=.45\textwidth]{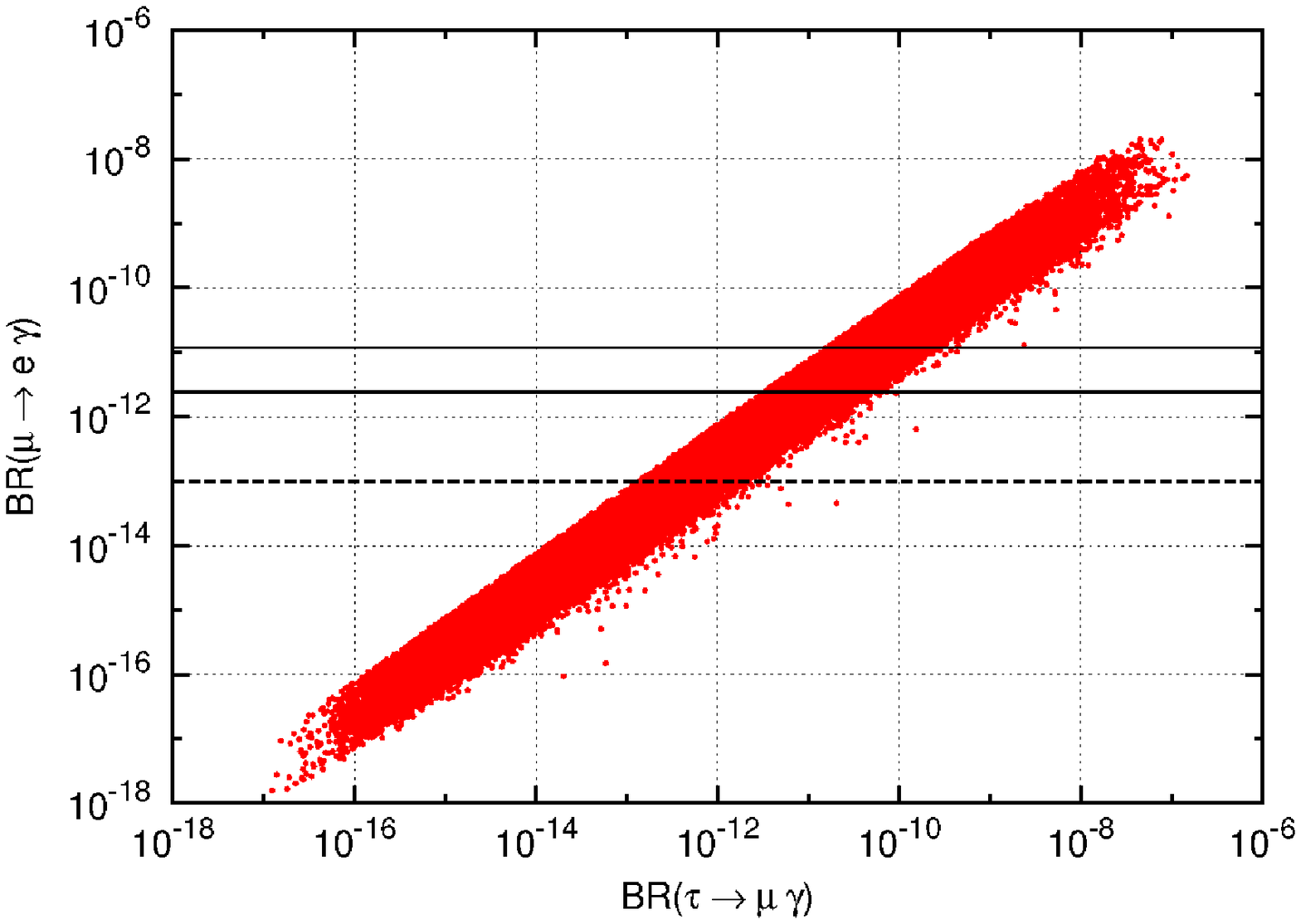}   
\caption{$R={\mathbf 1}$ scenario with hierarchical $M_R$. Left panel: ${\rm BR}(\mu\to e\gamma)$ vs.~$M_{R_3}$ 
(green points give $\delta a_\mu > 10^{-9}$, blue points $\delta a_\mu > 2\times 10^{-9}$); right panel:
 ${\rm BR}(\mu\to e\gamma)$ vs.~${\rm BR}(\tau\to \mu \gamma)$.}
\label{fig:R1plots}
\end{figure}

We now discuss the results of the numerical analysis of the scenario with $R={\mathbf 1}$ and hierarchical RH neutrinos.
We varied $M_{R_3}$ approximately in the range $3\times 10^{10}\div 3\times 10^{14}$ GeV and $U_{e3}$ was also randomly
varied within the 95\% CL range recently provided by T2K, $U_{e 3}\simeq 0.08\div 0.28$ \cite{Abe:2011sj}. 

In the left panel of Figure~\ref{fig:R1plots}, we plot  ${\rm BR}(\mu\to e\gamma)$ as a function of the heaviest RH neutrino 
$M_{R_3}$. The green points gives a SUSY contribution to the 
muon magnetic moment $\delta a_\mu > 10^{-9}$, the blue ones correspond to $\delta a_\mu > 2\times 10^{-9}$.
We see that the present bound ${\rm BR}(\mu\to e\gamma)< 2.4\times 10^{-12}$
already constrains $M_{R_3} \lesssim 6\times 10^{13}$ GeV (with $M_{R_3}\lesssim 4\times 10^{13}$ GeV for the blue points, 
for which the $(g-2)_\mu$ tension is lowered below 1$\sigma$). 
A negative result at MEG would further lower this bound to about $M_{R_3} \lesssim 5\times 10^{12}$ GeV
($M_{R_3}\lesssim 2\times 10^{12}$ GeV for the blue points).
On the other hand, we see that a positive signal at MEG would constrain the third RH neutrino mass in the following range:  
$10^{11}~{\rm GeV}~\lesssim M_{R_3}\lesssim 10^{14}~{\rm GeV}$
(for instance ${\rm BR}(\mu\to e\gamma)\simeq 10^{-12}$ would imply 
$5\times 10^{11}~{\rm GeV}~\lesssim M_{R_3}\lesssim 5\times 10^{13}~{\rm GeV}$).

In the right panel of Figure~\ref{fig:R1plots}, we show ${\rm BR}(\mu\to e\gamma)$ vs.~${\rm BR}(\tau\to \mu \gamma)$ for 
the same scenario. As we can see, the numerical results are consistent with the estimate for the ratio of the two branching ratios, $R_{\tau \mu}$, in Tab.~\ref{tab:seesaw-deltas}. 
As a consequence, the present bound on $\mu\to e\gamma$ already constrains ${\rm BR}(\tau\to \mu \gamma)$ to be at most 
$\mathcal{O}(10^{-10})$, beyond the sensitivity of SuperB~\cite{KEK} and SuperFlavour~\cite{superF} factories.
A positive signal for $\tau\to \mu \gamma$ at these facilities would then rule out this scenario.
This is a consequence of the T2K results, which prefer quite large values of $U_{e 3}$ and disfavour scenarios
with suppressed ${\rm BR}(\mu\to e \gamma)$ which usually require hierarchical RH neutrinos and vanishing $U_{e 3}$.

We do not display the case with $R={\mathbf 1}$ and almost degenerate RH neutrinos, which exhibits only a mild 
dependence on $U_{e 3}$ (as can be easily checked by means of the expressions of Tab.~\ref{tab:seesaw-deltas}) 
and predicts values of ${\rm BR}(\mu\to e\gamma)$ typically of the same order of the hierarchical case for the T2K 
range of $U_{e 3}$.\footnote{An exception is realised if the degree of degeneracy of $M_{R_i}$ is higher than 
5\% and the light neutrinos are almost degenerate as well, $m_{\nu_1}\simeq 0.1$ eV. This is the only setup of
the parameters for which, within the T2K range, ${\rm BR}(\mu\to e\gamma)$ can be suppressed with respect to the
hierarchical case up to 2 orders of magnitude.}

In models with a CMSSM-like spectrum, the mass splitting between selectrons and smuons 
($\Delta m_{\tilde \ell}/m_{\tilde \ell}$) is an LHC observable directly correlated 
to LFV in the $\mu$-$e$ as well as in the $\tau$-$\mu$ sector: 
$\Delta m_{\tilde \ell}/m_{\tilde \ell}\approx {\rm max}(\delta^e_{12},\delta^e_{23}/2)$~\cite{Buras:2009sg}. 
Such an inteplay has been recently addressed in the context of SUSY seesaw~\cite{Abada:2010kj}.
In our set-up, the mass splitting between LH selectron and LH smuon is always small,
$\Delta m_{\tilde \ell}/m_{\tilde \ell}\lesssim 10^{-3}$, as a consequence of the MEG bound and the T2K $U_{e3}$ range, whose combination prevents $(\delta^e_{LL})_{23}$ to reach sizeable values.
Even though resolving mass splitting at this level is definitvely beyond the reach of LHC7, it 
might be possible in the future $\sqrt{s}=14$ TeV run after some years of data taking~\cite{Allanach:2008ib,Buras:2009sg},
provided that $\mu\to e\gamma$ is in the reach of MEG.

\subsection{Flavoured CMSSM}

\subsubsection{Neutral Meson Mixing}

FCNC in neutral mesons pose one of the most stringent constraints to all models beyond the SM. The accurate measurement of the mass differences in the $K$, $D$, $B$ and $B_s$ sector, as well as that of the CPV parameters $\epsilon_K$ and $\sin2\beta$, suggest that any NP model contributing to flavour either is manifest at very high scales, or has strong flavour suppression~\cite{Isidori:2010kg}.

Recently, better theoretical predictions of non-perturbative parameters has lead to a small tension between $\epsilon_K$, $S_{\psi K_s}$ and $\Delta m_B/\Delta m_{B_s}$~\cite{Buras:2008nn,Altmannshofer:2009ne}. In addition, evidence for a large phase on the $B_s$ sector was reported in~\cite{Barberio:2008fa,Bona:2008jn}, which would be related to an anomalous dimuon charge asymmetry observed at D$\varnothing$~\cite{Abazov:2010hv,Abazov:2011yk}. Thus, it is possible that small hints in favour of NP might already be starting to appear in the flavour sector.

The relation between $\epsilon_K$, $S_{\psi K_s}$ and $\Delta m_B/\Delta m_{B_s}$ can be understood most easily through the following formula~\cite{Altmannshofer:2009ne}:
\begin{equation}
 |\epsilon_K|^{\textrm{SM}} = \kappa_\epsilon C_\epsilon \hat B_K |V_{cb}|^2|V_{us}|^2\left(
\frac{1}{2}|V_{cb}|^2 R_t^2 \sin2\beta\eta_{tt} S_0(x_t)+ R_t\sin\beta\left(
\eta_{ct} S_0(x_c,x_t)-\eta_{cc} x_c
\right)\right),
\end{equation}
where $R_t$ is a side of the Unitarity Triangle, defined through:
\begin{equation}
 R_t = \xi\frac{1}{\lambda}\sqrt{\frac{m_{B_s}}{m_B}}\sqrt{\frac{\Delta m_B^{SM}}{\Delta m_{B_s}^{SM}}},
\end{equation}
and $\sin2\beta$ is the SM contribution to $S_{\psi K_s}$:
\begin{equation}
\label{def_spsiks} 
\sin(2\beta+2\Phi_B) = S_{\psi K_s}.
\end{equation}
Here, $\Delta m^{SM}_F$ represents the SM prediction for the mass difference of the $F=B,B_s$ sectors, while $\Phi_B$ is a NP phase appearing in the $B$ sector. The definition of the rest of the parameters can be found in~\cite{Altmannshofer:2009ne}, and those we use in our predictions are shown in Table~\ref{tab:mesonpar}. Thus, a fixed value of $\sin2\beta$ and $\Delta m_B/\Delta m_{B_s}$ fixes $\epsilon_K$. As can be seen in~\cite{Barbieri:2011ci}, if one fits the CKM parameters without taking into account $\epsilon_K$, the predicted value shows a discrepancy with experiment beyond $2\sigma$.

\begin{table}[tbp]
 \begin{center}
\begin{tabular}{|c|rl||c|rl||c|rl||c|rl|}
\hline
$\hat B_K$ & $0.724\pm0.03$ & \cite{Aubin:2009jh} & $f_K$ & $0.1558\pm0.0017$ GeV & \cite{Laiho:2009eu} & $\eta_{tt}$ & $0.5765\pm0.0065$ & \cite{Buras:1990fn} & $\eta_B$ & $0.551\pm0.007$ & \cite{Buchalla:1996ys} \\
$\hat B$ & $1.22\pm0.05$ & \cite{Charles:2004jd} & $f_B$ & $0.194\pm0.009$ GeV & \cite{Ciuchini:2000de} & $\eta_{ct}$ & $0.496\pm0.047$ & \cite{Brod:2010mj} & $\kappa_\epsilon$ & $0.94\pm0.02$ & \cite{Buras:2010pza} \\
$\hat B_s$ & $1.28\pm0.04$ & \cite{Charles:2004jd} & $f_{B_s}$ & $0.239\pm0.01$ GeV & \cite{Ciuchini:2000de} & $\eta_{cc}$ & $1.87\pm0.76$ & \cite{Brod:2011ty} & & & \\
\hline
 \end{tabular}
 \end{center}
\caption{Parameters used in our simulations. The $B_i$ parameters for the SUSY contribution to our observables were taken from~\cite{Lubicz:2008am}.}
\label{tab:mesonpar}
\end{table}

On the other hand, the dimuon anomaly is related to the semileptonic CP asymmetries $A^d_{SL}$ and $A^s_{SL}$. The latter is in turn connected to the time-dependent CP asymmetry $S_{\psi\phi}$ through~\cite{Grossman:2009mn}:
\begin{equation}
 A^s_{SL} = - \frac{\Delta\Gamma_s}{\Delta M_{Bs}}\frac{S_{\psi\phi}}{\sqrt{1-S^2_{\psi\phi}}}
\end{equation}
The fit done in~\cite{Bauer:2010dga} shows that, in order to reproduce the observed anomaly, one requires both a larger $\Gamma^s_{12}$ than that predicted by the SM, and a non-zero NP phase in the $B_s$ sector, $\Phi_{B_s}$. The latter phase is connected to $S_{\psi\phi}$ in a way analogous to Eq.~(\ref{def_spsiks}):
\begin{equation}
\sin(2\beta_s-2\Phi_{B_s}) = S_{\psi\phi}.
\end{equation}
where $\sin2\beta_s\approx0.036\pm0.002$ is the SM prediction.\footnote{Note that the LHCb collaboration has recently carried out a combined analysis of $B_s\to\psi\phi$ and $B_s\to\psi f_0(980)$ with 0.3 fb$^{-1}$. The central value agrees with the SM prediction, albeit with large errors, $2\beta_s^{\rm exp}=0.03\pm0.16\pm0.07$~\cite{Spsiphi_Comb}.}

The MFV contributions of the CMSSM with and without right-handed neutrinos are known to be insufficient to solve the flavour tension~\cite{Buras:2009pj} nor the dimuon anomaly, although the latter can be slightly ameliorated in general MFV scenarios with large $\tan\beta$ and new phases~\cite{Lenz:2010gu}. If any of these models were the correct description of flavour physics in the MSSM, the tensions should disappear with inclusion of further data.

A Flavoured CMSSM, on the other hand, should be capable of providing larger contributions to FCNC. The examples mentioned in Section~\ref{FlavCMSSM}, in particular, can provide contributions to $\epsilon_K$ of the correct order of magnitude~\cite{Calibbi:2009ja}, and in some cases can even provide a large $\Phi_{B_s}$ phase~\cite{Altmannshofer:2009ne}.

\begin{figure}[tbp]
\includegraphics[scale=.6]{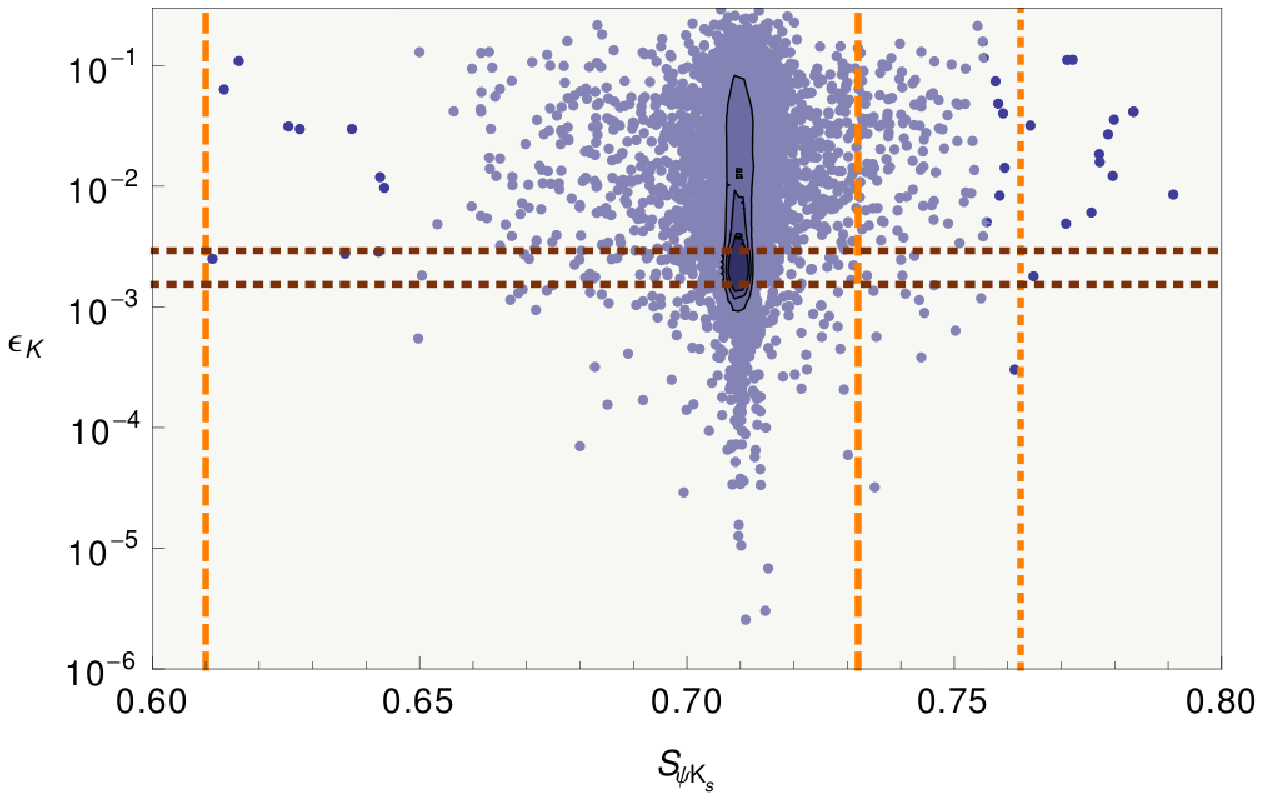} \qquad
\includegraphics[scale=.6]{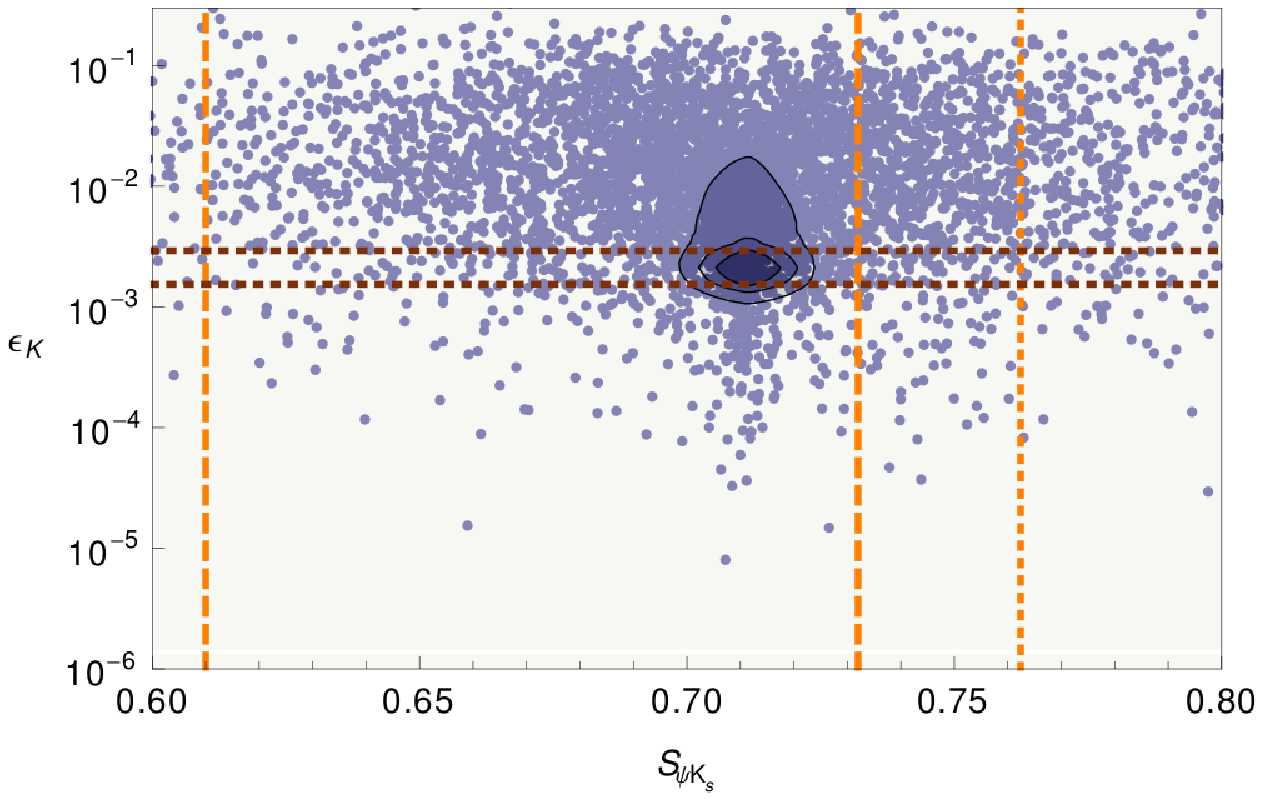}
\caption{Constraints due to $S_{\psi K_s}$ and $\epsilon_K$ in Model 1 (left) and Model 2 (right) from~\cite{Calibbi:2009ja}. Dotted (dashed) lines correspond to $3\sigma$ ($2\sigma$) constraints. All points satisfy LFV constraints.}
\label{fig:mesoncpv-rvv}
\end{figure}

Nevertheless, it turns out that the contributions to CPV observables can also exceed the requirements. In Figure~\ref{fig:mesoncpv-rvv}, we show typical values of $S_{\psi K_s}$ and $\epsilon_K$ for both of our examples. All points satisfy the LFV constraints, which shall be explained in detail in the next section. The dotted (dashed) lines correspond to $3\sigma$ ($2\sigma$) constraints, and include an approximate theoretical error, $\sigma^{\textrm{th}}_{S_{\psi K_s}}=0.02$ and $\sigma^{\textrm{th}}_{\epsilon_K}=0.23\times10^{-3}$. The contours again represent a qualitative measure of the density of points, which means that one should not compare contours from different models. We see that, for $\epsilon_K$, the contributions can easily exceed the $3\sigma$ bounds. Although for both models the main bulk of points is located within a region with no conflict with $\epsilon_K$, there is a very large number of points where the SUSY contribution is too large. In the second example, we see that it is possible to exceed also the bounds in $S_{\psi K_s}$. This means that solving the flavour tension can become a new, important source of constraints for any Flavoured CMSSM model.

\begin{figure}[tbp]
\includegraphics[scale=.6]{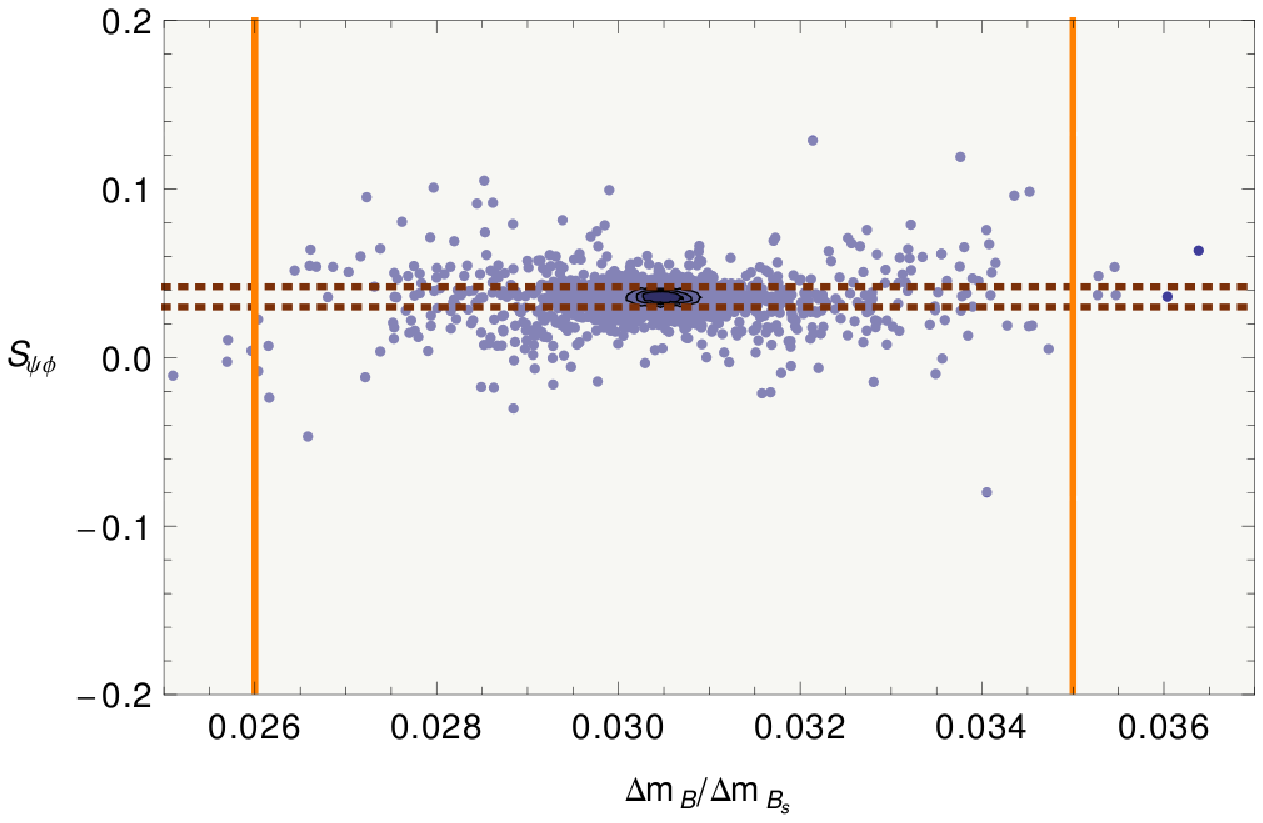} \qquad
\includegraphics[scale=.6]{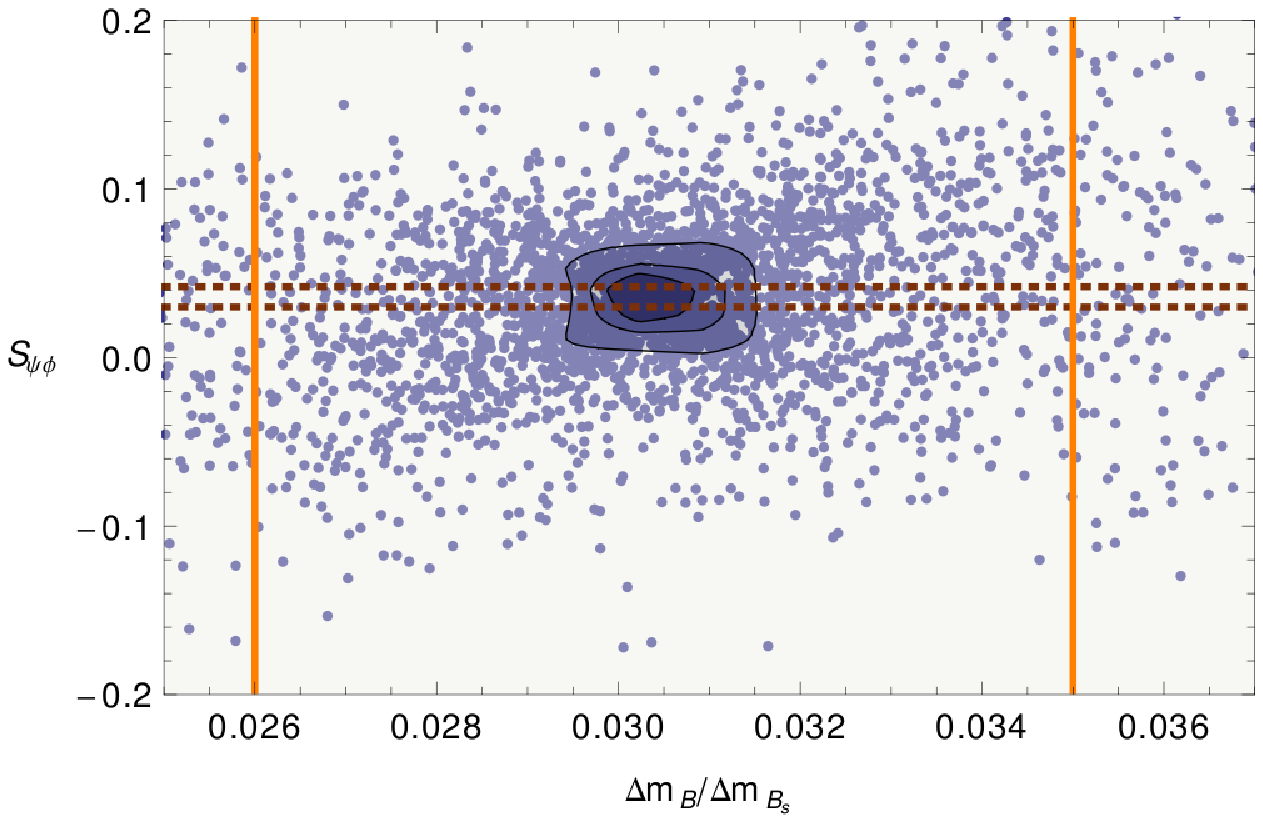}
\caption{Values of $S_{\psi\phi}$ and constraints due to $\Delta m_B/\Delta m_{B_s}$ in Model 1 (left) and Model 2 (right) from~\cite{Calibbi:2009ja}. The dashed (solid) lines correspond to $2\sigma$ ($1\sigma$) constraints.}
\label{fig:spsiphi-rvv}
\end{figure}

After applying the bounds on $S_{\psi K_s}$ and $\epsilon_K$, one can turn to $S_{\psi\phi}$. In Figure~\ref{fig:spsiphi-rvv} we show $S_{\psi\phi}$ vs $\Delta m_B/\Delta m_{B_s}$, where we have taken out all points that do not satisfy the $\epsilon_K$ and $S_{\psi K_s}$ constraints. The SM prediction for $S_{\psi\phi}$ at $3\sigma$ is shown by the brown dotted lines, while the $1\sigma$ bounds of $\Delta m_B/\Delta m_{B_s}$ are shown by the solid orange lines. We find that $\Delta m_B/\Delta m_{B_s}$ does not constrain any of the models.
Notice that in the second example, although we find it is very difficult to obtain a value larger than 0.1, it is not absolutely impossible. Thus, $S_{\psi\phi}$ could be used as a tool for distinguishing between Flavoured CMSSM models, provided that the experimentally measured value is somewhat larger than that of the SM. This is expected to be probed by LHCb in the near future.

If we study both Figures~\ref{fig:mesoncpv-rvv} and~\ref{fig:spsiphi-rvv} at the same time, we can see an interesting fact. Although both of our examples are based on the same symmetry, a small variation can cause important differences in the phenomenology. Furthermore, it is unlikely that the modification of a model shall affect only one single observable. As we can see in the Figures, although the second model allows us to achieve a larger $S_{\psi\phi}$, which is needed to solve the dimuon anomaly, this generates large deviations in $S_{\psi K_s}$, which is kept under much better control in the first model.

This is something to be expected of all Flavoured CMSSM models: as soon as one observable gets enhanced, one shall need to verify that all other important observables do not exceed their bounds. It would be desirable to follow patterns such as the one shown in these examples, where $S_{\psi\phi}$ is somewhat increased, and then $\epsilon_K$ and $S_{\psi K_s}$ receive very small contributions in the proper directions, such that their tension is cancelled.

\subsubsection{EDMs and LFV}

As in the previous CMSSM model with RH neutrinos, LFV processes can impose strong constraints on the parameter space of Flavoured CMSSM models. For the examples we have taken, it was shown in~\cite{Calibbi:2008qt} that both neutralino and chargino loops contributed significantly to the decay, being the situation particularly restrictive when $A_0\ne0$ due to an enhancement in the neutralino contribution. In addition, if the EDMs are generated by flavour-dependent phases, one could expect interesting correlations to arise between them and flavour-violating processes.

\begin{figure}[tbp]
\includegraphics[scale=.4]{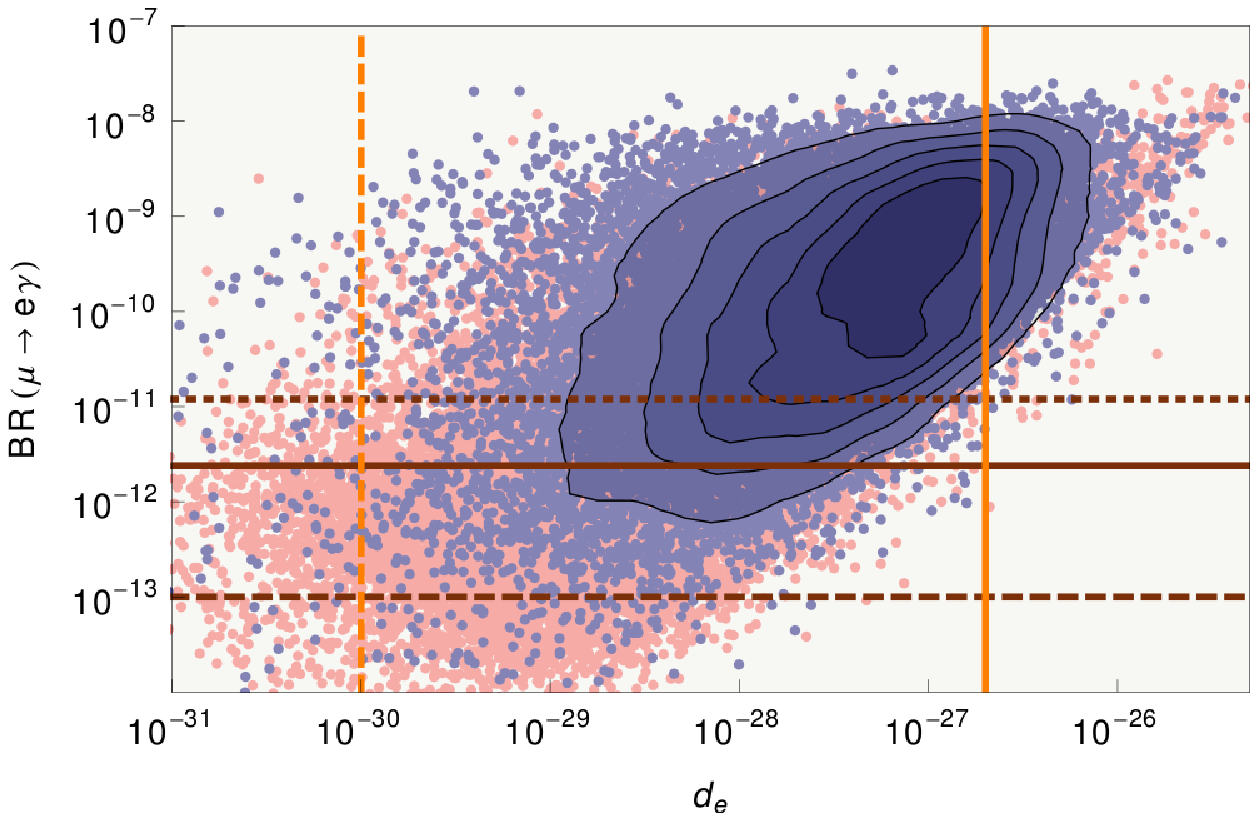}
\includegraphics[scale=.4]{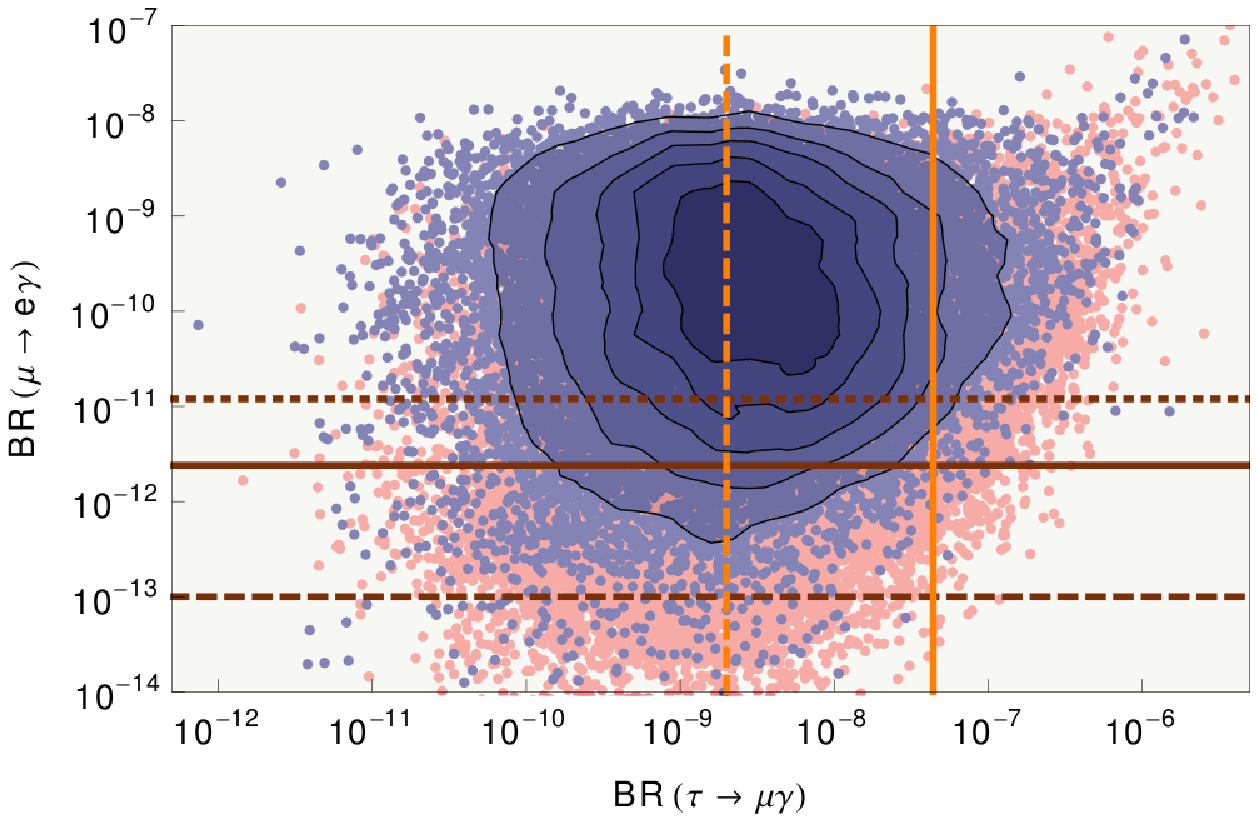}
\includegraphics[scale=.4]{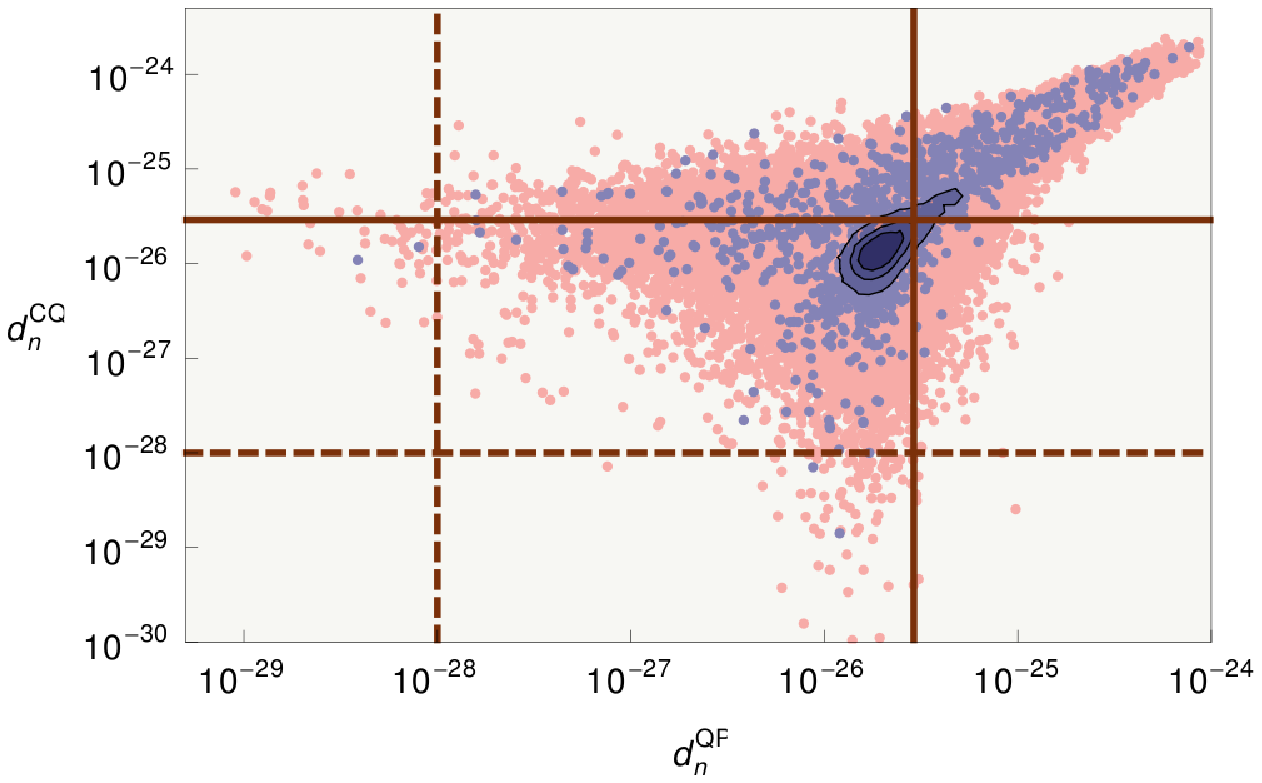}
\caption{Several prediction for LFV and EDMs, for the first example from~\cite{Calibbi:2009ja}. On the left panel, we have BR$(\mu\to e\gamma)$ vs $d_e$, on the central panel we show BR$(\mu\to e\gamma)$ vs BR$(\tau\to\mu\gamma)$ and on the right we compare $d_n^{CQ}$ and $d_n^{QP}$. On all plots, solid lines indicate current bounds, while dashed lines indicate prospects in future experiments. For BR$(\mu\to e\gamma)$, the dotted line shows the old MEGA bound. Pink points are ruled out by meson constraints.}
\label{fig:meg-edm-rvv}
\end{figure}

We show correlations between observables of interest in Figure~\ref{fig:meg-edm-rvv}. This time, when comparing both our examples, we do not find major differences between them, so we shall only show results for the first one. In all panels, points in pink give too large contributions to neutral meson processes, and are ruled out, while points in blue are allowed. Contours refer to the qualitative density of blue points. The solid lines give the current bounds for both processes, while the dashed line gives the future prospects.

On the left panel of Figure~\ref{fig:meg-edm-rvv}, we show the predictions for BR$(\mu\to e\gamma)$ vs $d_e$. Apart from the current and future bounds, we also show in dotted lines the old MEGA bound for $\mu\to e\gamma$. We find that, in general, $\mu\to e\gamma$ and the neutral meson bounds, such as $\epsilon_K$, work in opposite directions. Most points that satisfy the $\mu\to e\gamma$ bound are ruled out by the quark sector constraints, leaving only a relatively small number of points allowed. This is mainly due to the fact that $\epsilon_K$ prefers points with low $m_0$, which are in turn ruled out by the neutralino contribution to $\mu\to e \gamma$ whenever $|a_0|\gtrsim1$. We find that the points not ruled out by the current MEG bound do not have a too large value for $d_e$. Furthermore, we can see that the future prospects for MEG, combined with our conservative expectation for $d_e$ (based on~\cite{Lamoreaux:2001hb,cit:eedm}), can probe most of the still-allowed points in both models.

The central panel shows BR$(\mu\to e\gamma)$ vs BR$(\tau\to \mu\gamma)$. We see that both processes are useful for probing the parameter space. However, we find that very few points surviving the MEG constraint are ruled out by the current $\tau\to\mu\gamma$ bound. In addition, when looking at the future prospects, we can see that $\tau\to\mu\gamma$ is not as strong as $\mu\to e\gamma$ or $d_e$ in constraining the model.

Finally, the right panel shows the predictions for $d_n$ in both the CQ and QP models. In this panel, the pink dots also indicate points ruled out by LFV processes. Notice that, as we are comparing two neutron models, the only points effectively ruled out by $d_n$ lie on the upper right square limited by the solid lines. We find that, after applying all bounds, not many points are left. In addition, a future improvement in the measurement of $d_n$~\cite{cit:nedm} would be able to probe the square limited by the dashed lines, i.e.~all but the most fine-tuned points. Thus, if neutron EDM models prove to be consistent, we find that $d_n$ is the observable most sensitive to the predictions of these example models.

\subsubsection{Allowed Parameter Space}

To summarise, we find that a Flavoured CMSSM can be expected to saturate the neutral meson bounds, particularly in the CP violating sectors. Furthermore, the $\mu\to e\gamma$ branching ratio can rule out a sizeable amount of points that do satisfy the neutral meson bounds, providing a complementary constraint. Thus, although flavoured models such as those provided in the examples are not ruled out and can still provide interesting predictions, the allowed parameter space should be expected to be heavily constrained. 

\begin{figure}
\includegraphics[scale=.45]{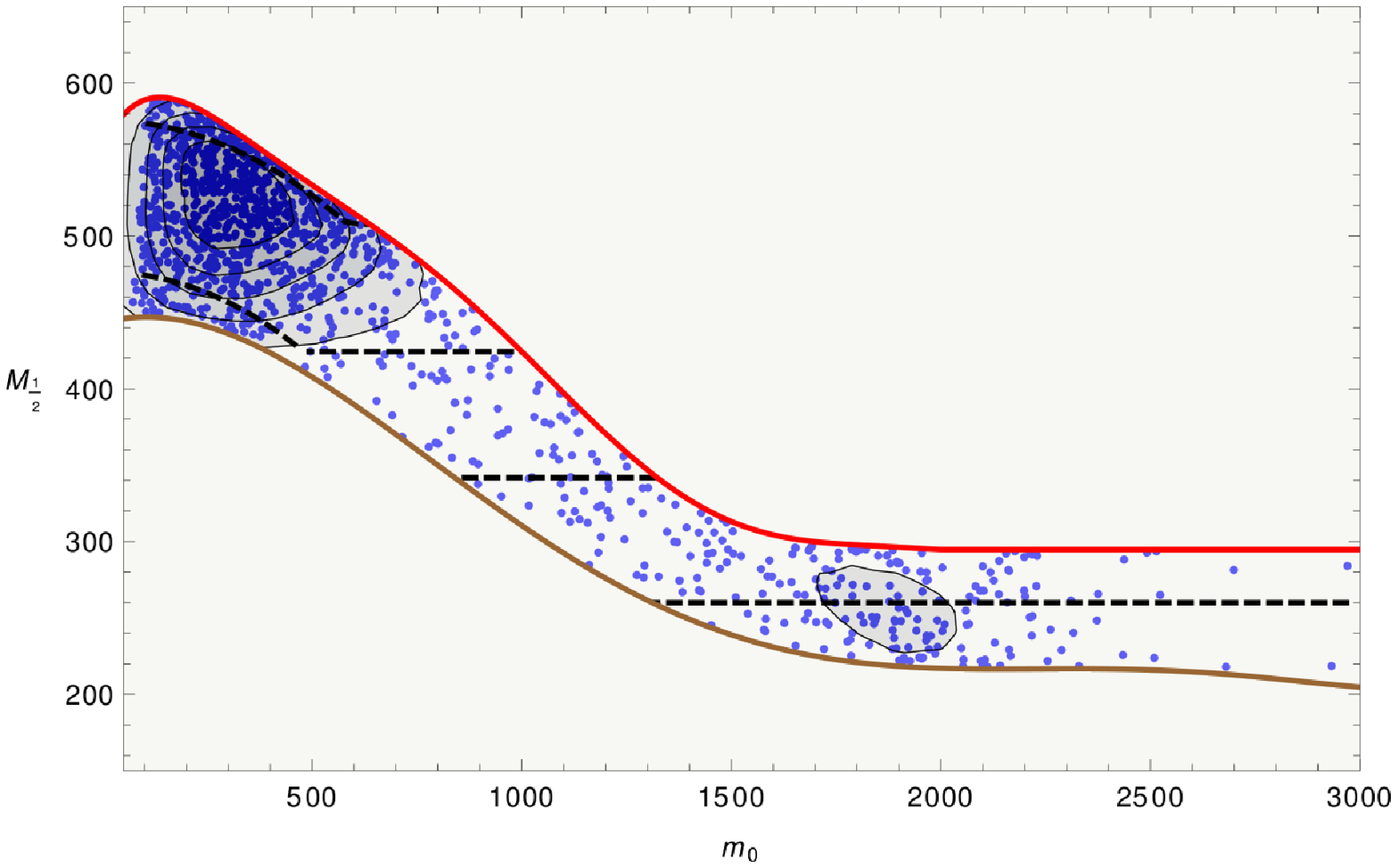} \quad \includegraphics[scale=.45]{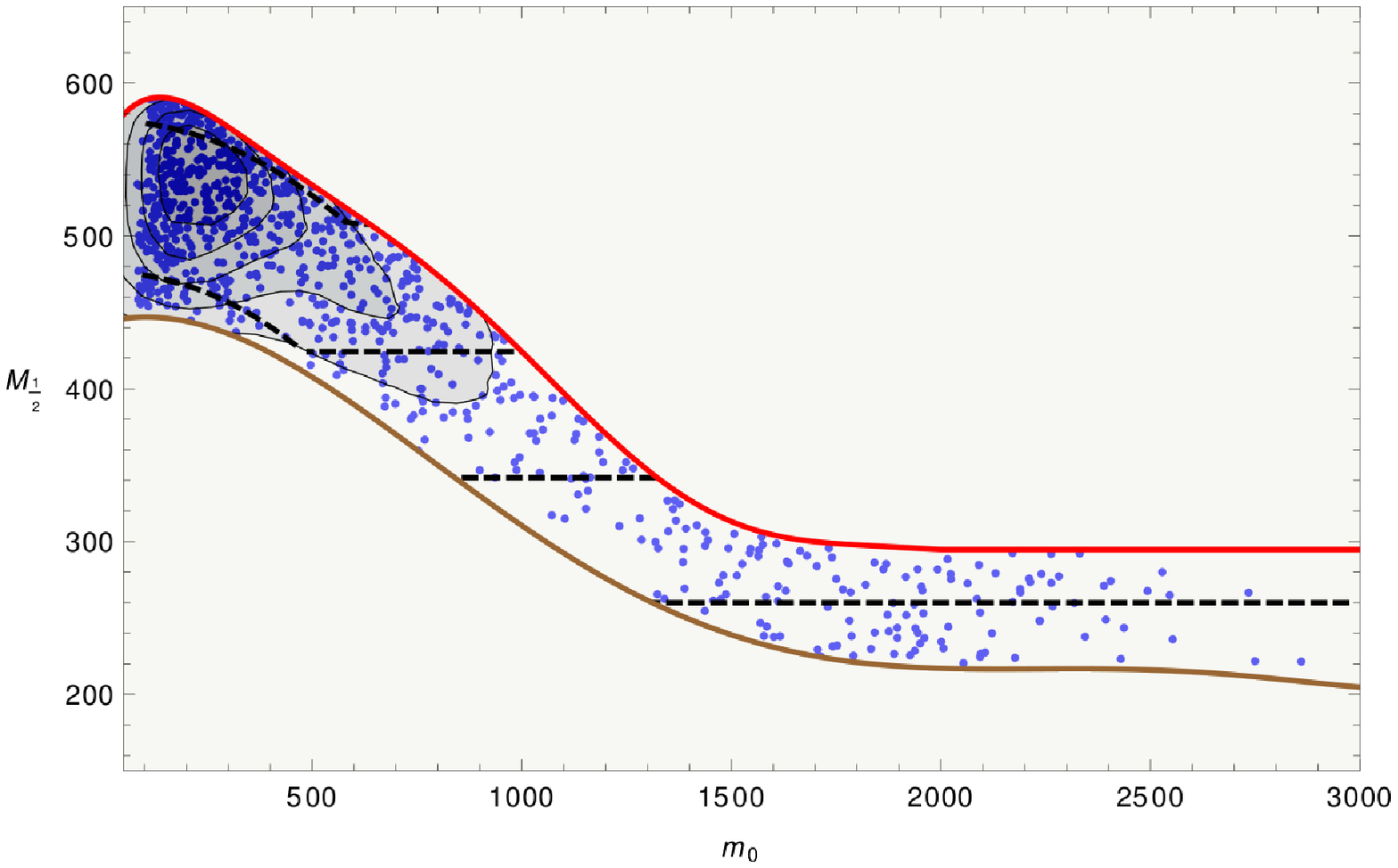}  \\
$\,$ \\
\includegraphics[scale=.45]{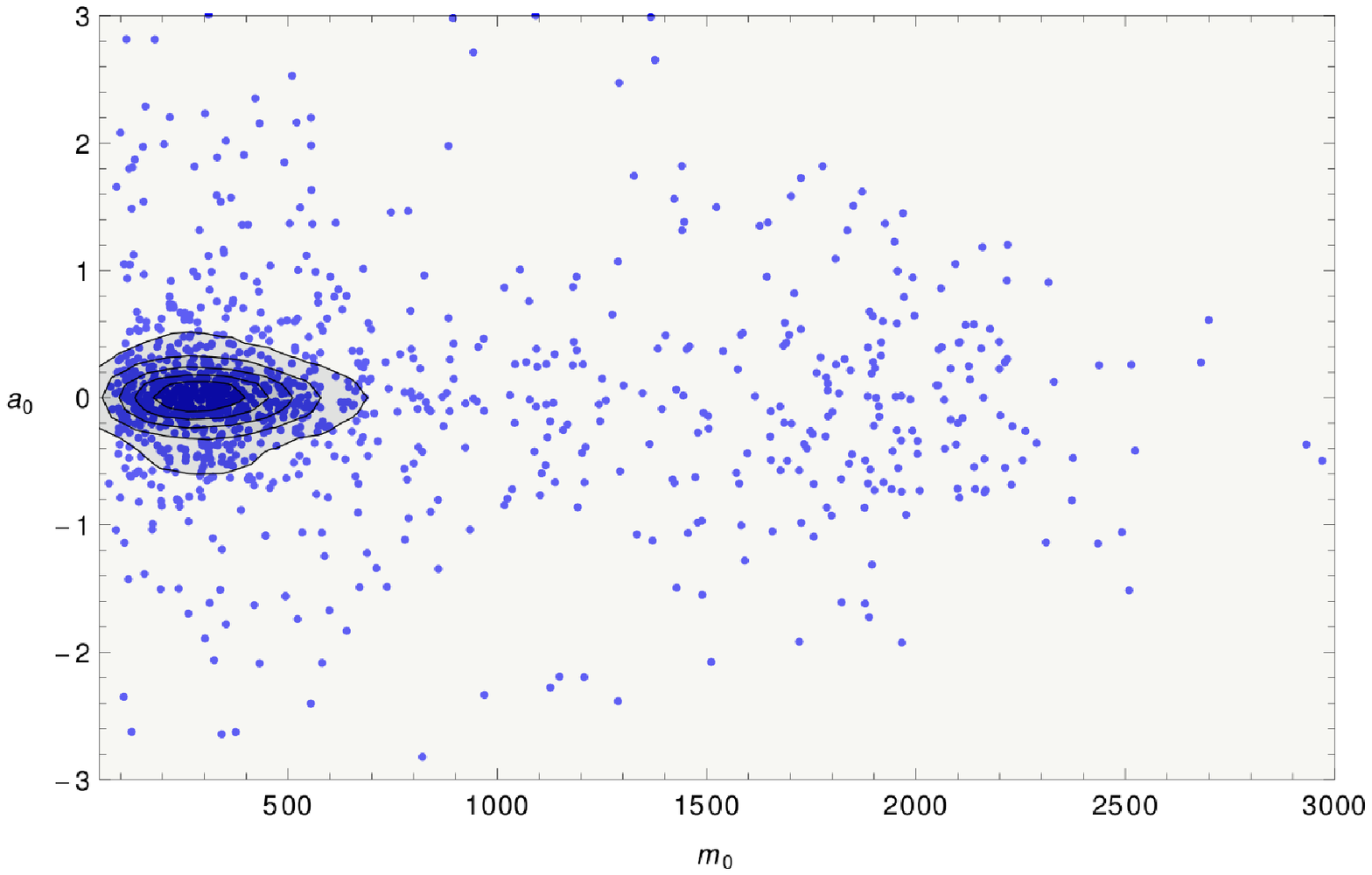} \quad \includegraphics[scale=.45]{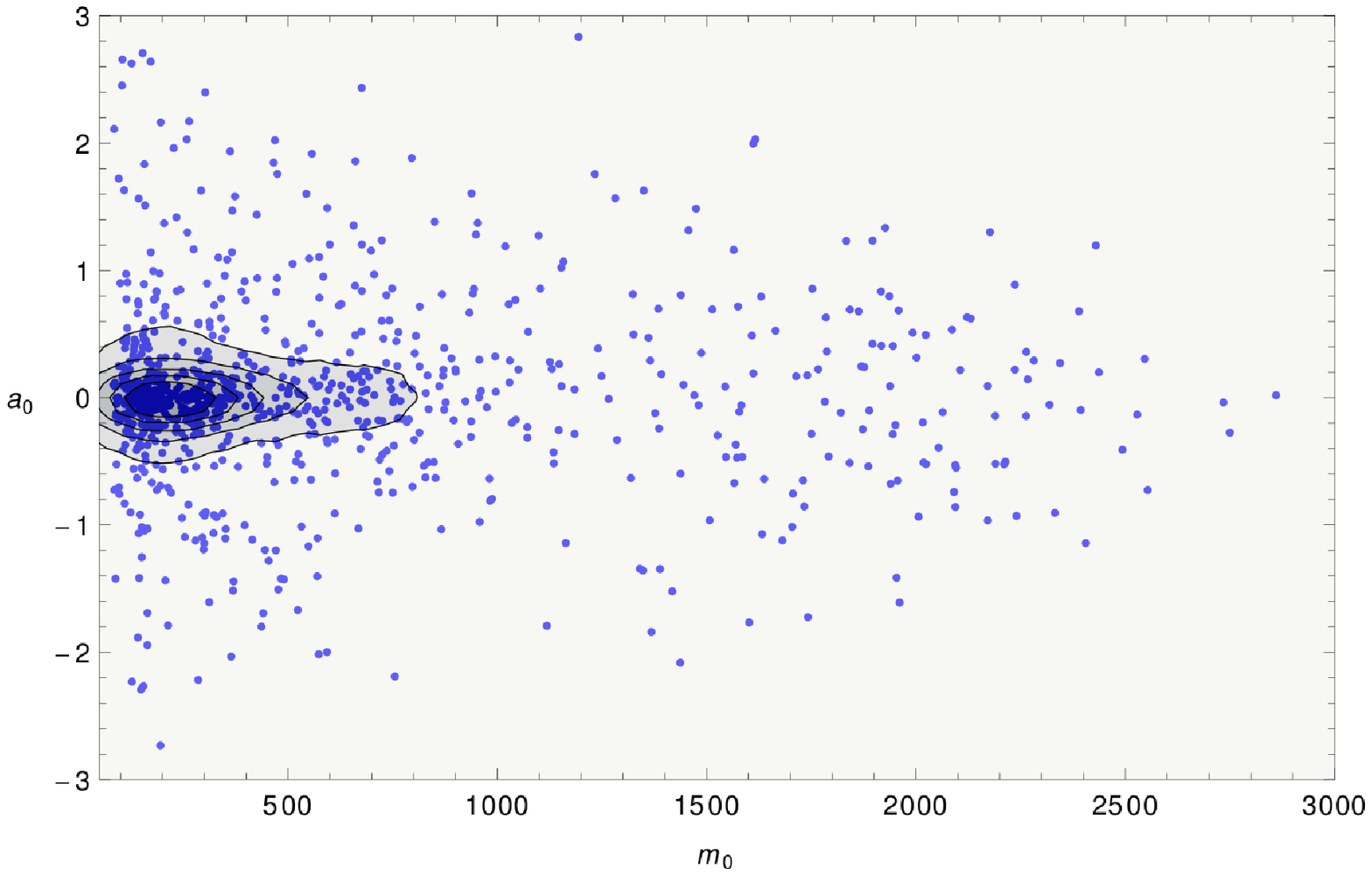} \\
$\,$ \\
\includegraphics[scale=.45]{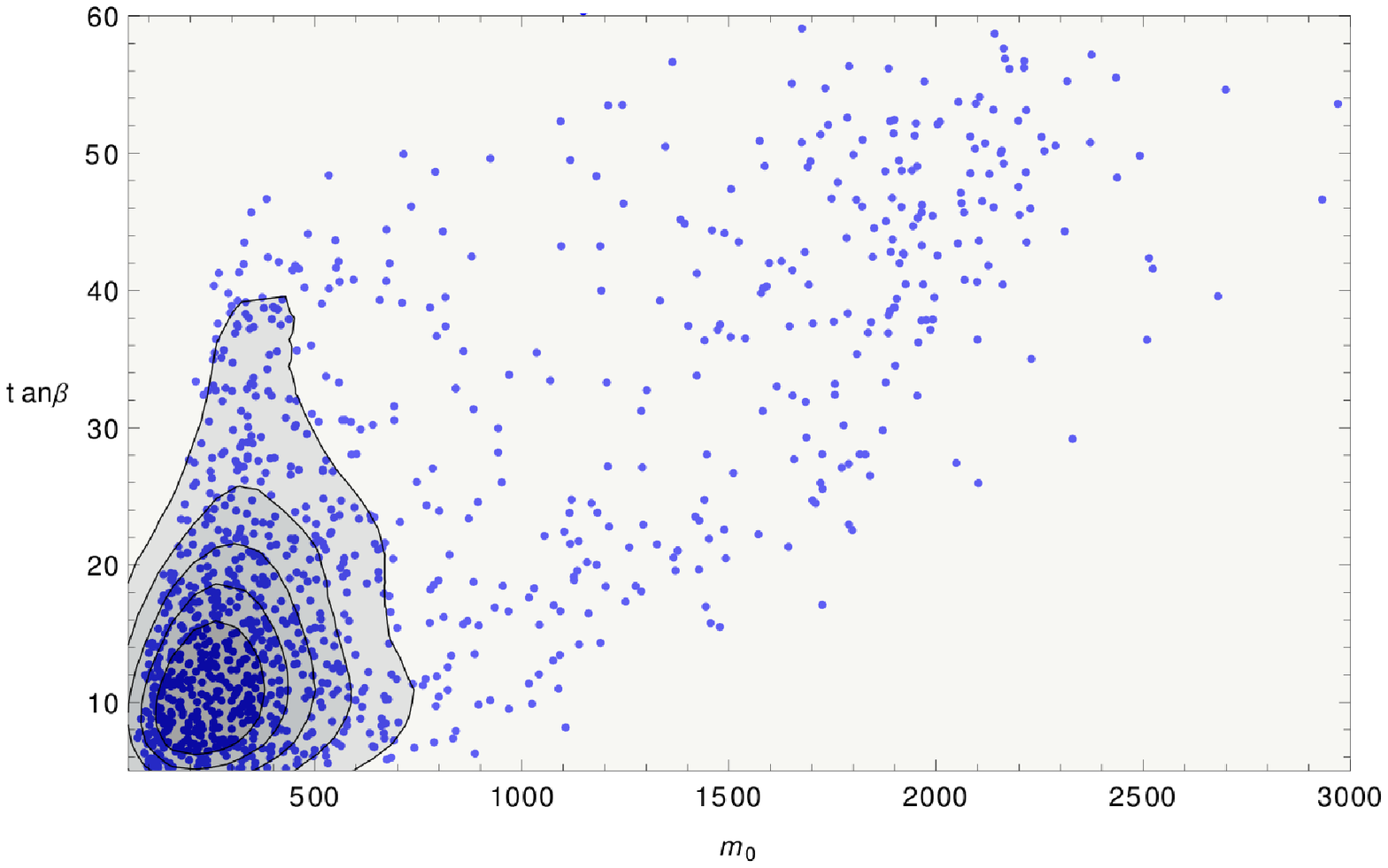} \quad \includegraphics[scale=.45]{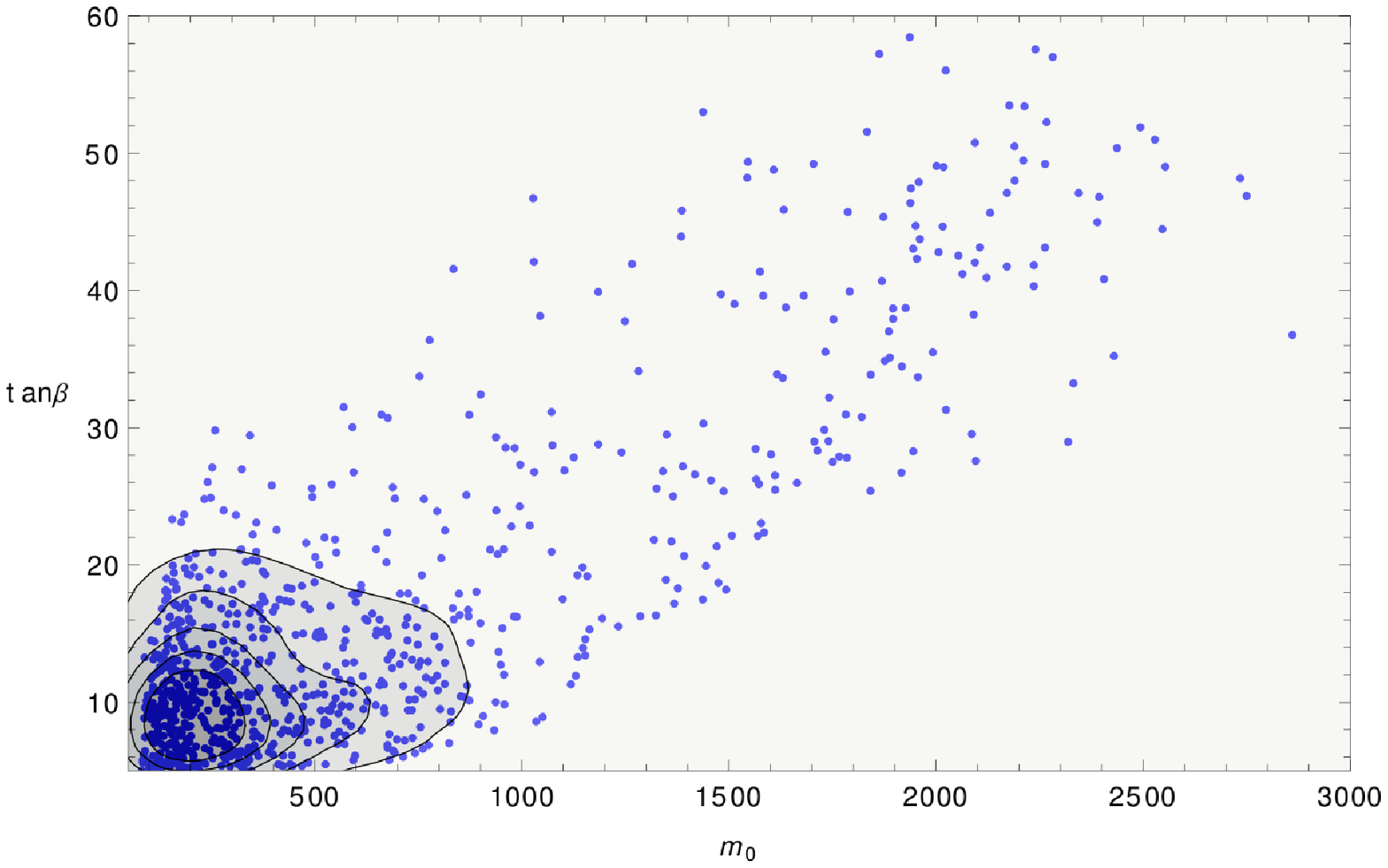} 
\caption{Area in the $m_0$--$M_{1/2}$ (top), $m_0$--$a_0$ (centre) and $m_0$--$\tan\beta$ (bottom) planes for Model 1 (left) and Model 2 (right), including CPV bounds in meson mixing and LFV constraints.}
\label{fig:indirectrvv}
\end{figure}

In Figure~\ref{fig:indirectrvv} we show the available parameter space for our example models. %The top (bottom) row shows the results for RVV1 (RVV2).
In the top row we show the $m_0$--$M_{1/2}$ plane, where we see that usually small values of $m_0$ are preferred, which due mainly to the $\epsilon_K$ constraint. The preference for small $m_0$ can be understood through the additional RG suppression factor $m_0^2/(m_0^2+6M^2_{1/2})$, which keeps $\epsilon_K$ under control. For the first example we see an additional higher-density area at larger $m_0$, which we interpret as being due to an interference between $(\delta^d_{12})_{LL}(\delta^d_{12})_{RR}$ and $(\delta^d_{12})^2_{RR}$ insertions, as explained in~\cite{Jones:2010}. Note, however, that for both models it is still possible to reach large values of $m_0$, even though they are not common.

The central row shows the preference for $a_0\rightarrow0$ in the $m_0$--$a_0$ plane. This is mainly due to the sizeable additional neutralino contribution that saturates $\mu\to e\gamma$. It must be noted, however, that as each term of the trilinears is multiplied by an arbitrary $\ord{1}$, the values of $a_0$ shown in this plot just indicate an approximate number around which the trilinears shall vary. Thus, they should not be compared to the respective plots for the CMSSM, unless $a_0=0$.

The bottom row shows the allowed points in the $m_0$--$\tan\beta$ plane. We find that, generally, small values of $\tan\beta$ shall be preferred, although large values can still be found. We see that the first model allows larger values of $\tan\beta$ than the second, which is mainly due to the fact that the latter includes a $\tan\beta$ enhancement to  $(\delta^{d,e}_{RR})_{i3}$ insertions, which can exceed the neutral meson bounds.

A valid question is whether these models survive if we demand all observables to satisfy the $2\sigma$ constraints, as we did in the CMSSM. The answer is affirmative, although the strong bounds considerably reduce the amount of points. The main result is that only those points with small $m_0$, moderate $\tan\beta$ and small $a_0$ (typically, $m_0 \lesssim 500$ GeV,
$\tan\beta\lesssim 20$ and $|a_0|\lesssim 1$) manage to survive. In particular, it is remarkable that, despite the stringent MEG and $\epsilon_K$ constraints, they can still provide a sizeable contribution to $a_\mu$, such that the tension with the experiments can be lowered below the 2$\sigma$ level.

To conclude, even though we have restricted our attention to only two models of our interest, we can say that for a Flavoured CMSSM we can expect an important interplay between collider and flavour observables. In particular, we confirm our earlier claim of~\cite{Calibbi:2008qt}, where we estimated that this combination of data would be able to fully probe the $SU(3)$ models if SUSY was light enough. Thus, although strongly constrained, these examples can still fit within an early SUSY discovery scenario, and give interesting and verifiable predictions for low-energy experiments.

\subsection{LHC Observables} 
\label{sec:lhc-obs}

In the previous sections, we have analysed possible signals in flavour observables in the region of the MSSM parameter 
space that would be observable as an excess in the jets plus missing energy 
channel at the LHC with 5 fb$^{-1}$. In this section we provide examples of the expected signals at LHC for three benchmark points 
in the ``observable'' region. 

As explained in section \ref{CMSSM-Pheno}, this observable region was estimated using the results of~\cite{Baer:2010tk}. Following that analysis, we define a point in the MSSM parameter space to be observable when the expected number of 
signal events $S$ after the cuts are,
\bea
S \geq {\rm max} [ 5 \sqrt{B}, 5 , 0.2~B ] \, ,
\eea
where  $B$ is the number of background events after the cuts. The set of cuts is optimised for each points using a grid of cuts in different variables to maximise $S/\sqrt{S+B}$~\cite{Baer:2010tk}.  The cuts we use in our optimisation procedure are the missing transverse energy, ${\slashed E_T}$, number of jets, $n({\rm jets})$, number of b-jets, $n(b)$, transverse energy of the leading jet, $E_T(j_1)$ and transverse energy of the second jet, $E_T(j_2)$. 

We select three benchmark points in the CMSSM 
parameter space (in the Seesaw, or Flavoured CMSSM, the LHC signal
would be practically identical):
\begin{enumerate}
\item Benchmark A: $m_0 = 1330$~GeV, $m_{1/2} = 270$~GeV, $\tan \beta= 55$, $A_0 =1830$~GeV.\\ $\Rightarrow$ $m_{\tilde g} = 712.6$~GeV, $m_{\tilde q_{1}} \simeq 1430$~GeV.
\item Benchmark B: $m_0 = 670$~GeV, $m_{1/2} = 385$~GeV, $\tan \beta= 22$, $A_0 =1210$~GeV. \\$\Rightarrow$ $m_{\tilde g} = 943.6$~GeV, $m_{\tilde q_{1}} \simeq 1055$~GeV. 
\item Benchmark C: $m_0 = 170$~GeV, $m_{1/2} = 535$~GeV, $\tan \beta= 14$, $A_0 =-510$~GeV.\\ $\Rightarrow$ $m_{\tilde g} = 1235.5$~GeV, $m_{\tilde q_{1}} \simeq 1140$~GeV.
\end{enumerate}
As we can see, Benchmark A corresponds to a relatively light gluino that while B and C correspond to progressively larger gluino masses. Squarks of the first generation are in all three benchmarks points heavier than 1 TeV, and in Benchmark A  $m_{\tilde{q}_1}\sim$1.4 TeV. Furthermore, only Benchmark C has 
squarks lighter than the gluino. These features determine the observable signal in the three different benchmark points. 
\begin{figure}
\includegraphics[scale=.4]{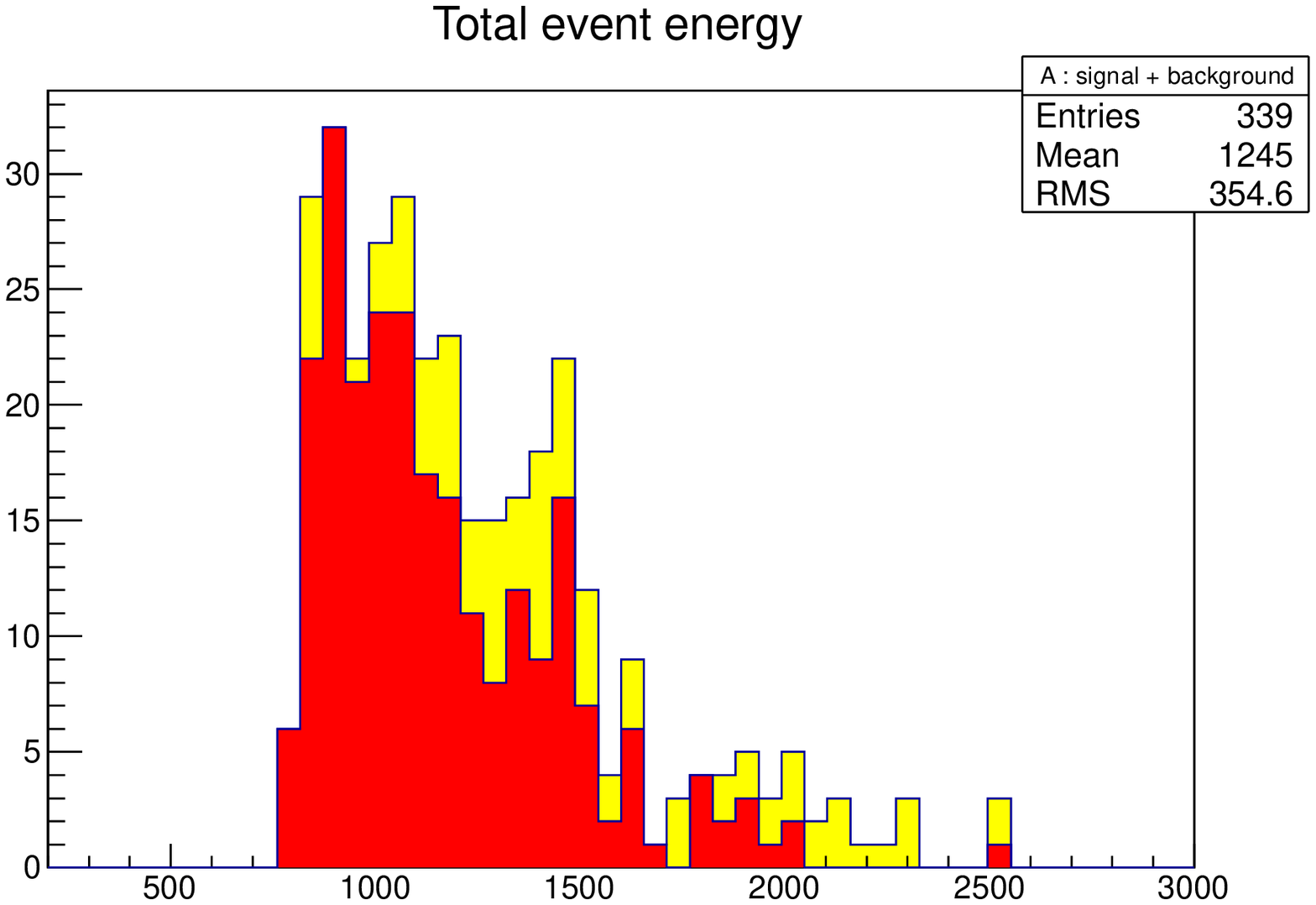} 
\includegraphics[scale=.4]{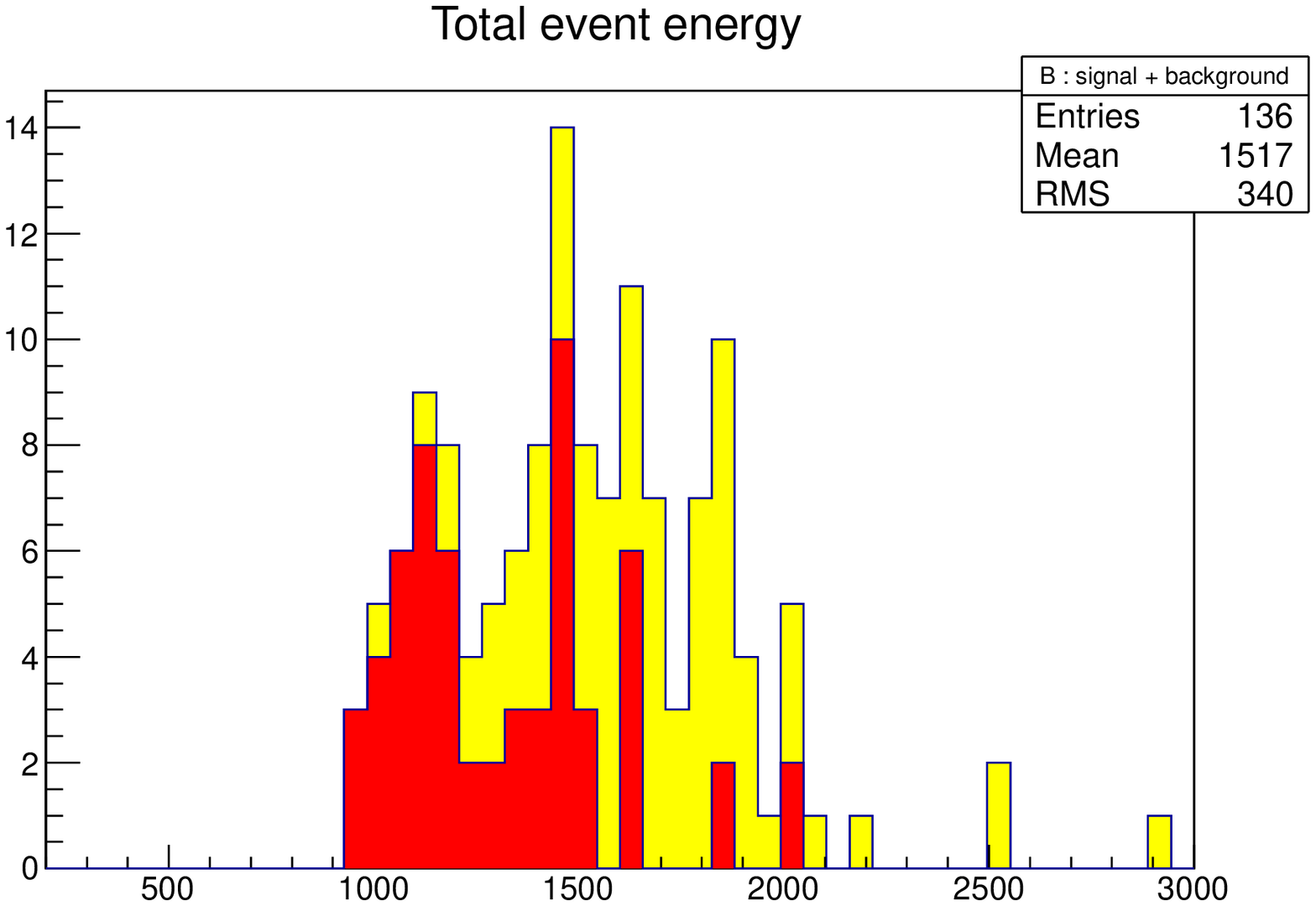}
\includegraphics[scale=.4]{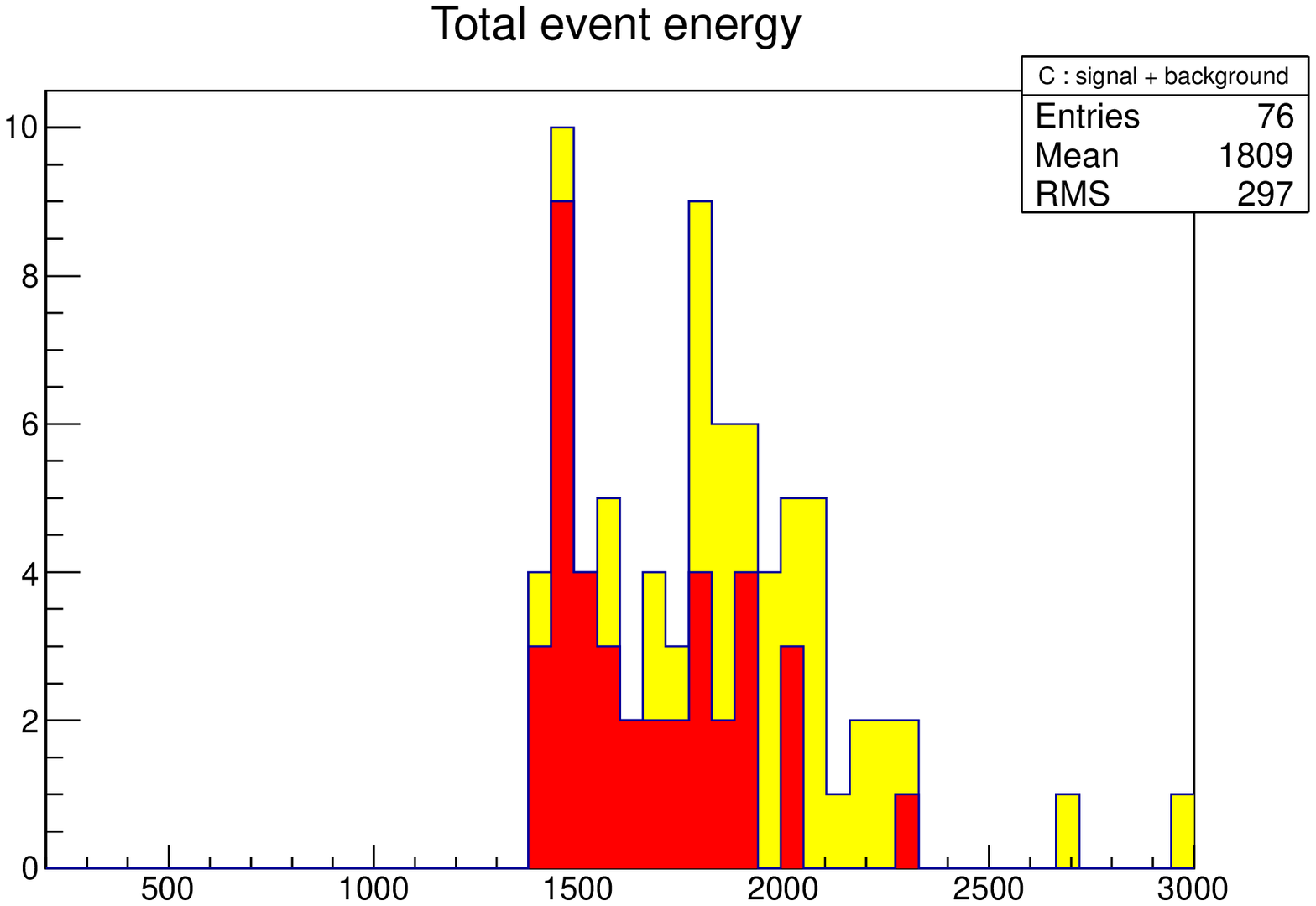}
\caption{Distribution of background and signal events as a function of $M_{\rm eff}$ after the corresponding cuts for benchmark points A (left), B (right) and C (centre).The background events are plotted in red and the signal+background events in yellow.}
\label{fig:meff}
\end{figure}

\begin{figure}
\includegraphics[scale=.4]{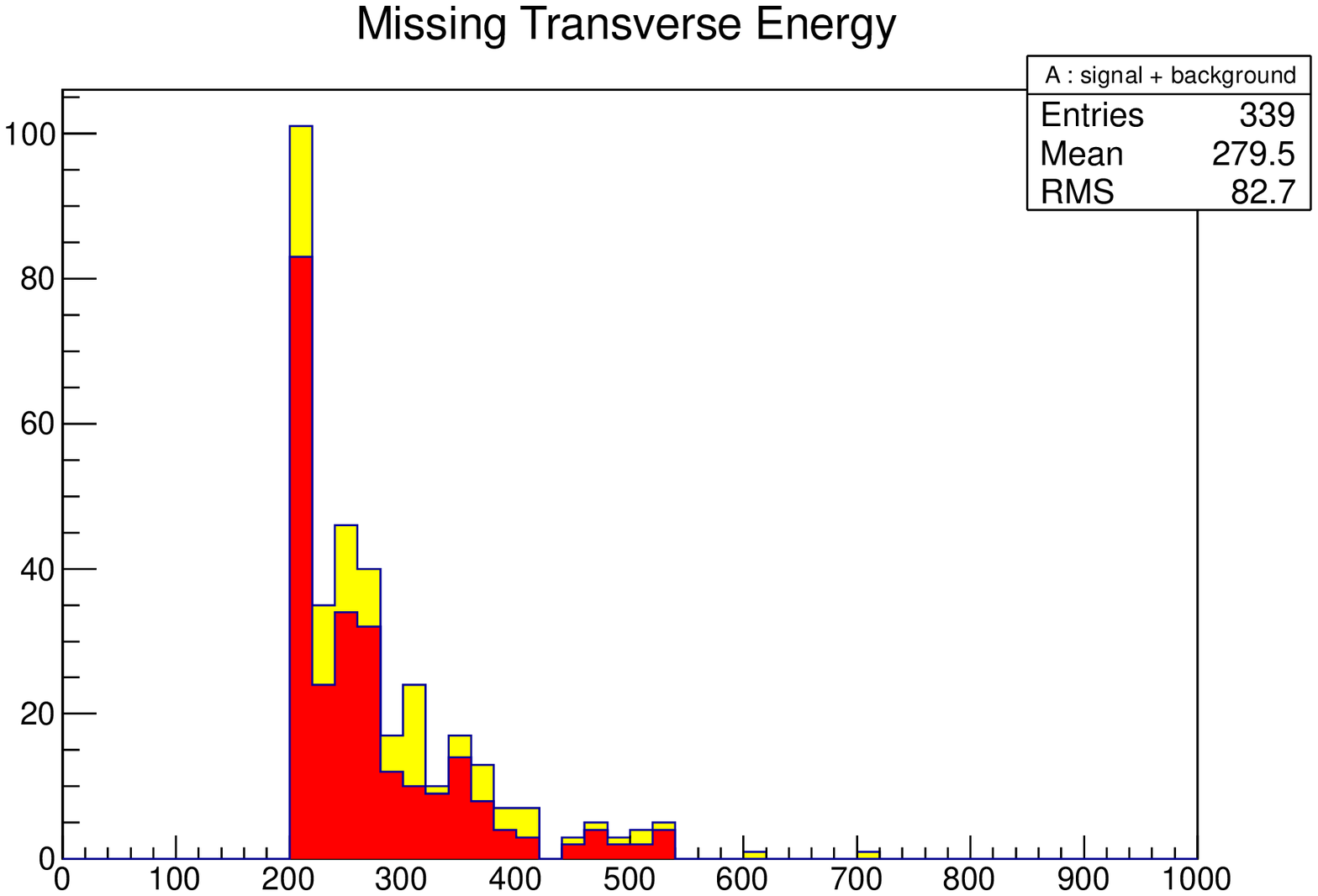} 
\includegraphics[scale=.4]{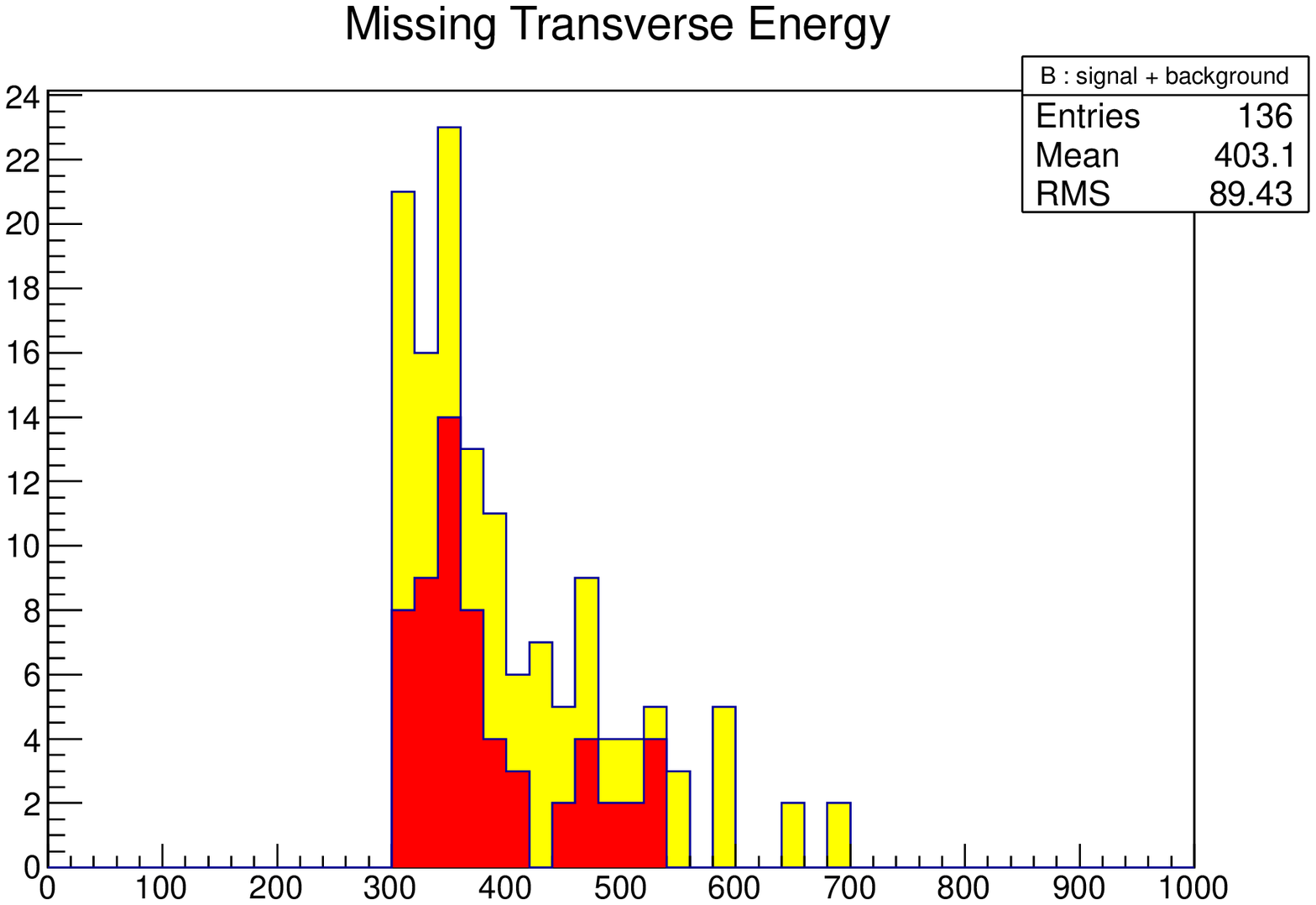}
\includegraphics[scale=.4]{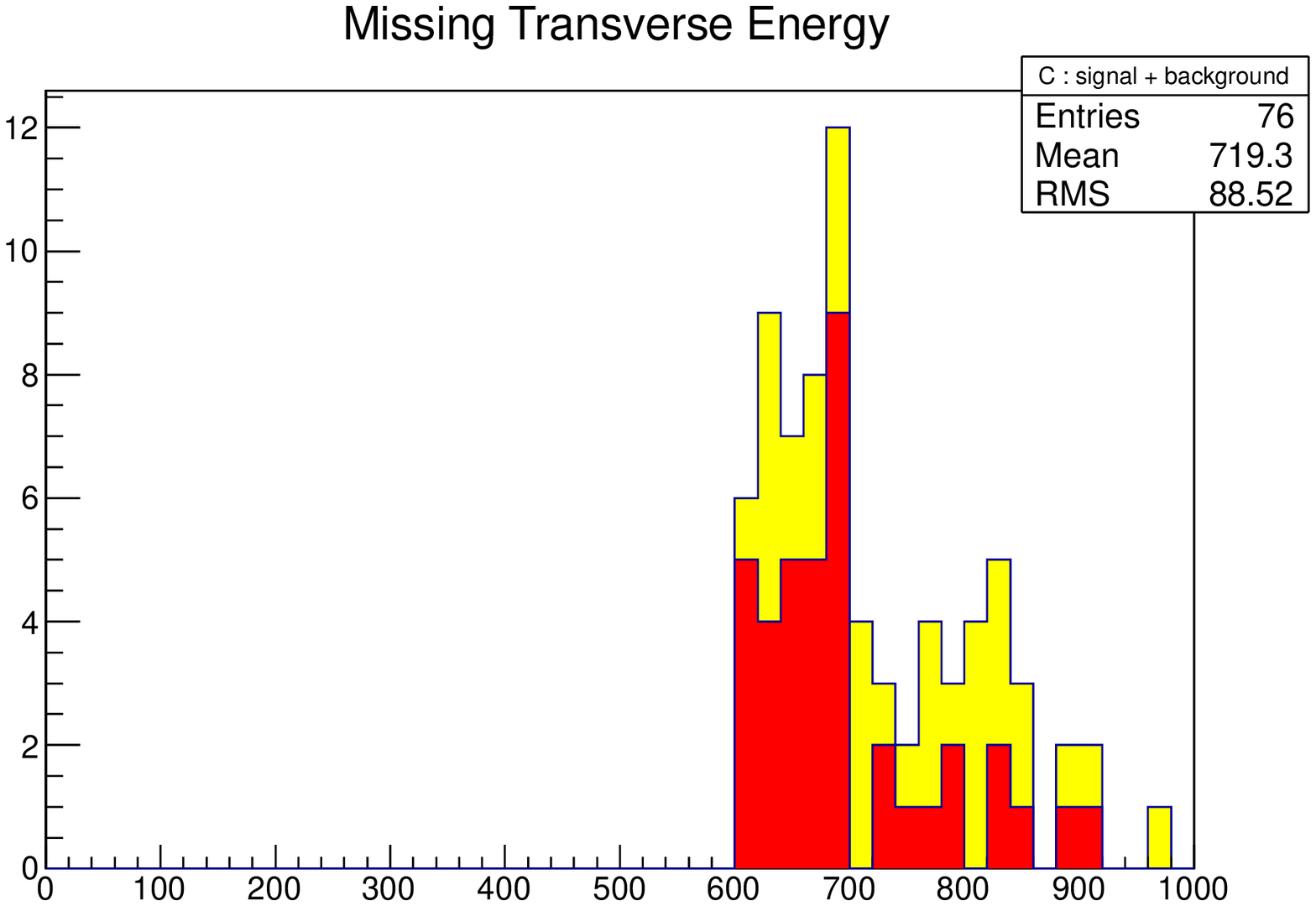}
\caption{Distribution of events as a function of $\slashed{E}_T$ after the corresponding cuts for benchmark points A (left), B (right) and C (centre).The background events are plotted in red and the signal+background in yellow.}
\label{fig:misset}
\end{figure}

\begin{figure}
\includegraphics[scale=.4]{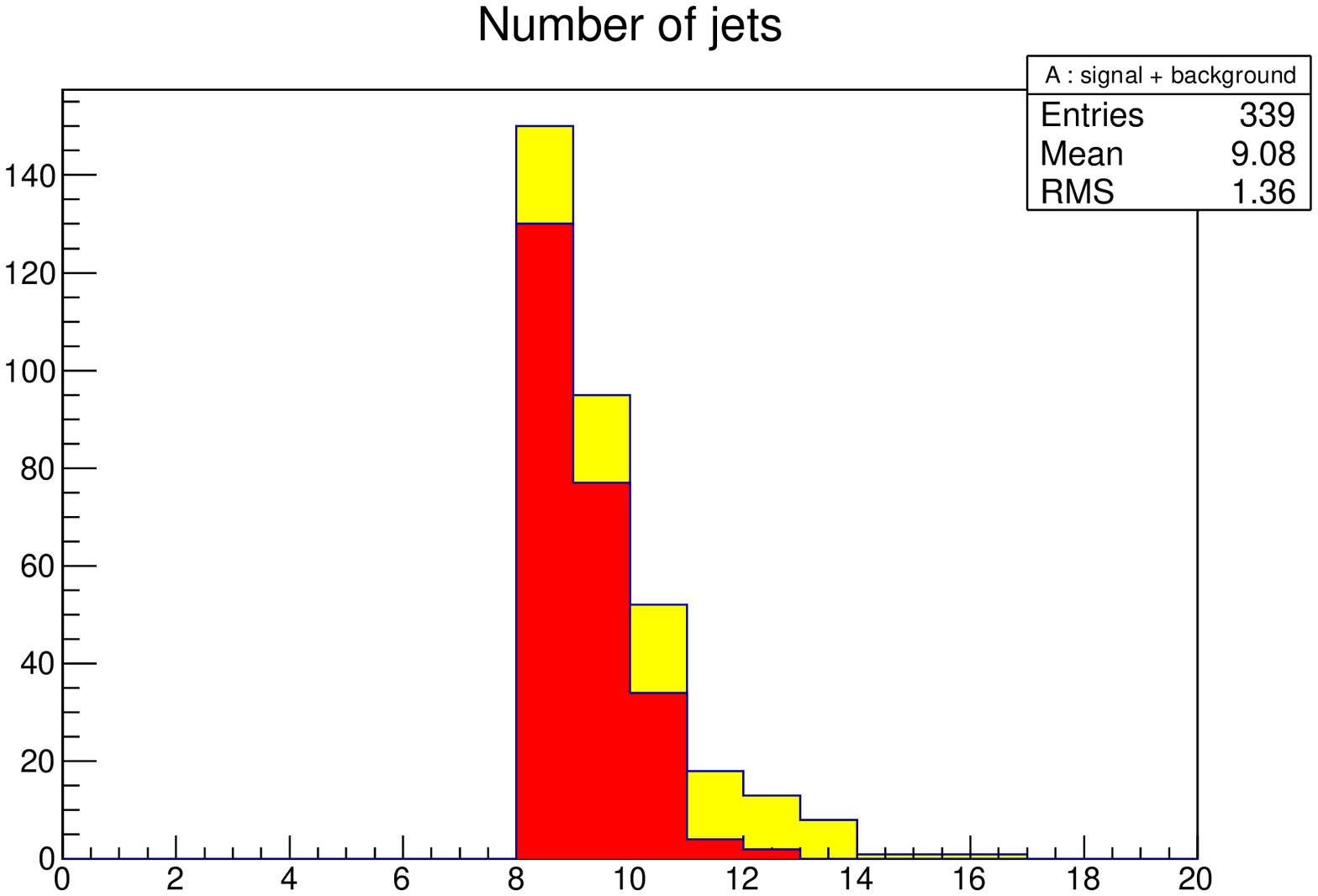} 
\includegraphics[scale=.4]{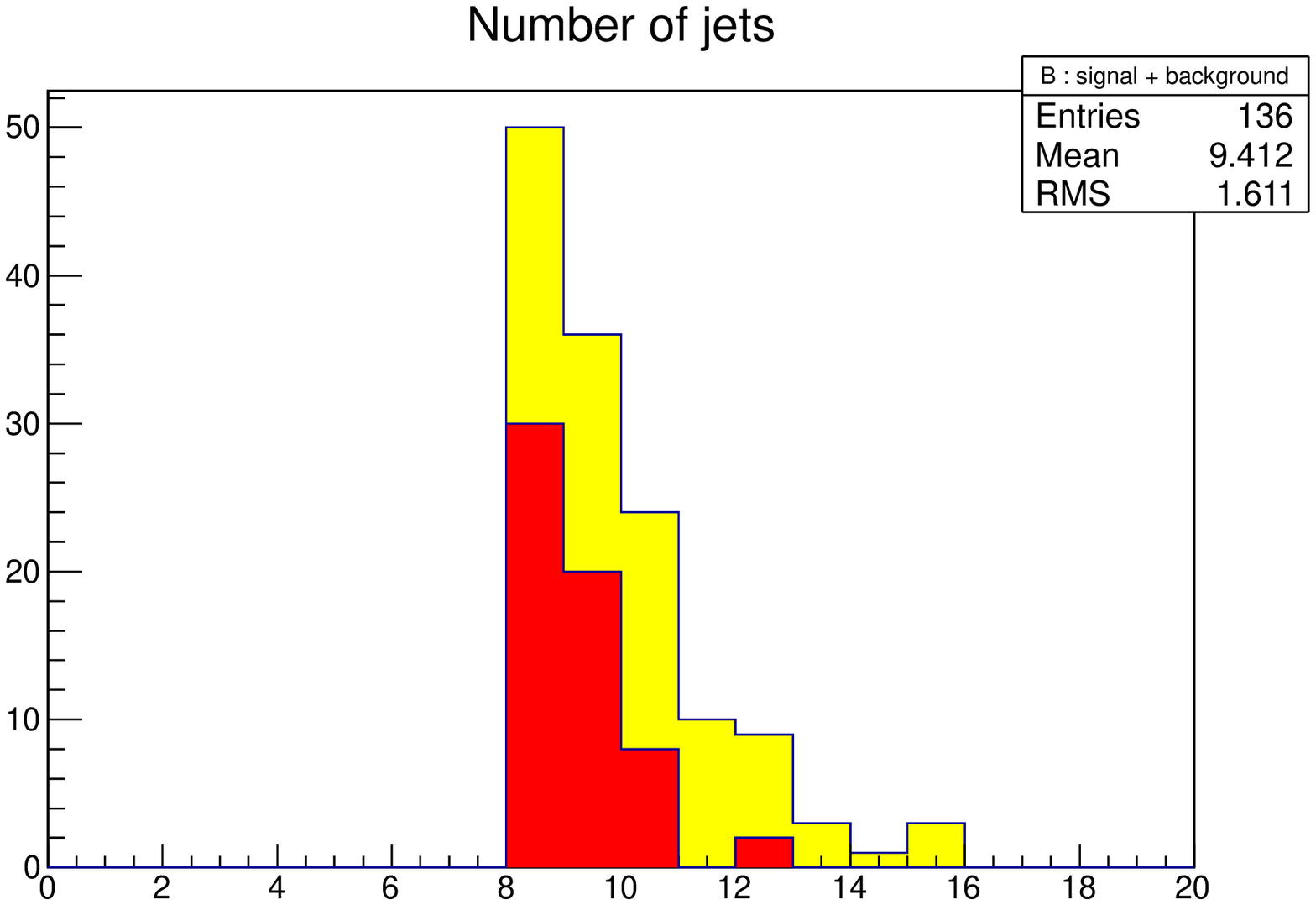}
\includegraphics[scale=.4]{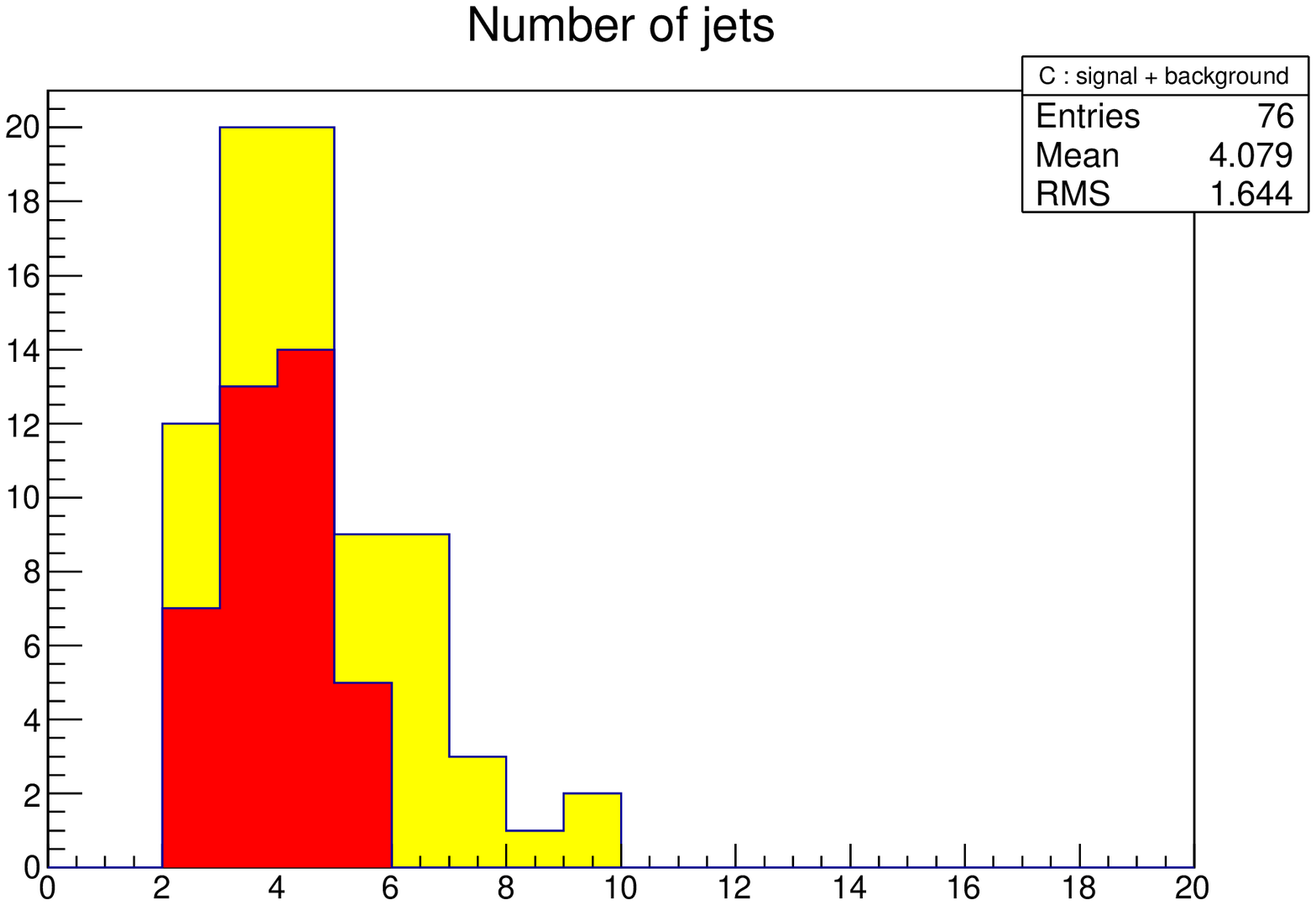}
\caption{Distribution of events as a function of $n({\rm jets})$ after the corresponding cuts for benchmark points A (left), B (right) and C (centre).The background events are plotted in red and the signal+background in yellow.}
\label{fig:jetnumb}
\end{figure}
  
We simulate the signal and background for these points using MadGraph \cite{Alwall:2011uj} and PYTHIA \cite{Sjostrand:2006za} and the detector effects with PGS \cite{PGS}. The simulated signal and background corresponds to a full 5 fb$^{-1}$ simulation.
In this framework, we choose, for each point, the set of cuts on $\slashed{E}_T$, $n({\rm jets})$, $n(b)$,  $E_T(j_1)$ and $E_T(j_2)$, (making a grid of the possible values of the cuts as in \cite{Baer:2010tk}) that maximise the significance  $S/\sqrt{S+B}$. Using this set of constraints, we are not making any requirement on the total energy, $M_{\rm eff}$, of the event. In a second round, we use the minimum value of $M_{\rm eff}$ that we obtain for signal and background in the first round to set a new cut on $M_{\rm eff}$ and choose again the values of the aforementioned cuts that maximise the significance, taking now into account the minimum value of $M_{\rm eff}$. Notice that adding this new cut on $M_{\rm eff}$, would have no effect on the points selected by the previous set of constraints, but now, we are scanning again the constraints above in the subset of points with a minimum value of $M_{\rm eff}$ and choosing a new set of cuts that maximise $S/\sqrt{S+B}$. The final result of the number of $S$, $B$, significance and the set of applied cuts for each point are the following:
\begin{enumerate}
\item Cuts point A: $\slashed{E}_T \geq 200$~GeV, $n({\rm jets}) \geq 8$, $n(b) = 0$, $E_T(j_1)\geq 100$~GeV, $E_T(j_2)\geq 80$~GeV and $M_{\rm eff} \geq 800$~GeV. $\Rightarrow$ $S = 92$, $B = 247$, Signif.= 5.0
\item Cuts point B: $\slashed{E}_T \geq 300$~GeV, $n({\rm jets}) \geq 8$, $n(b) = 0$, $E_T(j_1)\geq 50$~GeV, $E_T(j_2)\geq 50$~GeV and $M_{\rm eff} \geq 900$~GeV. $\Rightarrow$ $S = 76$, $B = 60$, Signif.= 6.5
\item Cuts point C: $\slashed{E}_T \geq 600$~GeV, $n({\rm jets}) \geq 2$, $n(b) = 0$, $E_T(j_1)\geq 300$~GeV, $E_T(j_2)\geq 110$~GeV and $M_{\rm eff} \geq 1400$~GeV. $\Rightarrow$ $S = 37$, $B = 39$, Signif.= 4.2 
\end{enumerate}
In Figures \ref{fig:meff}, \ref{fig:misset} and  \ref{fig:jetnumb}, we plot the distribution of background and signal events in $M_{\rm eff}$, $\slashed{E}_T$ and $n({\rm jets})$ for all three points after applying the corresponding cuts. Although the statistics is still limited for the three points, these Figures give
an idea of the expected signal at LHC with 5 fb$^{-1}$.

\begin{table}[t]
\centering
\begin{tabular}{|c|ccc|}
\hline
  &  Point A & Point B & Point C \\
\hline
$\sigma_{\tilde{g}\tilde{g}}$&  155 & 10 & 0.5 \\  
$\sigma_{\tilde{q}\tilde{q}}$&  1.5 & 28 & 14 \\  
$\sigma_{\tilde{q}\tilde{q}^*}$&  0.1 & 4 & 2 \\  
$\sigma_{\tilde{q}\tilde{g}}$&  33 & 43 & 7.5 \\  
$\sigma_{\tilde{t}\tilde{t}^*}$&  0.003 & 0.1 & 0.05 \\  
$\sigma_{\tilde{b}\tilde{b}^*}$&  0.003 & 0.1 & 0.05 \\  
\hline
$\sigma_{\rm tot}$ & 190 & 85 & 24 \\
\hline
\end{tabular} 
\caption{SUSY production cross-section (expressed in fb) for different channels. \label{tab:xsec}} 
\end{table}
From the analysis of $M_{\rm eff}$ in Figure~\ref{fig:meff}, we see that in points A and B signal plus background have a broad distribution of events centred roughly at $M_{\rm eff}\sim 1500$~GeV and $M_{\rm eff}\sim 1500$~GeV respectively and for point C the events are centred around $M_{\rm eff}\sim 2000$~GeV. However, it is hard to make a more precise statement on the value of $M_{\rm eff}$ due to the limited statistics. In fact, the excess of events above the expected background is of order 100 events for points A and B and of order 30 for point C. For these signal events, points A and B are hard to distinguish in this plot. From Figures~\ref{fig:misset} and  \ref{fig:jetnumb}, we see that points A and B are still very similar and it is difficult to distinguish them with these observables and limited statistics. In Figure~\ref{fig:jetnumb}, we can see that the jet number is very different for point C and the rest. In points A and B, we have on average a large number of jets, while point C has a smaller number of jets. Although, this is partly an effect of the imposed cut on number of jets ($n({\rm jets}) \geq 8$ for A and B and $n({\rm jets}) \geq 2$ for C), it is clear that we have events in A and B with a much larger number of jets than for point C. This is a clue on the nature of the coloured sparticles produced in the collision. We must take into account that the main SUSY production channels are either two gluinos, two first-generation squarks or a gluino plus a first generation squark. Clearly, if the event consists in the production of a pair of gluinos, we will have on average two additional jets with respect to the production of two squarks and analogously for the case of production of a squark and a gluino. Therefore, from this, we would expect that points A and B correspond mainly to the production of gluino pairs while point C, where there is a significant excess in two and three-jet events, this excess must correspond to the production of a pair of first generation squarks 
(cfr.~the SUSY production cross-sections for our three benchmark points, as computed by PROSPINO 2.0~\cite{prospino}, in table~\ref{tab:xsec}).
Given that this two and three-jet excess is a significant part ($\sim 1/3$) of total excess of events, we would conclude that the production of first generation squarks in this point is important and the masses of first generation squarks are probably lighter or at least of the order of the gluino mass. 
Although it is difficult to make quantitative statements, using the approximate relation $M_{\rm eff}\sim 1.6~ M_{\rm SUSY}$, with $M_{\rm SUSY}$ the lightest coloured particle mass, $M_{\rm SUSY} = {\rm min}(m_{\tilde g},m_{\tilde q_1})$, we would roughly expect that $m_{\tilde g}\sim 940$ GeV for points A and B and  $m_{\tilde q_1}\sim 1250$ GeV for point C.
Finally, the analysis of the missing $E_T$ in Figure~\ref{fig:misset} provides also some very interesting information on the mass splitting from the initially produced coloured sparticle and the LSP. The largest possible missing energy corresponds to the two LSPs carrying away, each of them, one half of the mass splitting between the initially produced sparticle and the LSP. Therefore, for the points A and B, where the maximum missing $E_T$ is of order of 700 GeV, we would conclude that the mass difference between the gluino and the lightest neutralino is 700 GeV, and thus we could roughly estimate $m_{\tilde{\chi}^0_1} \sim 240$ GeV. Similarly, for point C, where the maximum missing $E_T$ is 900 GeV, we would estimate a lightest neutralino mass of $m_{\tilde{\chi}^0_1} \sim 350$ GeV.  

Even though we can obtain much information from our analysis, we see that it is very difficult to distinguish points A and B with collider observables at 5 fb$^{-1}$. 
At this point, one can turn to flavour for more information. For instance, from the previous sections, we know that the measurement of BR$(B_s \to \mu^+ \mu^-)$ can separate regions in the parameter space. In fact, a 2 fb$^{-1}$ search at LHCb would show a $3\sigma$ evidence for this decay in point A, while leaving point B and C consistent with the SM expectations. Thus, $B_s\to\mu^+\mu^-$ can favour one region of the parameter space over another in light of collider data.

\begin{table}[tbp]
\renewcommand{\arraystretch}{1.3}
 \begin{center}
\begin{tabular}{|c||c|c|c|c|c|}
\hline
Benchmark & BR$(b\to s\gamma)$ & $\delta a_\mu$ & BR$(B_d\to\mu^+\mu^-)$ & BR$(B_s\to\mu^+\mu^-)$ & $R(B^+\to\tau^+_\nu)$ \\
\hline
A & $3.00\times10^{-4}$ & $1.06\times10^{-9}$ & $2.01\times10^{-10}$ & $6.36\times10^{-9}$ & $0.60$ \\
B & $2.89\times10^{-4}$ & $1.05\times10^{-9}$ & $1.20\times10^{-10}$ & $3.80\times10^{-9}$ & $0.97$ \\
C & $2.92\times10^{-4}$ & $1.07\times10^{-9}$ & $1.21\times10^{-10}$ & $3.81\times10^{-9}$ & $0.99$ \\
\hline
 \end{tabular}
 \end{center}
\caption{Central values for some flavour observables in each benchmark point. All points satisfy $2\sigma$ constraints once experimental and theoretical errors are included.}
\label{tab:bench}
\end{table}

Moreover, flavour provides more tools than just point differentiation. The flavour phenomenology can be re-introduced to the collider observations as a way of roughly testing the coherence of our main assumption, namely, that the SUSY spectrum is close to that described by the CMSSM. Thus, it is possible to provide some ``flavour feedback'' to colliders.

For instance, requiring the $2\sigma$ flavour constraints on $b\to s\gamma$ and $(g-2)_\mu$ to be satisfied (which is true for all benchmark points, as seen in Table~\ref{tab:bench}), evidence for $B_s\to\mu^+\mu^-$ leads us to favour relatively large values of $\tan\beta$, $A_0$ and $m_0$, as was shown in Figure~\ref{fig:sigma2}. Combining this with the rough $M_{\rm eff}$ contours, one can expect to obtain some information on $M_{1/2}$. For instance, in point A, as the large $m_0$ leads to large squark masses, we would require a light gluino to be responsible for the $M_{\rm eff}$ measurement, as was made evident from our collider analysis. This gives us direct information on the value of $M_{1/2}$.

Once this information is obtained, one can then check the consistency with rest of the collider information. As we have mentioned, the fact that the gluino is the dominant source of the excess in jets has a direct repercussion on the number of jets we should observe, and this should be consistent with both $M_{\rm eff}$ and flavour data. Also, the endpoint of the ${\slashed E_T}$ distribution would give us an idea of the splitting between this gluino and the LSP, which tests directly the assumption of gaugino universality.

On the other hand, we could have no evidence whatsoever for $B_s\to\mu^+\mu^-$. In this case, inspection of Figure~\ref{fig:sigma2} does not allow us to make such strong statements as those done for point A. We find that larger values of $M_{1/2}$ are favoured, but we cannot expect to be able to reach a conclusion regarding $m_0$, unless we're in a very special scenario with very large or very low $M_{\rm eff}$.

Still, even if $B_s\to\mu^+\mu^-$ is not observed, the interplay with $b\to s\gamma$ and $(g-2)_\mu$ does give information about $\tan\beta$ and $A_0$. For example, if $M_{\rm eff}$ is low enough, one can determine the sign of $A_0$, as well as give upper and lower bounds on $\tan\beta$. This is the case in point B. For too high $M_{\rm eff}$, one only obtains an upper bound on $m_0$ due the the LHC reach, and from this it is possible to establish an upper bound on $\tan\beta$. Nevertheless, it shall be unfeasible to carry out any self-consistency tests with only this amount of information.

Moreover, notice that the different models considered in this paper, the CMSSM, the Seesaw CMSSM or the Flavoured CMSSM, have a very similar spectrum and would give identical information in collider observables. Flavour here becomes again useful, as one can take this information and go to indirect searches, mainly $\mu\to e\gamma$, neutral meson mixing and electric dipole moments. Clearly, if a positive signal is found in any of these indirect observables, the CMSSM has to be abandoned and we have to increase the number of parameters in the model.

\section{Conclusions}
\label{conclusion}

We have discussed the interplay between LHC and flavour and CP violation experiments in testing supersymmetric models.  
Under the assumption that a hint for SUSY particles will be indeed found at LHC7 (after analysing 5 fb$^{-1}$ of data), 
and taking into account the exclusion bounds already provided by ATLAS and CMS,
we have studied the consequent SUSY predictions for flavour and CPV observables. This analysis has been performed for a set of phenomenologically motivated SUSY models,
namely the CMSSM, a SUSY seesaw and a Flavoured CMSSM, i.e.~an extension of the CMSSM with non-trivial flavour structures controlled by the 
same dynamics responsible for the SM fermion masses and mixing. In particular, we focused on the capability of flavour experiments to discriminate
among different models and to constrain or give information on the parameters of a given model.  
The outcome of our study can be summarised as follows.
\begin{itemize}
\item As expected, LHC experiments and flavour observables are complementary in probing the SUSY parameter space.
      In particular, the imposition of flavour constraints (especially BR($b\to s\gamma$) and $(g-2)_\mu$) 
      can give information on the SUSY parameters that are not directly constrained by jets plus $\slashed E_T$ searches at LHC (such as $A_0$ and $\tan\beta$). 
\item Positive or negative results of $B_s\to \mu^+\mu^-$ searches at LHCb and
      CMS can further disentangle different regimes of the parameters, in some cases selecting very restricted allowed values. 
      In general $B_s\to \mu^+\mu^-$ seems to be a crucial test for SUSY models, in case non-standard signals are observed at LHC.
\item The interplay of LHC and the MEG experiment will be crucial to probe SUSY seesaw scenarios, providing indirect information on the neutrino Yukawa mixing
      and/or the RH neutrino mass scale. In particular large-mixing, large-Yukawa scenarios are already ruled out by the recent MEG limits on BR($\mu\to e\gamma$).
      Neutrino oscillation and $(g-2)_\mu$ experiments can further increase our capability to access the seesaw parameters and
      thus to test indirectly very high-energy scales.
\item In some simple seesaw scenarios, the preferred range for $U_{e3}$ recently reported by T2K implies small rates for the $\tau\to \mu \gamma$ decay. 
      Therefore, such scenarios can be excluded by an evidence for $\tau\to \mu \gamma$ at the Super B factories.
\item The Flavoured CMSSM we considered is still a viable model for addressing the SM and the SUSY flavour problems at the same time. Interesting    correlations
      among different observables (such as $\epsilon_K$ and $\mu\to e\gamma$) tend to restrict the allowed parameter space also in this case. LFV and EDM experiments
      will completely test the model at least within the region accessible at LHC7 with 5 fb$^{-1}$, providing an important cross-check of the LHC findings. 
      On the other hand, a large phase of the $B_s$ mixing would disfavour this kind of scenarios.
\item The observability at LHC7 of the parameter space region we studied and the features of the signals have been checked for three benchmark points
      in section~\ref{sec:lhc-obs} by means of numerical simulations. We have shown that set-ups which provide similar signatures at LHC7 can
      be actually distinguished by means of flavour observables (especially $B_s\to \mu^+\mu^-$) and the interplay between collider and flavour
      signatures can provide useful information in the attempt of determining the fundamental parameters of the model.      
\end{itemize}

As a final conclusion, although CMSSM-like models are currently considered somewhat disfavoured in light of the latest ATLAS and CMS results, we find that they can still provide very interesting phenomenology and correlations in both the collider and flavour sectors. The current run of the LHC is still exciting for SUSY, and it may be possible to find surprises just around the corner.

\acknowledgments
We thank Massimo Passera and Jonas Lindert for useful discussions, Werner Porod for the assistance provided with SPheno, and Joachim Brod for pointing out a mistake in Table~\ref{tab:mesonpar}. L.C., J.J.P., A.M.~and O.V.~thank 
the Galileo Galilei Institute for Theoretical Physics for the hospitality and the INFN for partial support during the workshop that gave birth to this work. 
L.C., J.J.P., R.N.H.~and O.V.~thank the Universit\`a di Padova for the hospitality and support provided during their visit. L.C. and J.J.P. are grateful 
to the Universitat de Val\`encia for hospitality and financial support during several stages of this work. J.J.P.~would also like to thank the 
Max-Planck-Institut f\"ur Physik for the hospitality and support provided during his visit. A.M. acknowledges the  contribution of the CARIPARO Excellence Project "LHC and Cosmology". We acknowledge further support from the MICINN-INFN agreements 
ACI2009-1049 and INFN2008-016. O.V. and R.N.H. acknowledge  partial support by MEC and FEDER (EC), Grants No. FPA2008-02878 and FPA2011-23596 and by 
the Generalitat Valenciana under the grant PROMETEO/2008/004.

\end{document}